\documentclass[12pt,oneside]{book}
\usepackage{amsmath}
\usepackage{amsthm}
\usepackage{graphics}
\usepackage{epsfig,amssymb,latexsym,verbatim}
\usepackage{graphicx}
\usepackage{multirow}
\usepackage{multicol}
\usepackage{float}
\usepackage{booktabs}
\usepackage{fancyhdr}
\usepackage{enumerate,verbatim}
\usepackage[sort&compress]{natbib}
\usepackage{rotating}
\usepackage{hyperref}
\usepackage{subfigure}
\usepackage{ulem}
\usepackage{url}
\usepackage{threeparttable}
\usepackage{xr}
\usepackage{multirow}
\usepackage{multicol}
\usepackage{wrapfig}
\usepackage{lscape}
\usepackage{rotating}
\usepackage{epstopdf}
\usepackage{subfigure}

\usepackage{amsfonts}
\usepackage[usenames,dvipsnames]{color}
\usepackage{times,color}
\usepackage{amsbsy}
\usepackage{amsthm}
\usepackage[sf,small,center, compact]{titlesec}
\usepackage[lined,algonl]{algorithm2e}

\definecolor{red}{rgb}{1,0,0}
\definecolor{green}{rgb}{0,1,0}
\definecolor{blue}{rgb}{0,0,1}

\def\red{\textcolor{red}}
\definecolor{lxy}{RGB}{180,0,180}

\floatstyle{ruled}
\newfloat{algorithm}{tbp}{loa}
\floatname{algorithm}{Algorithm}
\newcommand{\thetheorem}{{\thesection. \arabic{theorem}}}
\newcommand{\thelemma}{{\thesection. \arabic{lemma}}}
\newcommand{\theproposition}{{\thesection. \arabic{proposition}}}

\begin{document}

\renewcommand{\baselinestretch}{1.2}

\markboth{\hfill{\footnotesize\rm  G. Wang, Z. Gu, X. Li, S. Yu, M. Kim, Y. Wang, L. Gao, and L. Wang}\hfill}
{\hfill {\footnotesize\rm } \hfill}

\renewcommand{\thefootnote}{}

\fontsize{10.95}{14pt plus.8pt minus .6pt}\selectfont
\vspace{0.8pc} \centerline{\Large\bf Comparing and Integrating US COVID-19 Data}
\centerline{\Large\bf  from Multiple Sources with Anomaly Detection}
\centerline{\Large\bf and Repairing} \vspace{.4cm}
\centerline{
Guannan Wang$^{a}$, Zhiling Gu$^{b}$, Xinyi Li$^{c}$, Shan Yu$^{d}$, Myungjin Kim$^{b}$,} 
\centerline{Yueying Wang$^{b}$, Lei Gao$^{b}$, and Li Wang$^{b}$}

\vspace{.4cm}
\centerline{\it $^{a}$College of William \& Mary, USA, $^{b}$Iowa State University, USA} 
\centerline{\it $^{c}$Clemson University, USA}
\centerline{\it and $^{d}$University of Virginia, USA}
\vspace{.55cm} \fontsize{9}{11.5pt plus.8pt minus .6pt}\selectfont
\footnote{\emph{Address for correspondence}: Li Wang (lilywang@iastate.edu) 
}

\begin{quotation}
\noindent \textit{Abstract:} Over the past few months, the outbreak of Coronavirus disease (COVID-19) has been expanding over the world. A reliable and accurate dataset of the cases is vital for scientists to conduct related research and for policy-makers to make better decisions. We collect the United States COVID-19 daily reported data from four open sources: the New York Times, the COVID-19 Data Repository by Johns Hopkins University, the COVID Tracking Project at the Atlantic, and the USAFacts, then compare the similarities and differences among them. To obtain reliable data for further analysis, we first examine the cyclical pattern and the following anomalies, which frequently occur in the reported cases: (1) the order dependencies violation, (2) the point or period anomalies, and (3) the issue of reporting delay. To address these detected issues, we propose the corresponding repairing methods and procedures if corrections are necessary. In addition, we integrate the COVID-19 reported cases with the county-level auxiliary information of the local features from official sources, such as health infrastructure, demographic, socioeconomic, and environmental information, which are also essential for understanding the spread of the virus.

\vspace{9pt} \noindent \textit{Key words and phrases:} 
Anomaly detection; 
Coronavirus; 
Count time series; 
Data comparison; 
Data integration; 
Outlier correction.
\end{quotation}

\fontsize{10.95}{14pt plus.8pt minus .6pt}\selectfont

\thispagestyle{empty}
\setcounter{chapter}{1} \setcounter{equation}{0} 
\renewcommand{\theequation}{\arabic{equation}} 
\renewcommand{\thesection}{\arabic{section}}
\setcounter{section}{0} 

\section{Introduction}
\label{SEC:intro}
Since the first infected case reported in December 2019, the outbreak of Coronavirus disease (COVID-19) has unfolded across the globe. In the US, coronavirus has infected more than five million people and killed over 160,000 people, as of the time of writing. While essential public health, economic and social science research in measuring and modeling COVID-19 and its effects is underway, reliable and accurate datasets are vital for scientists to conduct related research and for governments to make better decisions \cite{killeen2020county}. Unfortunately, errors could occur in the data collection process, especially under such a pandemic. In this work, we focus on the data collection, comparison, data inconsistency detection, and the corresponding curating.

Living through unprecedented times, governments must rely on timely, reliable data to make decisions to mitigate harm and support their citizens. Every day, several volunteer groups and organizations work very hard on collecting data on COVID-19 from all the counties and states in the US. There are four primary sources, including (1) the New York Times (NYT) \cite{NYT:20}, (2) the COVID Tracking Project at the Atlantic (Atlantic) \cite{COVIDTrack}, (3) the data repository by the Center for Systems Science and Engineering (CSSE) at Johns Hopkins University (JHU) \cite{JHUCSSE}, and (4) USAFacts \cite{USAFact:20}. Although these sources usually obtain their confirmed infectious and death cases data from the government agencies, the counts still vary due to the time of their collection as well as several other issues. However, these differences can be critical for real-time analysis. In this work, we first collect and compare the COVID-19 daily reported data from the above four open resources. 

We observe a seven-day cyclical pattern for daily new cases and new deaths at the state and national level in the US. To test if the observed patterns are not accidental, we conduct a seasonality hypothesis test at the county, state and national level for the infected and death count time series from March 15 to July 25, 2020.

The COVID-19 data pose unique data quality challenges due to its spatiotemporal nature, and the problem of delayed-reporting and under-reporting. In this paper, we provide some anomaly and outlier detection techniques in the context of time series. After the anomaly detection, we explore various methods to  repair the problematic data. To be more specific, the entire data cleaning procedure has been divided into two categories: (1) manual cleaning, and (2) automatic cleaning. On the one hand, manual cleaning has very high accuracy; on the other hand, it is challenging to implement due to the high cost in time and effort. In this paper, we propose some data repairing methods to address the aforementioned issues. We summarize the background of these methods and give details on the implementation of the repairing procedure for COVID-19 reported data with manual and/or automatic cleaning methods. Although other researchers also mentioned some similar data problems for COVID-19 in the literature, to the best of our knowledge, our work is the first one that focuses on how to address these issues and repair the COVID-19 data.

Furthermore, it has been observed that the local characteristics, such as socioeconomic inequity, may also contribute to the spread of epidemic \cite{Stiglitz,ahmed2020inequality}. For example, the intrinsic local community characteristics might influence and shape the spread of COVID-19, such as mobility, demographics, and socioeconomic status. The availability of census data thus leads us to include all the epidemic data, control measures, and local information while modeling the infections, deaths, and recoveries. To facilitate research in identifying the significant factors that affect the disease spread pattern and predict future infections and deaths, we also collect and combine local auxiliary information at the county level in the US from reliable sources. 

To help users better visualize the epidemic data, we developed multiple \texttt{R} shiny apps embedded into a COVID-19 dashboard launched on March 27, 2020. Currently, we provide both infectious and death maps and time series of the US. Moreover, we provide a short-term (seven-day) forecast \cite{App_1} (updated daily) and a long-term (four-month) projection \cite{App_2} (updated weekly) of the COVID-19 infected and death count at both the county level and state level. For public usage, a Github repository (\url{https://github.com/covid19-dashboard-us/cdcar}) is established to provide daily updated and cleaned data.  An \texttt{R} package {\texttt{cdcar}} is also created for anomaly detection and repairing. In summary, we expect the proposed methods to have the following scientific merits. (i) Before choosing and integrating the data sources for analysis, it is important to understand how the data were collected and preprocessed. Therefore, in this article, we first investigate the similarities and dissimilarities among multiple data sources. (ii) Noticing the anomalies in the epidemic data, we develop several anomaly and outlier detection techniques in the context of count time series. Meanwhile, we further discover the reasons for these anomalies. (iii) After the anomaly detection, we introduce several methods to repair the problematic data and the corresponding historical data. Then we obtain our database by integrating the cured data with many local characteristics. (iv) The proposed methods and the data are built into an R package, which is publicly available through GitHub.

The rest of the paper is organized as follows. Section \ref{SEC:data} introduces the data related to the study of COVID-19, including a detailed description of the epidemic data, policy data, demographic characteristics, healthcare infrastructure, socioeconomic status, environmental factor and mobility data. Section \ref{SEC:comparison} discusses the comparison of the epidemic data from different sources. Section \ref{SEC:features} describes the cyclical pattern, types of anomalies of the COVID-19 reported time series, and how to perform the anomaly detection. Section \ref{SEC:repair} outlines methods for data repairing. Section \ref{SEC:tech} describes how to implement the proposed data comparison, anomaly detection and repairing procedure, and provide the details of the usage notes. Section \ref{SEC:discussion} concludes the paper with a discussion.

\setcounter{chapter}{2} \renewcommand{\theproposition}{{2.\arabic{proposition}}} 
\renewcommand{\thesection}{\arabic{section}} 
\renewcommand{\thesubsection}{2.\arabic{subsection}}
\renewcommand{\thetable}{\arabic{table}}
\renewcommand{\thefigure}{\arabic{figure}}
\setcounter{section}{1}
\section{Data}
\label{SEC:data}

We collect the epidemic data up to county level in the US along with control measures and other local information, such as socioeconomic status, demographic characteristics, healthcare infrastructure, and other essential factors to analyze the spatiotemporal dynamic pattern of the spread of COVID-19. Our data covers about 3,200 county-equivalent areas from 50 US states and the District of Columbia. A live version of the data analysis will be continually updated on our dashboard (\url{https://covid19.stat.iastate.edu}) and our Github repository (\url{https://github.com/covid19-dashboard-us/cdcar}). The sources and introductions for these data are detailed in Table \ref{tab:data}.

\begin{table}[htbp]
	\caption{Sources of datasets  \label{tab:data}} 
	\centering
	\scalebox{0.73}{\begin{tabular}{ll}
			\hline
			Data Type & Source \\
			\hline
			COVID-19 Related Time-series &  \\
			~~~~~~Infections Data &  \cite{NYT:20,COVIDTrack,JHUCSSE,USAFact:20} \\
			~~~~~~Fatality Data &  \cite{NYT:20,COVIDTrack,JHUCSSE,USAFact:20} \\
			~~~~~~Recovery Data &  \cite{COVIDTrack} \\
			Dates of COVID-19 Related Policies &  \\
			~~~~~~Declarations of State Emergency & \cite{BusiInsi:20}   \\
			~~~~~~Shelter-in-place or Stay-at-home Order & \cite{NYTSAH2020} \\
			Mobility Data\\
			~~~~~~Bureau of Transportation Statistics & \cite{TripofDistance}\\
			American Community Survey (ACS) Data &  \\
			~~~~~~2010-2018 Demographic and Housing Estimates & \cite{ACS1018} \\
			~~~~~~2005-2009 ACS 5-year Estimates & \cite{ACS0509} \\
			2012 Economic Census & \cite{EC12} \\
			2010 US Decennial Census & \cite{DC10} \\
			Homeland Infrastructure Foundation-level Data & \cite{HIFLD} \\
			USA Counties Database & \cite{uscounties}  \\
			US Census Bureau Gazetteer Files & \cite{gazetteer} \\
			\hline
	\end{tabular}}
\end{table}

\subsection{Epidemic Data}
\label{SUBSEC:epidata}
The daily counts of cases and deaths of COVID-19 are crucial for understanding how this pandemic is spreading. Thanks to the contribution of the data science communities across the world, multiple sources are providing the COVID-19 data with different precision and focus. In our article, we consider the reported cases from the following four sources: the NYT \cite{NYT:20}, the Atlantic \cite{COVIDTrack}, the JHU \cite{JHUCSSE}, and the USAFacts \cite{USAFact:20}. To clean the data, we first fetch data from the above four sources and compile them into the same format for further comparison and cross-validation. Then, we use the algorithms discussed in Section \ref{SEC:features} to detect the anomalies in the data sources and choose the one with the least anomalies for further repair.

\subsection{Other Factors}
\label{SUBSEC:otherfactors}
When analyzing the reported cases of COVID-19, many other factors may also contribute to the  temporal or spatial patterns; see the discussions in \cite{wang:2020:comparing}. For example, local features, like socioeconomic and demographic factors, can dramatically influence the course of the epidemic, and thus, the spread of the disease could vary dramatically across different geographical regions. Therefore, these datasets are also supplemented with the population information at the county level in our repository. We further classify these factors into the following six groups.

\subsubsection{2.2.1 Policy Data} \label{SUBSUBSEC:policydata}

In a race to stunt the spread of COVID-19, federal, state and local governments have issued various executive orders. Government declarations are used to identify the dates that different jurisdictions implemented various social distancing policies (emergency declarations, school closures, bans on large gatherings, limits on bars, restaurants and other public places, the deployment of severe travel restrictions, and ``stay-at-home'' or ``shelter-in-place'' orders). For example, President Trump declared a state of emergency on March 13, 2020, to enhance the federal government response to confront COVID-19. Later in the past spring, at least 316 million people in at least 42 states, the District of Columbia and Puerto Rico were urged to stay home.

Since the late April, all 50 states in the US began to reopen successively, due to the immense pressures of the crippled economy and anxious public. A state is categorized as ``reopening'' once its stay-at-home order lifts, or once reopening is permitted in at least one primary sector (restaurants, retail stores, personal care businesses), or once reopening is permitted in a combination of smaller sectors. We compiled the dates of executive orders by checking national and state governmental websites, news articles, and press releases.

\subsubsection{2.2.2 Demographic Characteristics}
\label{SUBSUBSEC:demochar}

In the demographic characteristics category, we consider the factors describing racial, ethnic, sexual, and age structures. These variables are extracted from the 2010 Census \cite{DC10}, and the 2010--2018 American Community Survey (ACS) Demographic and Housing Estimates \cite{ACS1018}.

\subsubsection{2.2.3 Healthcare Infrastructure}
\label{SUBSUBSEC:healthinfra}
We also incorporate several features related to the healthcare infrastructure at the county level in the datasets, including the percent of persons under $65$ years without health insurance, the local government expenditures for health per capita, and total bed counts per $1,000$ population.

\subsubsection{2.2.4 Socioeconomic Status}
\label{SUBSUBSEC:sociostat}

We consider diverse socioeconomic factors in the county level datasets. All of these factors collected from 2005--2009 ACS five-year estimates \cite{ACS0509}.

\subsubsection{2.2.5 Environmental Factor}
\label{SUBSUBSEC:environfac}

We also collect environmental factors that might affect the spread of epidemics significantly, such as the urban rate and crime rate.

\subsubsection{2.2.6 Mobility}
\label{SUBSUBSEC:mobil}

Another category of factors in the literature that affects the spread of infectious diseases significantly is the mobility; for example, movements of people from neighborhoods. We collect the mobility data from the Bureau of Transportation Statistics. 

\subsection{Geographic Information}
\label{SUBSUBSEC:geoinf}
The longitude and latitude of the geographic center for each county in the US are available in Gazetteer Files \cite{gazetteer}.  

\setcounter{chapter}{3} \renewcommand{\thetheorem}{{\arabic{theorem}}} 
\renewcommand{\thelemma}{{\arabic{lemma}}} 
\renewcommand{\theproposition}{{\arabic{proposition}}} 
\renewcommand{\thesection}{\arabic{section}} 
\renewcommand{\thesubsection}{3.\arabic{subsection}}
\setcounter{section}{2}

\section{Comparison of the Epidemic Data} \label{SEC:comparison}

In this subsection, we assess the similarities and differences of the reported infection and death cases from the previously mentioned four sources. The data collection sources and release times are indicated for each of the sources to help determine which factors may have an effect on the outcome of the assessment. The NYT released daily data at the national, state, and county levels at noon of the following day before July 18, after which the release time changed to midnight. The Atlantic releases daily state-level data along with testing, hospitalization, and recovery information, updated in the afternoon of the following day. The COVID-19 Data Repository by the CSSE at JHU provides both state and county-level data daily. JHU released data at midnight on and before April 22 and then changed the release time to the early morning of the following day. USAFacts collects the county-level data in the evening and releases them in the early morning of the following day (by 9 a.m. PST) \cite{USAFacts:20}. Table \ref{tab:comparison} summarizes the differences among the four sources of data based on how the data are collected and compiled.

\begin{table}[!ht]
	\caption{A summary of the comparison among four sources.     \label{tab:comparison}} 
	\centering
	\begin{threeparttable} 
		\begin{tabular}{p{4.6cm}p{1.3cm}p{1.3cm}p{1.3cm}p{1.3cm}}\hline
			Source& NYT & Atlantic & JHU & USAFacts  \\ \hline
			Infected \& death$^{*}$& 1,2,3 & 1,2 & 1,2,3 & 1,2,3 \\ \hline
			Recovered& 0 & 1,2 & 1,2,3$^{**}$& 0\\ \hline
			Tested & 0 & 1,2 & 1,2& 0\\ \hline
			Hospitalized & 0 & 1,2 & 1,2 &0 \\\hline
			Islands$^{***}$ & 2,3 & 2 & 2,3 &  0 \\ \hline
			Unallocated$^{****}$ & 3 & 0 & 3 & 3\\ \hline
			Place of infection$^{\#}$ & r,p & r,p & unknown & unknown \\ \hline
			Place of fatality & r,r+p,p & unknown & unknown & unknown \\ \hline
			Probable infected$^{\#\#}$ & y & y& y & unknown\\ \hline
			Probable death &    y & y$^{\#\#\#}$ & y & unknown\\ \hline
		\end{tabular}
		\begin{tablenotes}
			\small
			\item Note: ${*}$: Country Level $= 1$, State Level $= 2$, County Level $= 3$. USAFacts only provides county-level data for downloading. ${**}$: JHU pulls the number of people recovered data in the state-level from the Atlantic. ${***}$: Whether the source includes Puerto Rico, American Samoa, Guam, Northern Mariana Islands, Virgin Islands. ${****}$: Whether the dataset has unallocated/unassigned information, which is useful to match state-level and county-level data. ${\#}$: How does the dataset assign the cases to a place. $p$ indicates that the source assigns the counts according to the place of infection/fatality. $r+p$ indicates the source assigns both the deaths occur in the specific location, and the residents' deaths that occur outside the location. Specifically, this is related to New York City death data, see details in Section \ref{SEC:comparison}. $r,r+p,p$ indicates multiple standards exist. $unknown$ indicates the information is not found.  ${\#\#}$: Whether the dataset includes both confirmed and probable cases when probable data is available. $y$ means yes. NYT releases daily live data for probable and confirmed cases separately, but historical data is unavailable. ${\#\#\#}$: Colorado started to report the number of deaths where COVID-19 is listed as a contributing cause on the death certificates since May 16. This number is significantly lower than deaths among infected. The Atlantic uses deaths where COVID-19 is listed on death certificates, while the other three sources use deaths among infected. This unveils different definitions of probable deaths applied by the four sources.
		\end{tablenotes}
	\end{threeparttable} 
\end{table}

\begin{figure}
	\centering
	\begin{tabular}{cc}
		\includegraphics[width = .4 \textwidth]{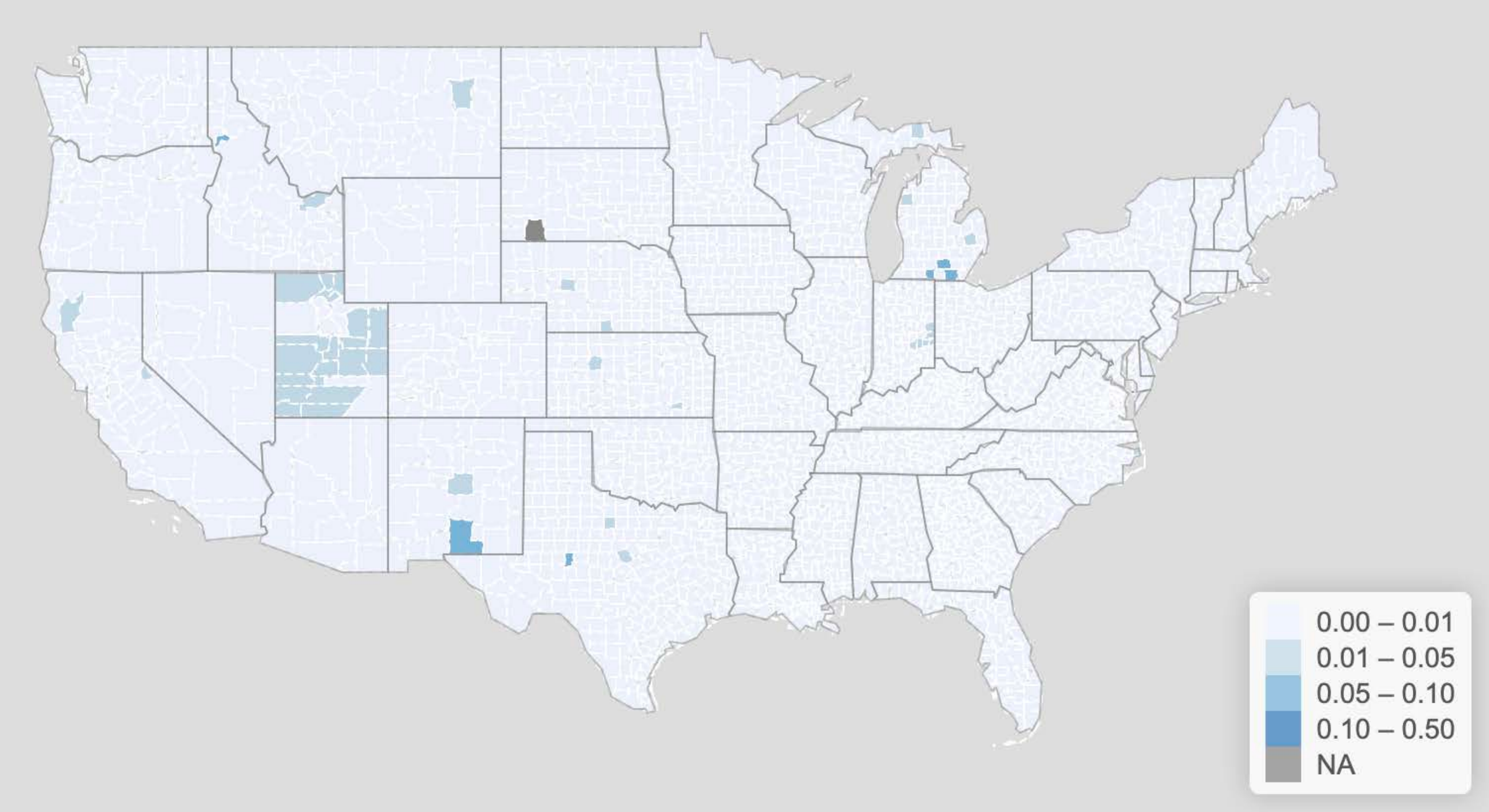}& 
		\includegraphics[width = .41 \textwidth]{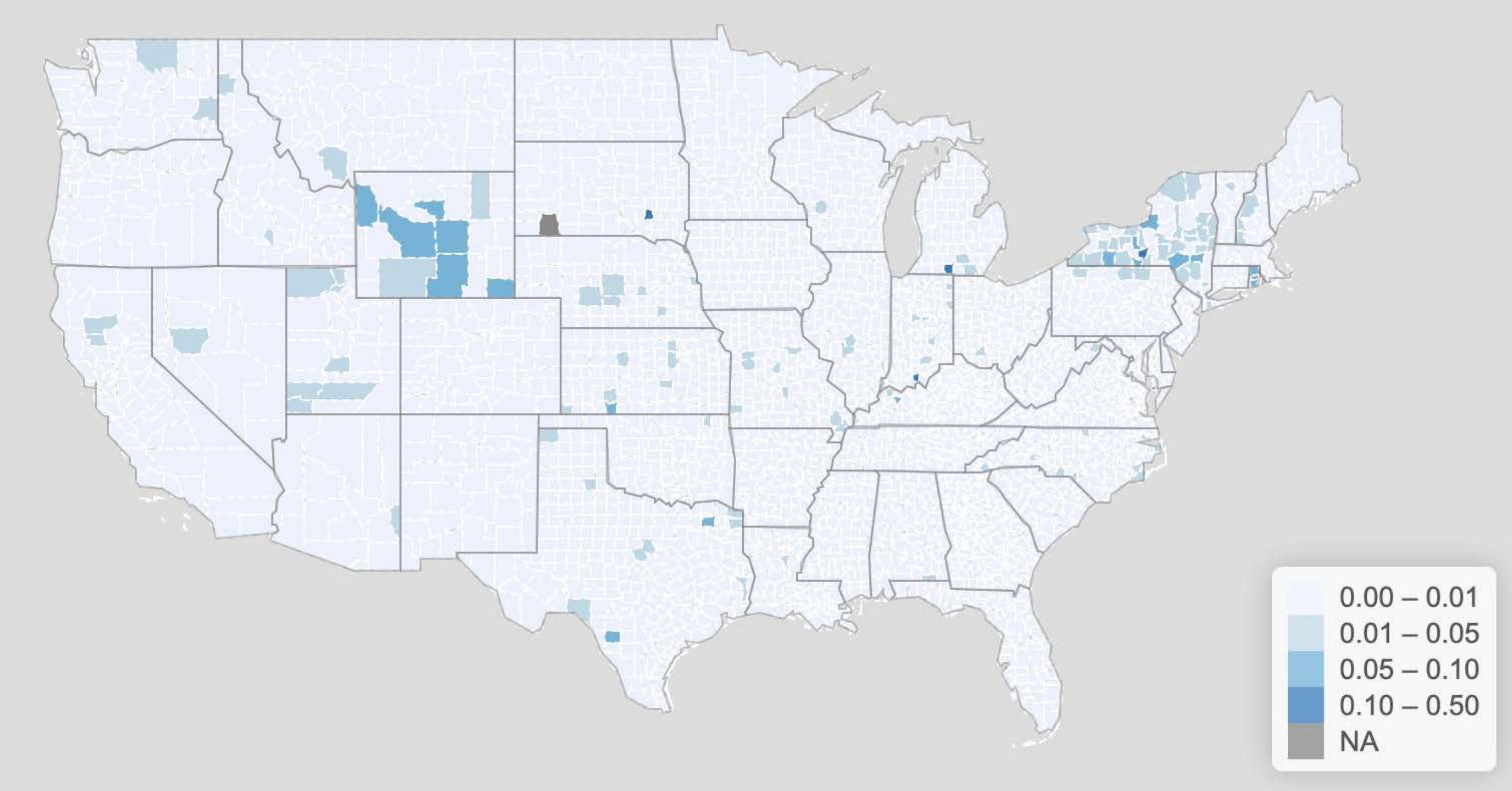} \\
		(a) Infection (NYT vs JHU) & (b) Death (NYT vs JHU)\\
		\includegraphics[width = .4 \textwidth]{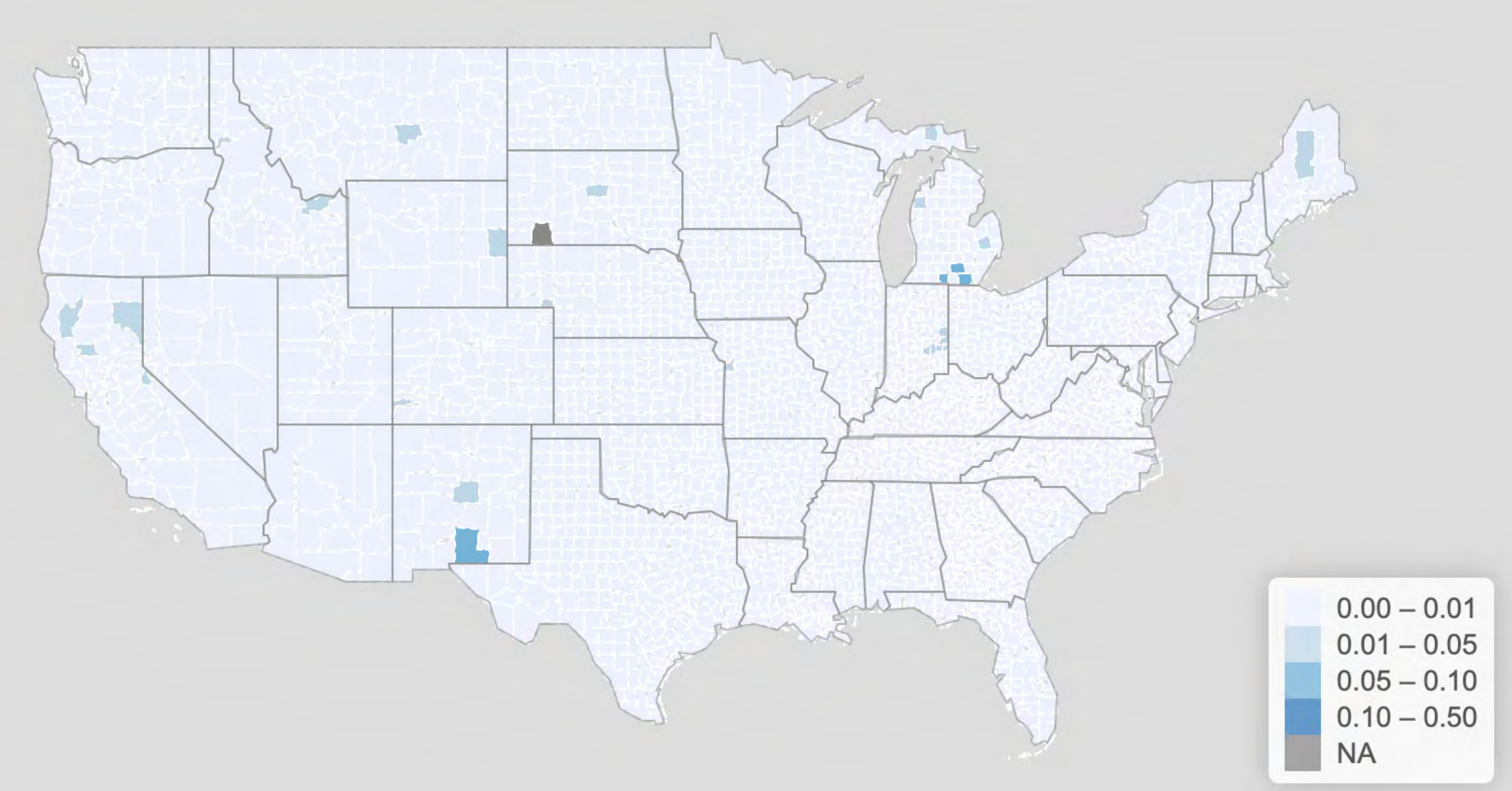} & 
		\includegraphics[width = .4 \textwidth]{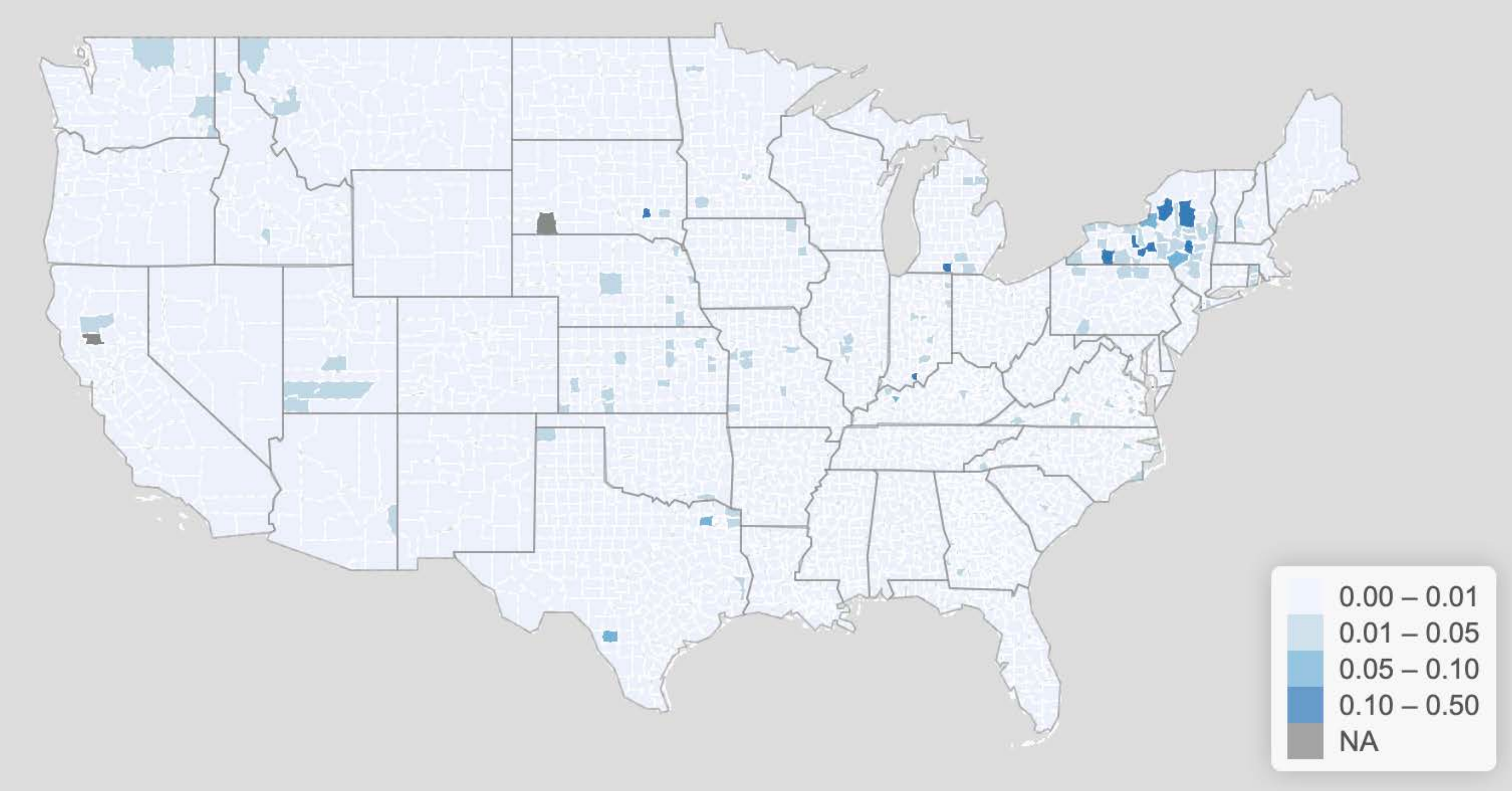}\\
		(c) Infection (NYT vs USAFacts) & (d) Death (NYT vs USAFacts)\\
		\includegraphics[width = .4 \textwidth]{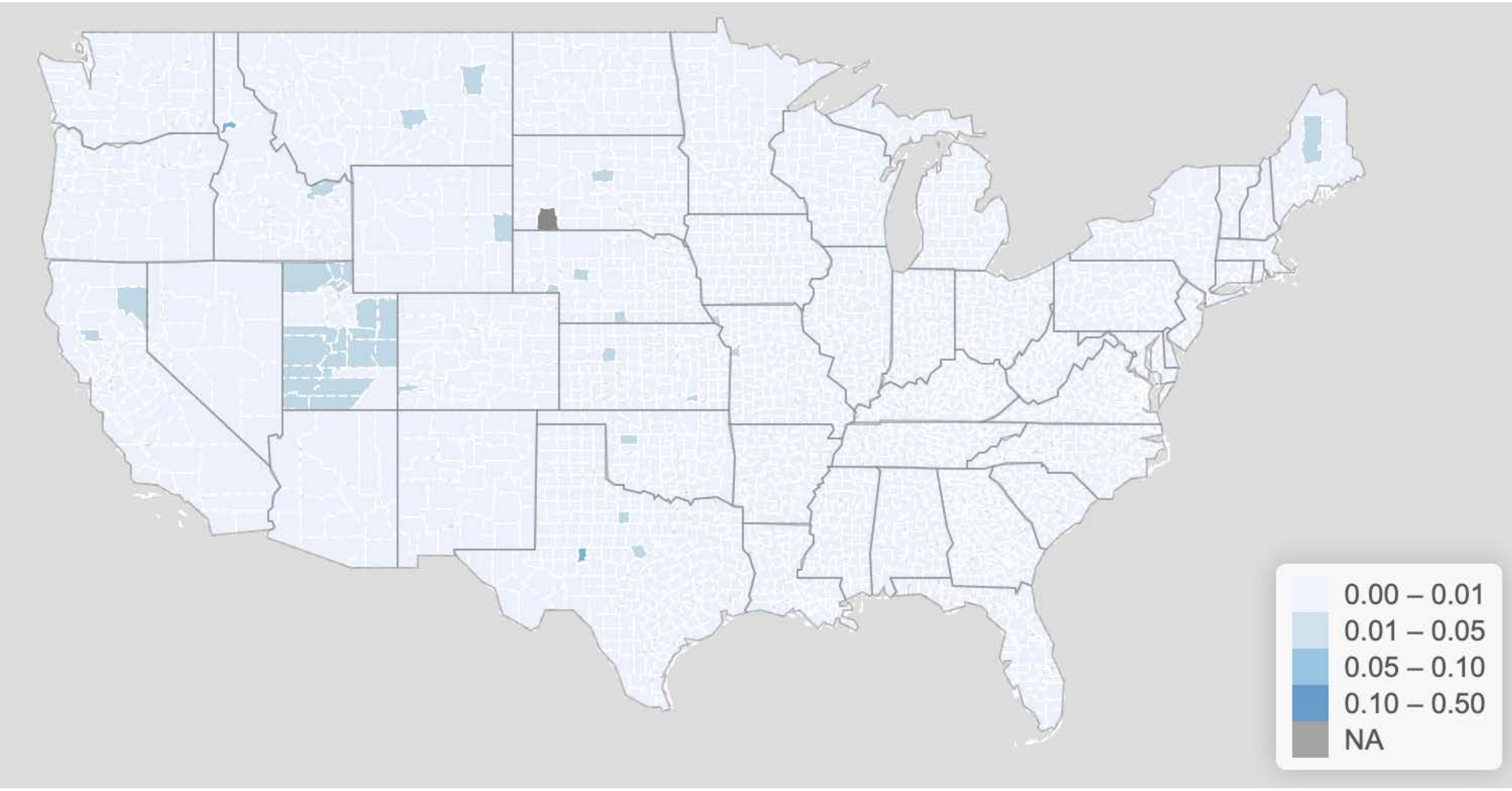}&
		\includegraphics[width = .4 \textwidth]{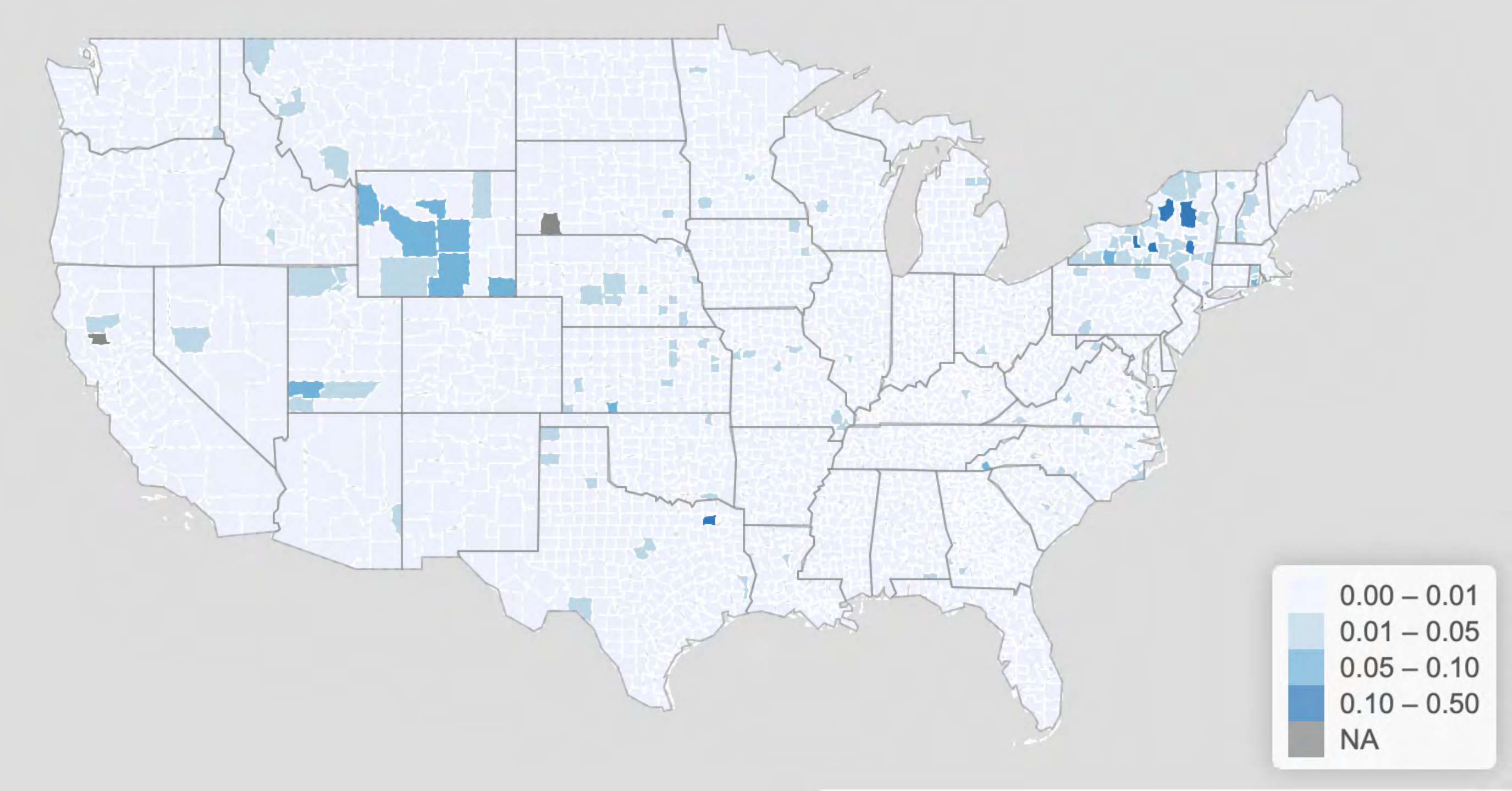}\\
		(e) Infection (JHU vs USAFacts) & (f) Death (JHU vs USAFacts)\\
	\end{tabular}
	\caption{County maps of the dissimilarity measure as of July 25, 2020.}
	\label{fig:county.tstat.compare} 
\end{figure}

\begin{figure}[!ht]
	\centering
	\begin{tabular}{cc}
		\includegraphics[width = .3 \textwidth]{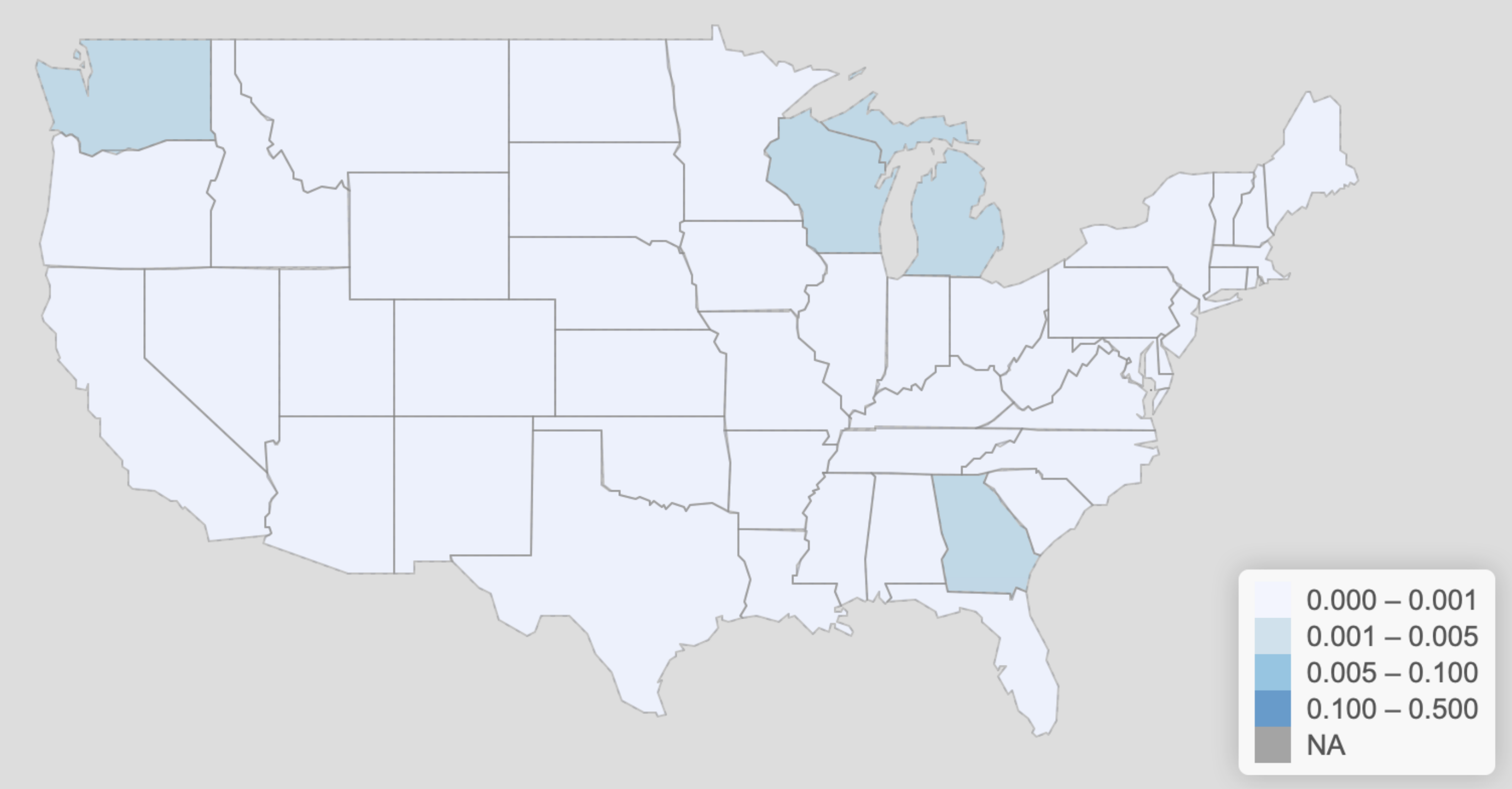}& 
		\includegraphics[width = .3 \textwidth]{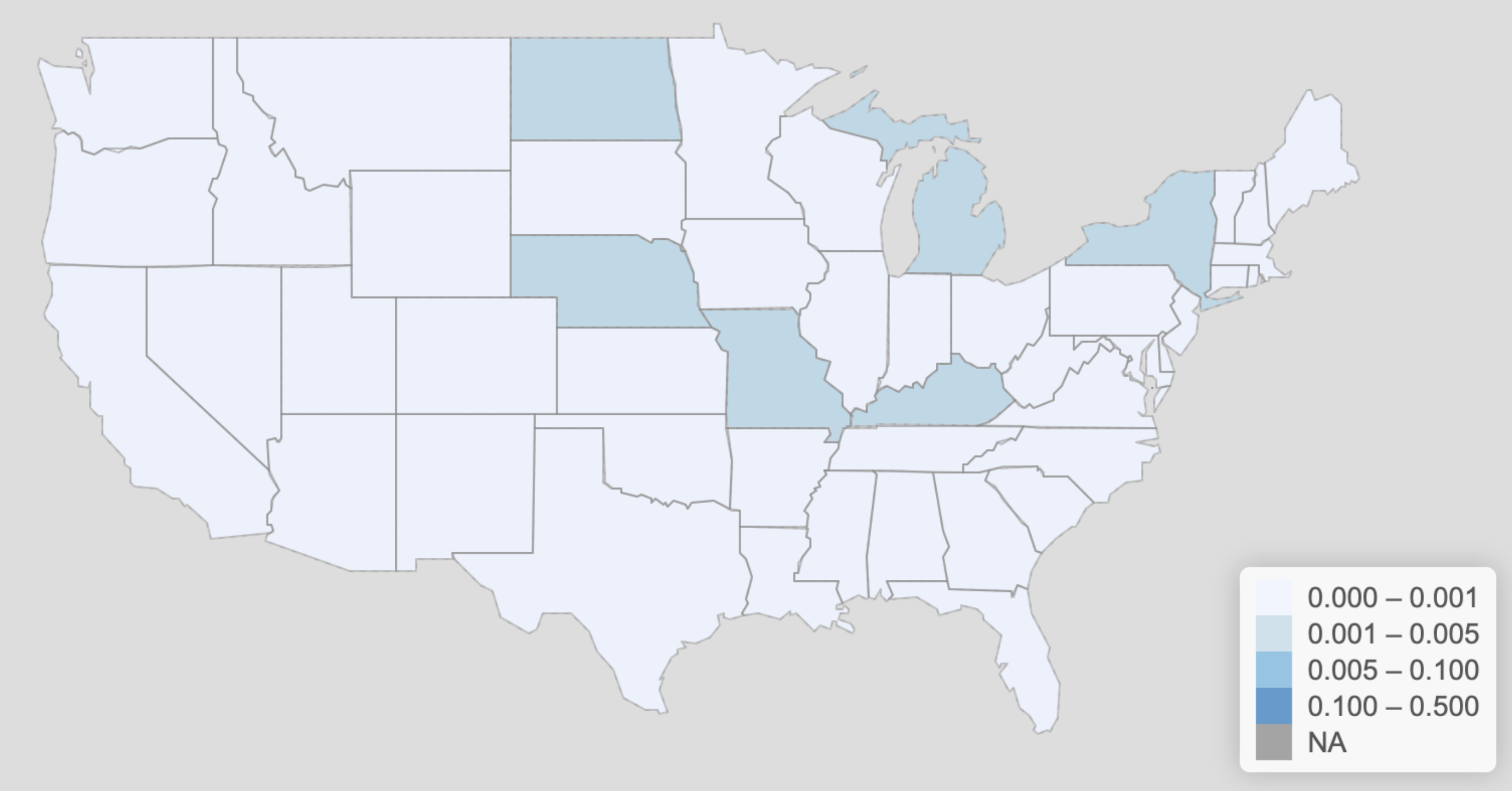} \\
		(a) Infection (NYT vs JHU) & (b) Death (NYT vs JHU)\\
		\includegraphics[width = .3 \textwidth]{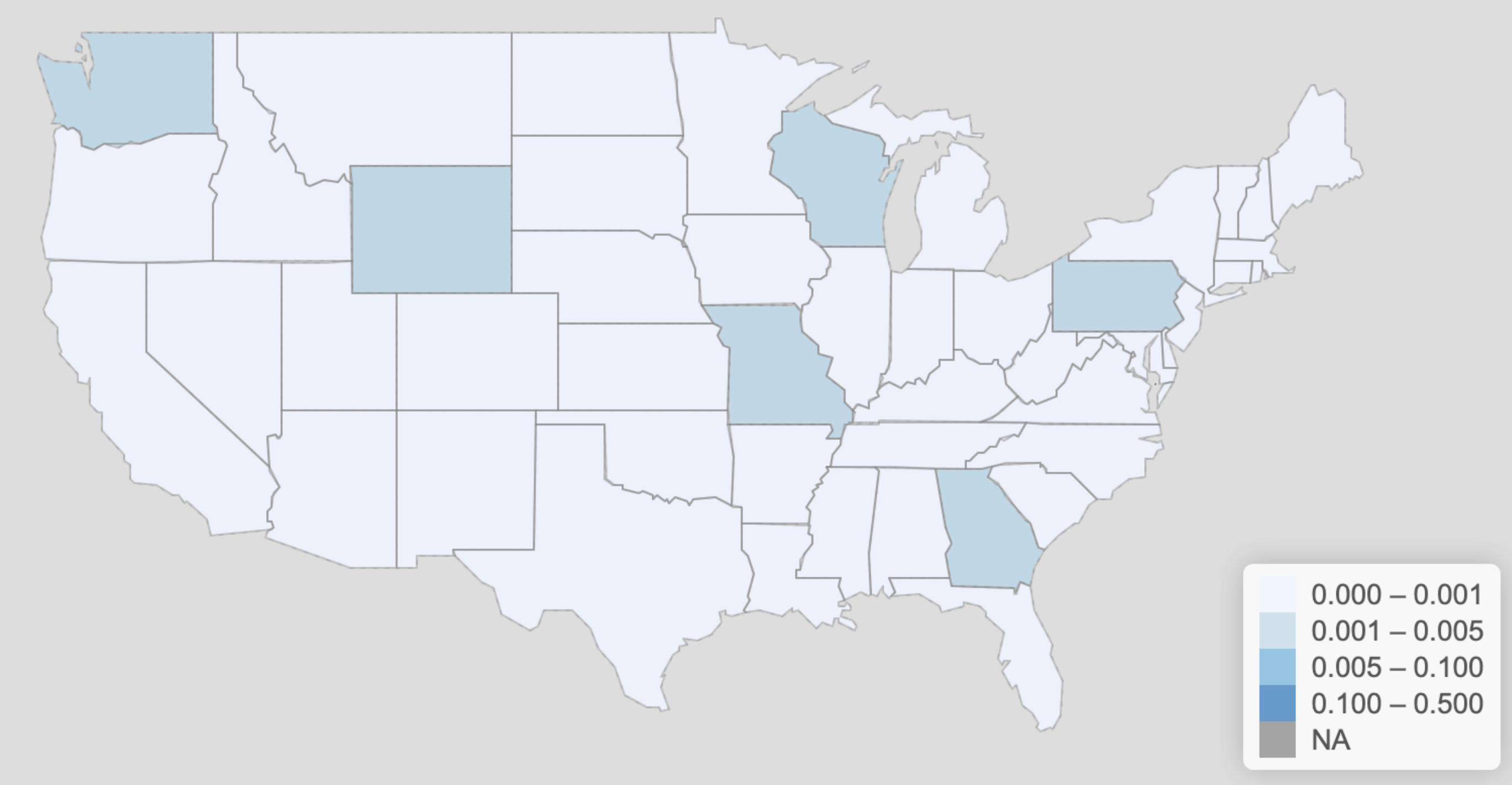} & 
		\includegraphics[width = .3 \textwidth]{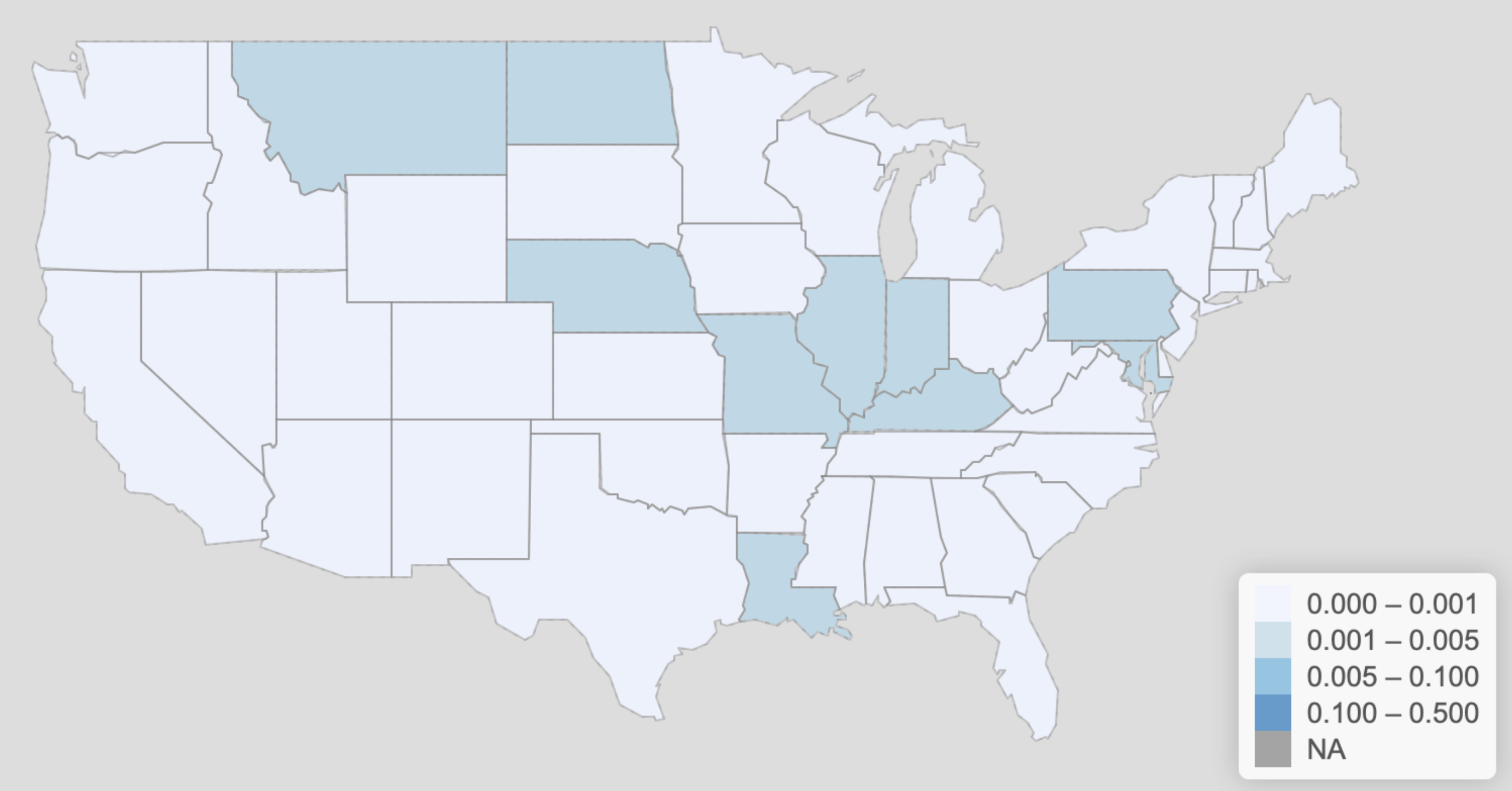}\\
		(c) Infection (NYT vs USAFacts) & (d) Death (NYT vs USAFacts)\\
		\includegraphics[width = .3 \textwidth]{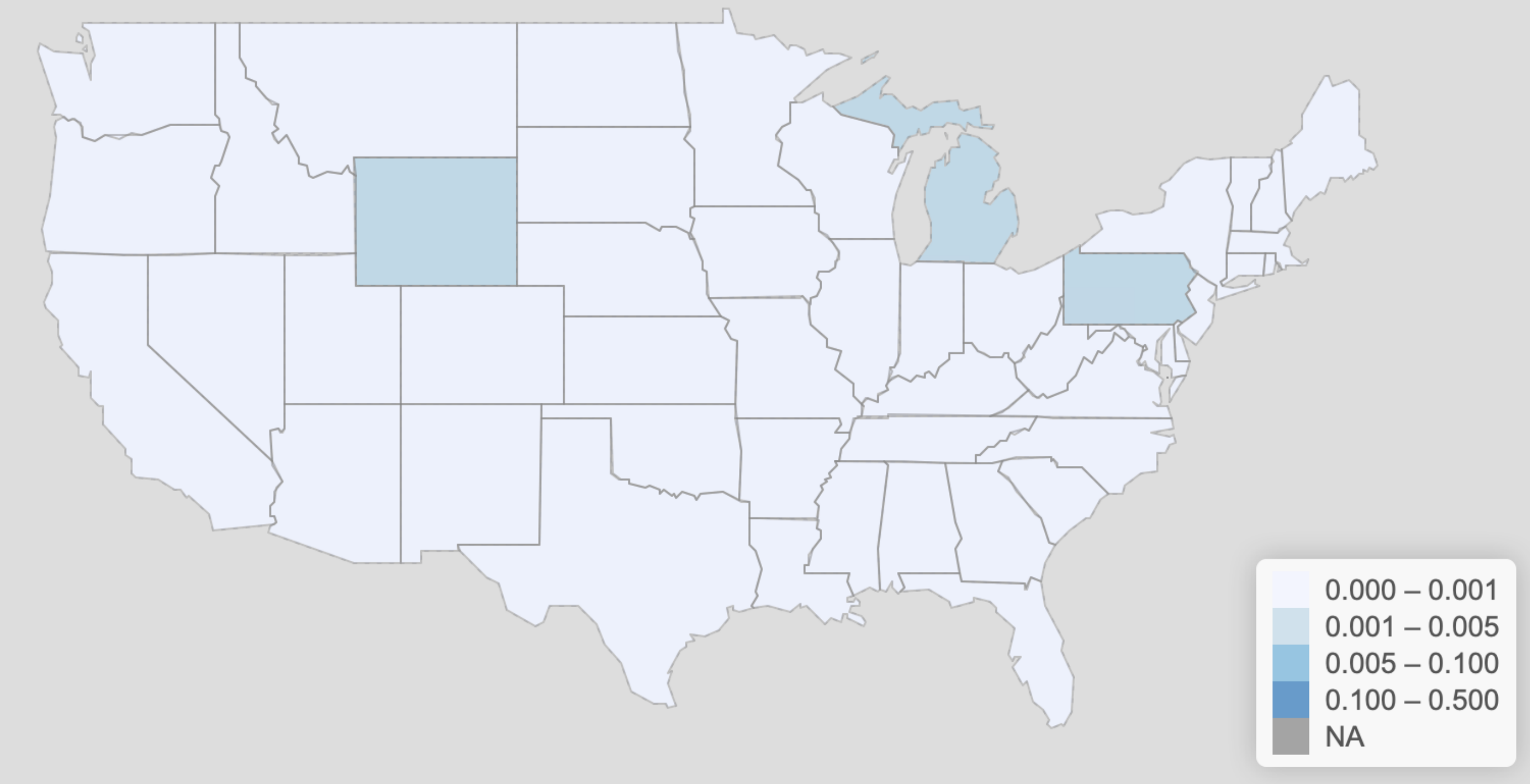}&
		\includegraphics[width = .3 \textwidth]{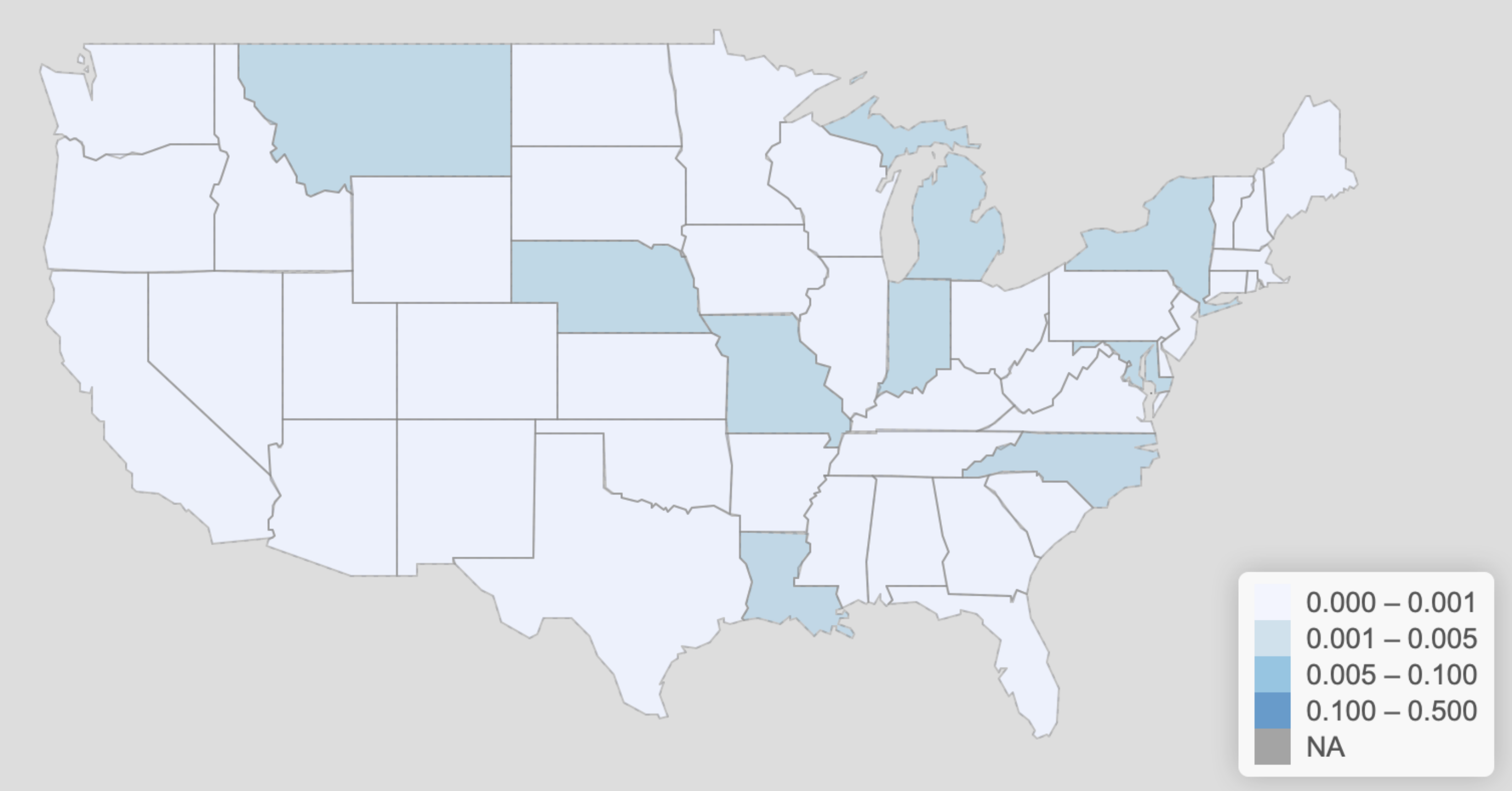}\\
		(e) Infection (JHU vs USAFacts) & (f) Death (JHU vs USAFacts)\\
		\includegraphics[width = .3 \textwidth]{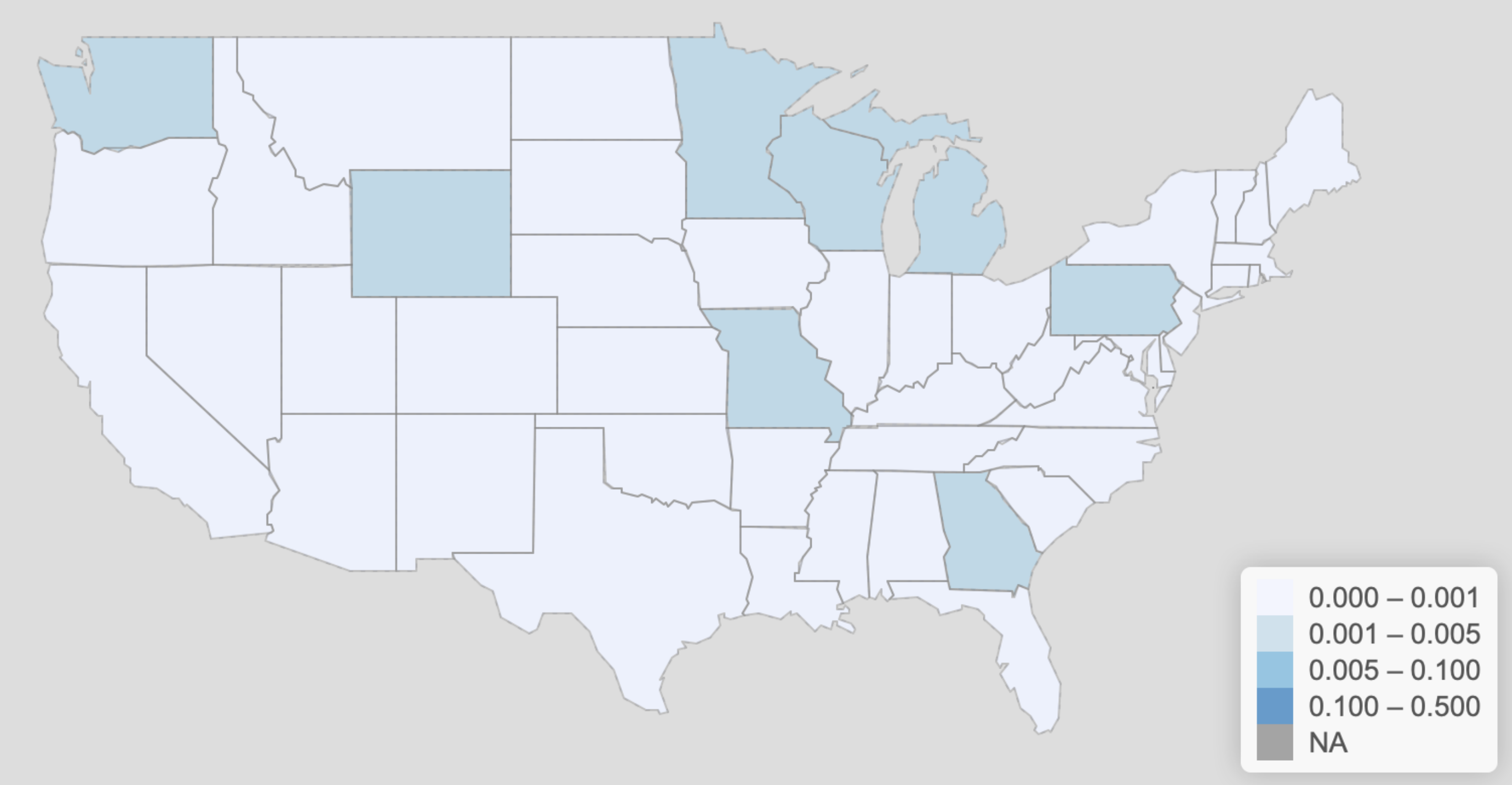}&
		\includegraphics[width = .3 \textwidth]{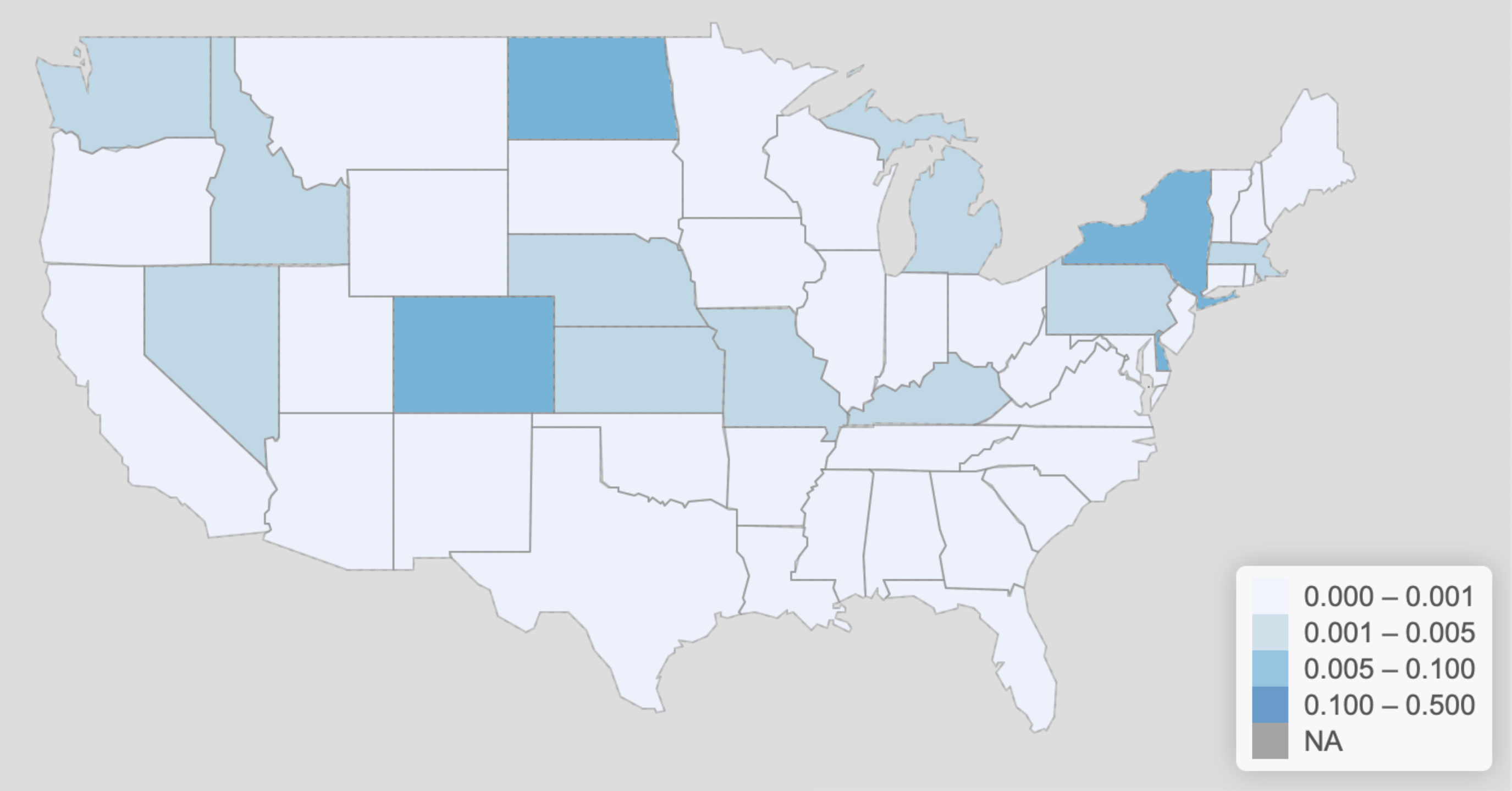}\\
		(g) Infection (NYT vs Atlantic) & (h) Death (NYT vs Atlantic)\\
		\includegraphics[width = .3 \textwidth]{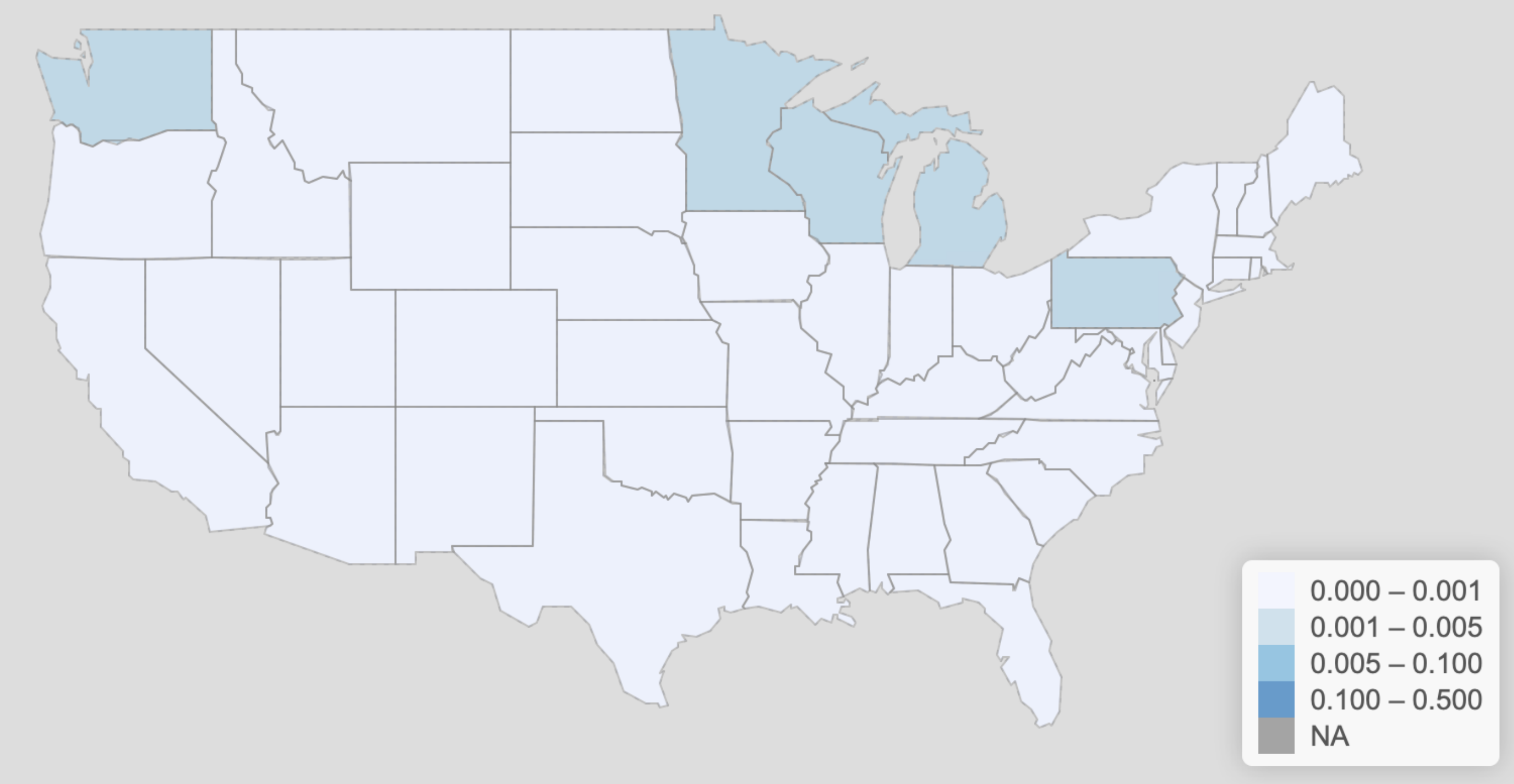}&
		\includegraphics[width = .3 \textwidth]{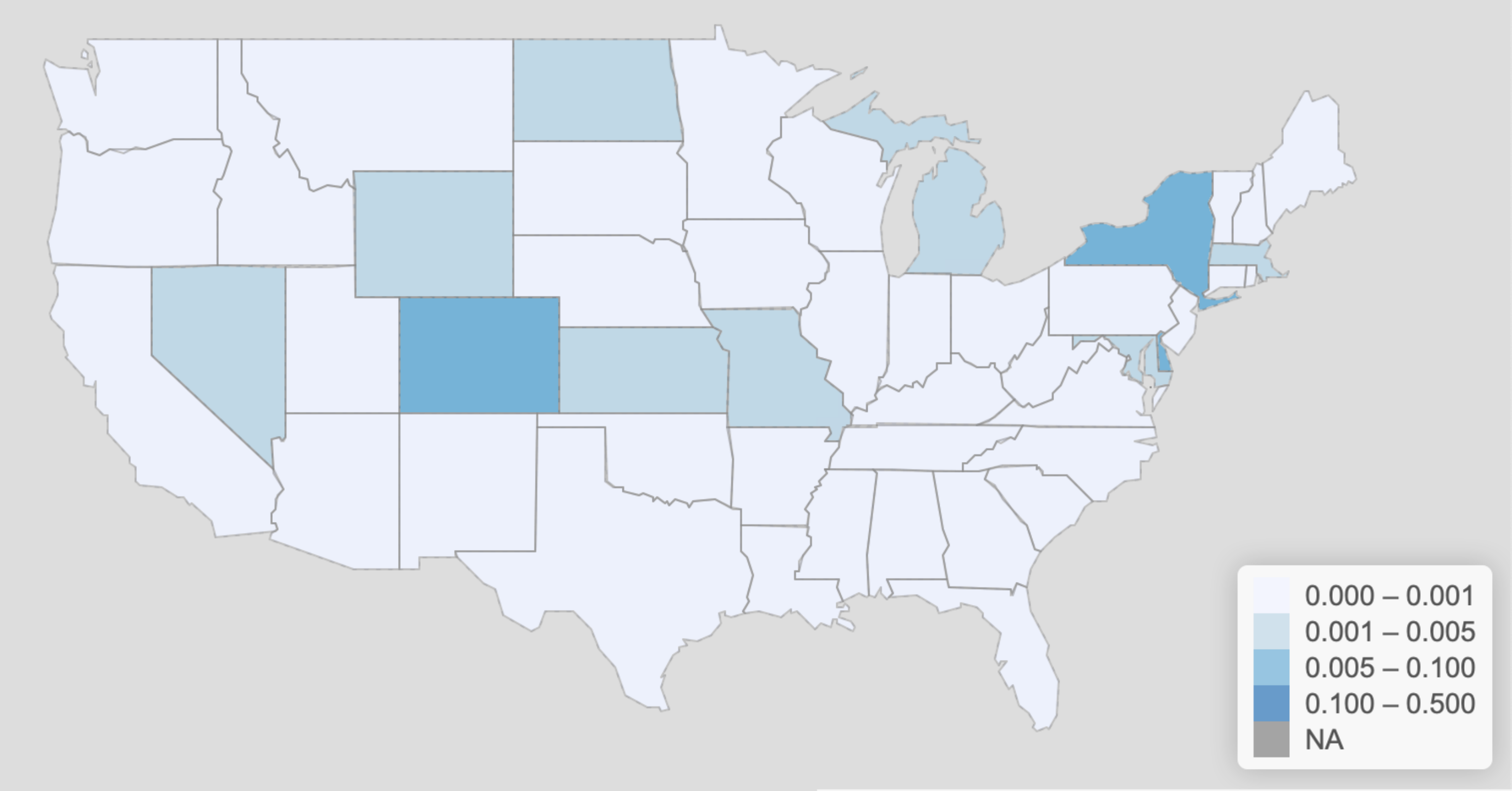}\\
		(i) Infection (JHU vs Atlantic) & (j) Death (JHU vs Atlantic)\\
		\includegraphics[width = .3 \textwidth]{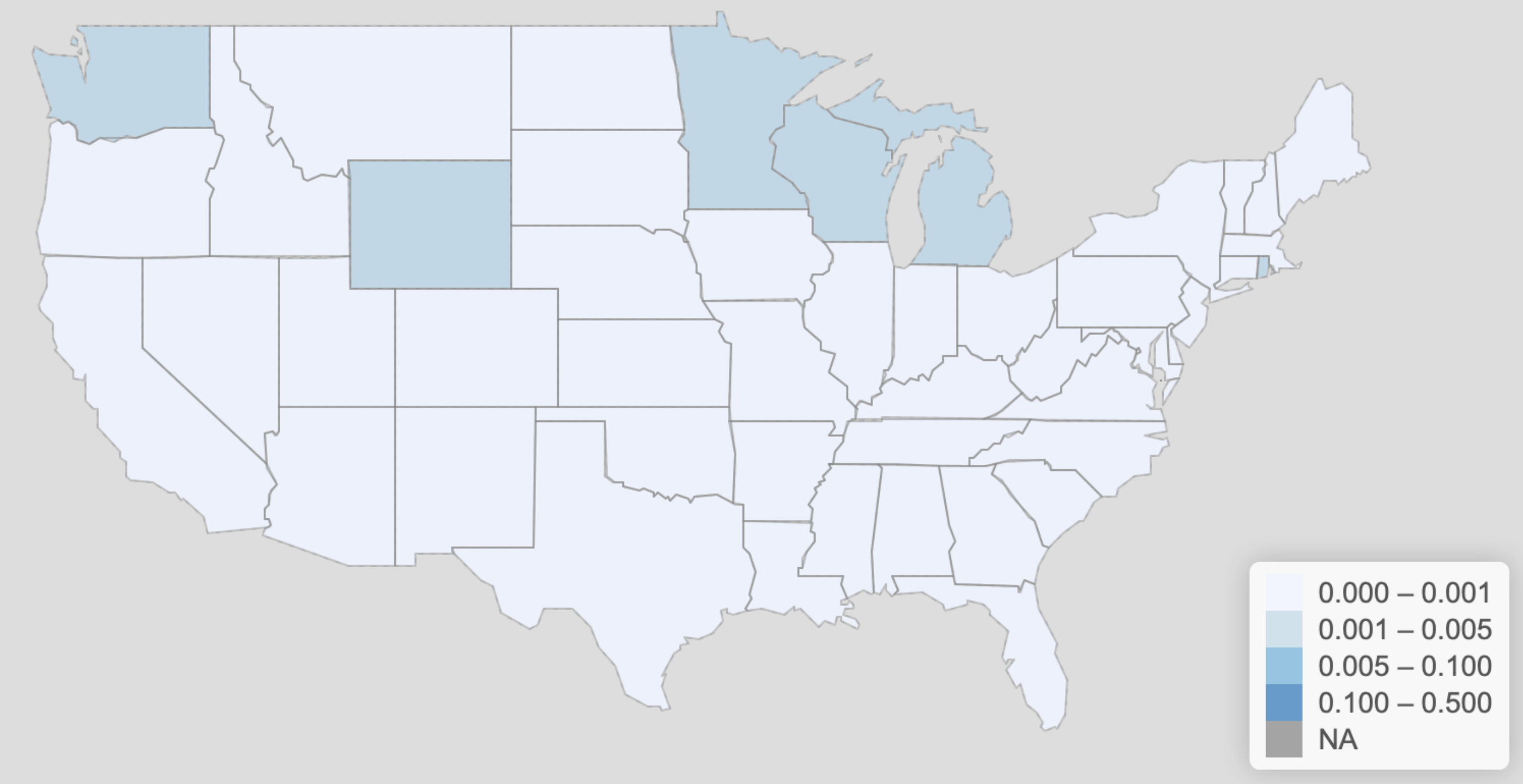}&
		\includegraphics[width = .3 \textwidth]{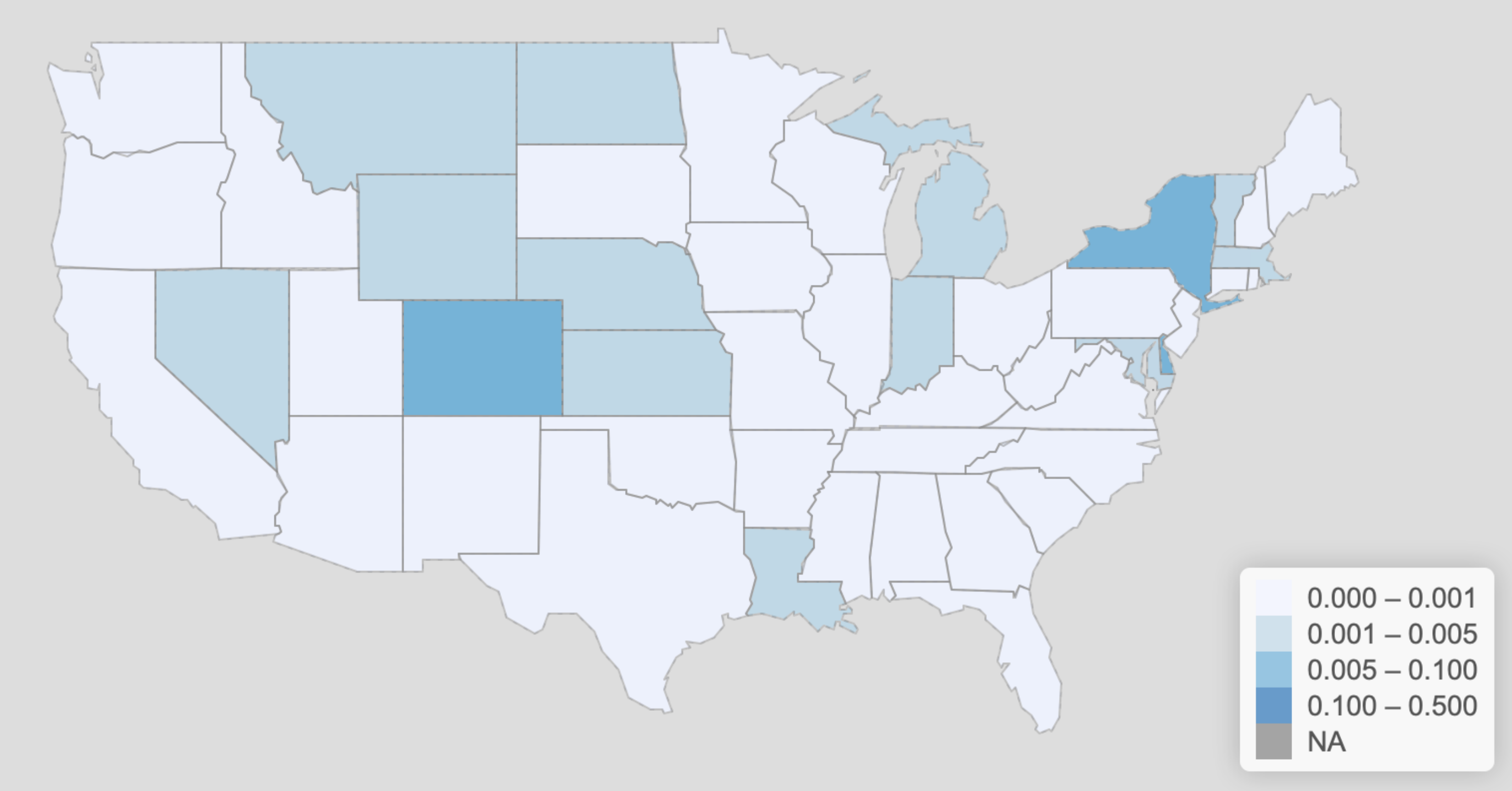}\\
		(k) Infection (USAFacts vs Atlantic) & (l) Death (USAFacts vs Atlantic)\\
	\end{tabular}
	\caption{State maps of the dissimilarity measure as of July 25, 2020.}
	\label{fig:state.tstat.compare} 
\end{figure}

Let $K$ be the number of all available sources in the comparison. For the county level comparison, $K = 3$ since the Atlantic does not provide county level data, while for the state level, $K = 4$. Let $T$ be the number of days observed, or the length of each time series. Let $n$ be the number of counties or states. For source $k$, $k=1,\ldots,K$, let $Y_{it}^{(k)}$ be the cumulative number of the reported cases of location $i$ on day $t$, where $i=1,\ldots,n$, $t=1,\ldots,T$. In the following, we define a dissimilarity measure to assess the difference between two time series: $\mathbf{Y}_i^{(k)}=\{Y_{it}^{(k)}\}_{t=1}^{T}$ and $\mathbf{Y}_i^{(k^{\prime})}=\{Y_{it}^{(k^{\prime})}\}_{t=1}^{T}$, for any $1\leq k\neq k^{\prime}\leq K$. Let $\overline{Y}_{it} = K^{-1}\sum_{k = 1}^{K} {Y}_{it}^{(k)}$, then the difference between $\mathbf{Y}_i^{(k)}$ and $\mathbf{Y}_i^{(k^{\prime})}$ is defined as:  
\begin{equation}
d(k,k^{\prime}) \equiv  d(\mathbf{Y}_i^{(k)}, \mathbf{Y}_i^{(k^{\prime})}) :=
\begin{cases}
\frac{1}{T} \|\mathbf{Y}_{i}^{(k)}- \mathbf{Y}_{i}^{(k^{\prime})}\|/\overline{Y}_{iT}, & \overline{Y}_{iT} >0\\
0, & \overline{Y}_{iT} = 0
\end{cases},
\label{eqn:measure}
\end{equation}
where $\overline {Y}_{iT}$ is used to mitigate the variability of the currently observed counts. Equation (\ref{eqn:measure}) provides a measurement that effectively detects the counties and states with the most discrepancy between each pair of sources and is meaningful in the comparison between different locations. In Figure \ref{fig:county.tstat.compare} we present the county map for infected and death counts collected from three data sources. In Figure \ref{fig:state.tstat.compare} we present the state map for infected and death counts collected from four data sources. Areas in dark blue in these two figures are determined to be different between the corresponding pair of two sources. In the rest of this section, we look further into the underlying reasons for dissimilarity at the county and state levels. 

We list the ten most dissimilar counties in a pairwise comparison of the three sources, in terms of infection and death counts, in Tables \ref{tab:top10counties_infect} and  \ref{tab:top10counties_death}, respectively. The specific reasons for these dissimilarities vary over locations. For the state of New York, the difference between sources is caused by different geographical assignments. NYT and JHU combine Kings, Queens, Bronx, and Richmond counties with New York City while USAFacts does not use that combination. For the state of Utah, JHU combines counties to jurisdictions to be consistent with the official state source, while NYT and USAFacts provide county-level data. For Guam, NYT includes the data reported from USS Theodore Roosevelt, while JHU and USAFacts do not. In Michigan, NYT considers federal and state prison inmates' data when reporting at the county-level, while the other two sources do not. For the state of Alaska, NYT and JHU include non-resident cases while USAFacts does not. For some states, such as Kentucky, Texas, Pennsylvania, Washington, Georgia and Tennessee, the official county-level data is subject to frequent adjustments, which can lead to discrepancies when one source corrects the errors while other sources do not. In summary, the county-level dissimilarities between data sources are mostly caused by different geographical rules, non-resident data, prison inmates data, and differed efforts in correcting the historical data.

\begin{table}[!ht]
	\caption{ {Top 10 counties with the largest value of the dissimilarity measure of the infectious counts between pairs of sources (as of July 25, 2020).}  \label{tab:top10counties_infect}}    
	\centering
	\begin{tabular}{p{3.6cm}p{3.6cm}p{3.6cm}}\hline
		NYT vs JHU &     NYT vs USAFacts & JHU vs USAFacts\\ \hline
		BristolBay, AK & BristolBay, AK & Lewis, ID\\
		Dillingham, AK & Dillingham, AK & Dukes, MA \\
		Lewis, ID     &  Branch, MI& Bronx, NY\\
		Dukes, MA  & Jackson, MI & Kings, NY\\
		Branch, MI& Otero, NM & New York, NY\\
		Jackson, MI& Bronx, NY & Queens, NY\\
		Lenawee, MI &Kings, NY &  Richmond, NY \\
		Otero, NM&  New York, NY & Sterling, TX \\
		Sterling, TX &Queens, NY & Emery, UT\\
		Piute, UT & Richmond, NY &  Piute, UT\\ \hline
	\end{tabular}
\end{table}

\begin{table}[!ht]
	\caption{Top 10 counties with the largest value of dissimilarity measure of the death counts between pairs of sources (as of July 25, 2020). \label{tab:top10counties_death} } 
	\centering
	\begin{tabular}{p{3.6cm}p{3.6cm}p{3.6cm}}\hline
		NYT vs JHU &     NYT vs USAFacts & JHU vs USAFacts\\ \hline
		Crawford, IN & Glenn, CA & Glenn, CA\\
		McLean, KY &  Crawford, IN  & Hamilton, NY\\
		Branch, MI & Bronx, NY & Bronx, NY\\
		Oswego, NY & Cortland, NY &  Cortland, NY\\
		Delaware, NY & Kings, NY &Lewis, NY\\
		Seneca, NY &Lewis, NY &Queens, NY\\
		Tompkins, NY & Schoharie, NY & Richmond, NY\\
		Davison, SD & Seneca, NY &Schoharie, NY \\
		Hopkins, TX & Tompkins, NY & Seneca, NY\\
		Teton, WY & Davison, SD & Kings, NY\\ \hline
	\end{tabular}
\end{table}

\begin{figure}
	\centering
	\begin{tabular}{cc}
		\includegraphics[width = .38 \textwidth]{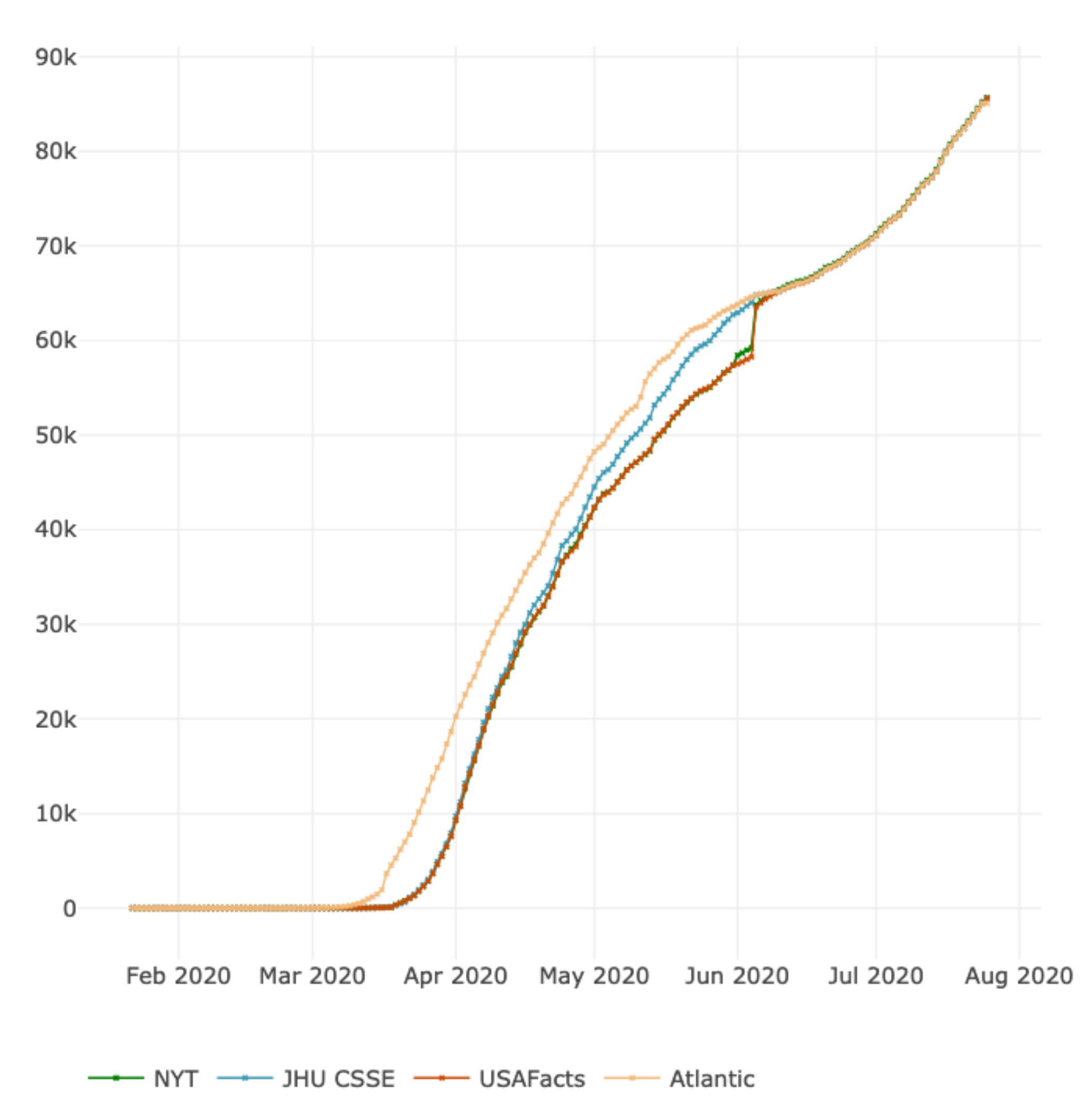} & 
		\includegraphics[width = .38 \textwidth]{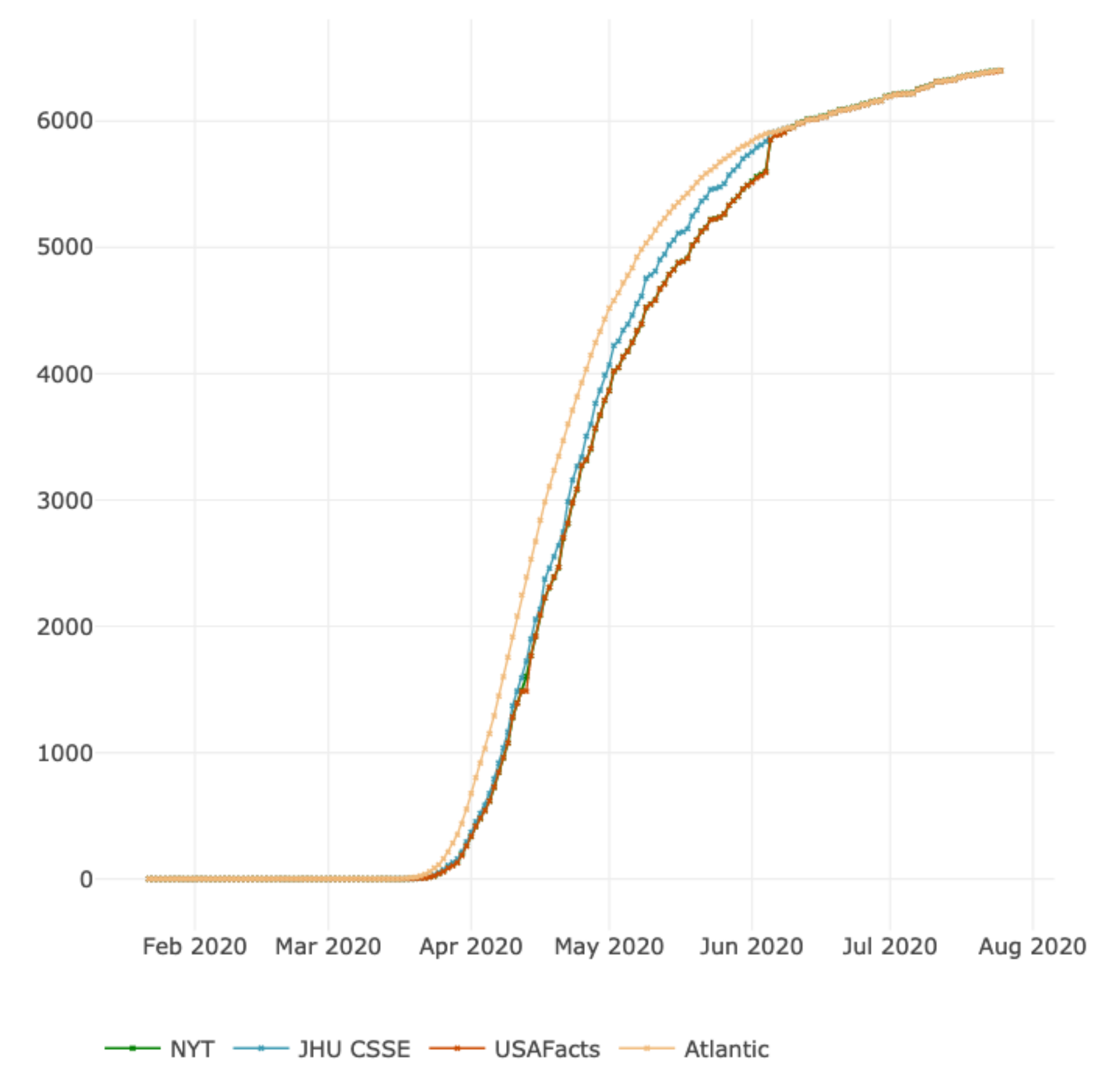}\\
		(a) Michigan infection & (b) Michigan death\\
		\includegraphics[width = .38 \textwidth]{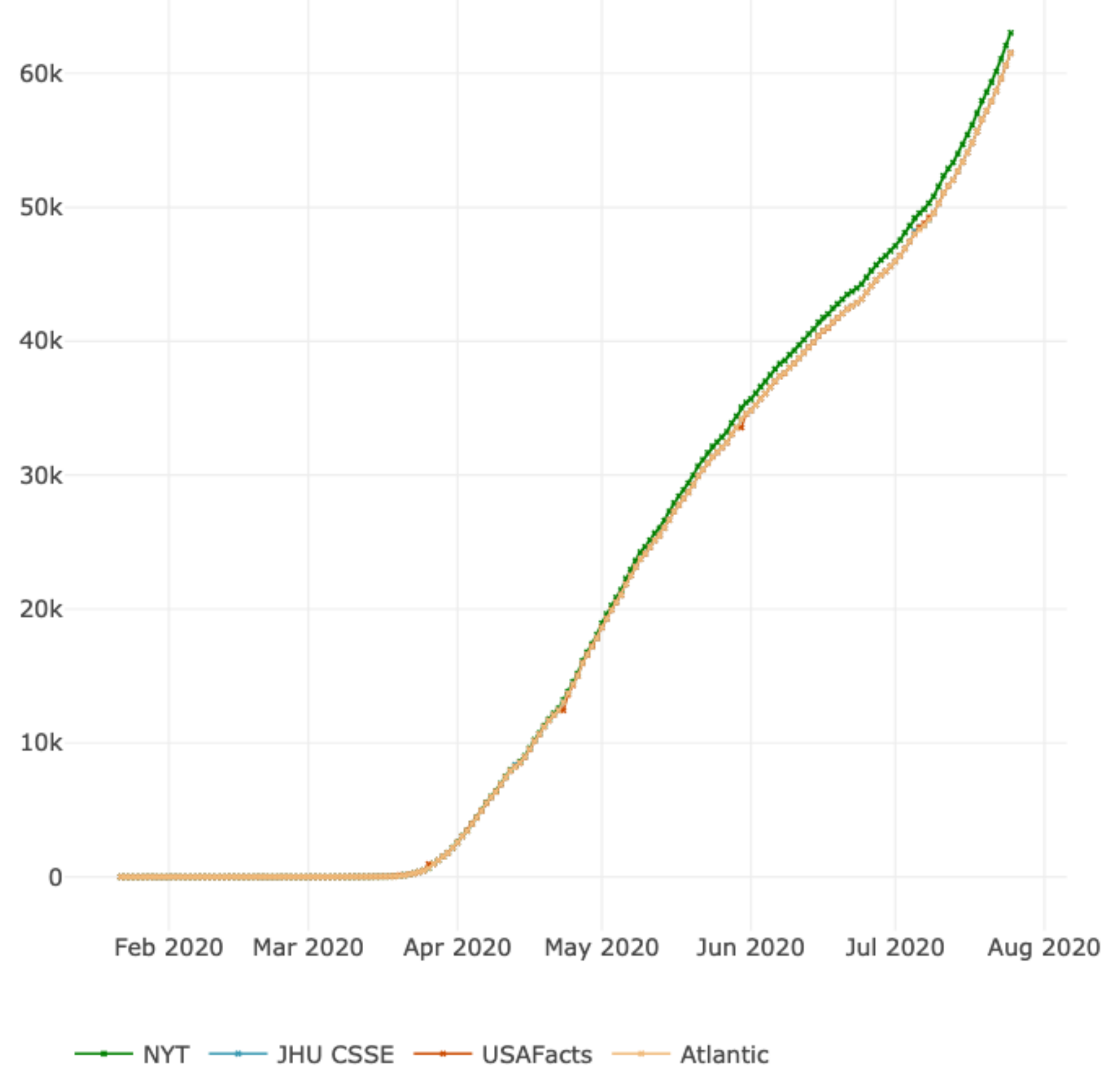} & 
		\includegraphics[width = .38 \textwidth]{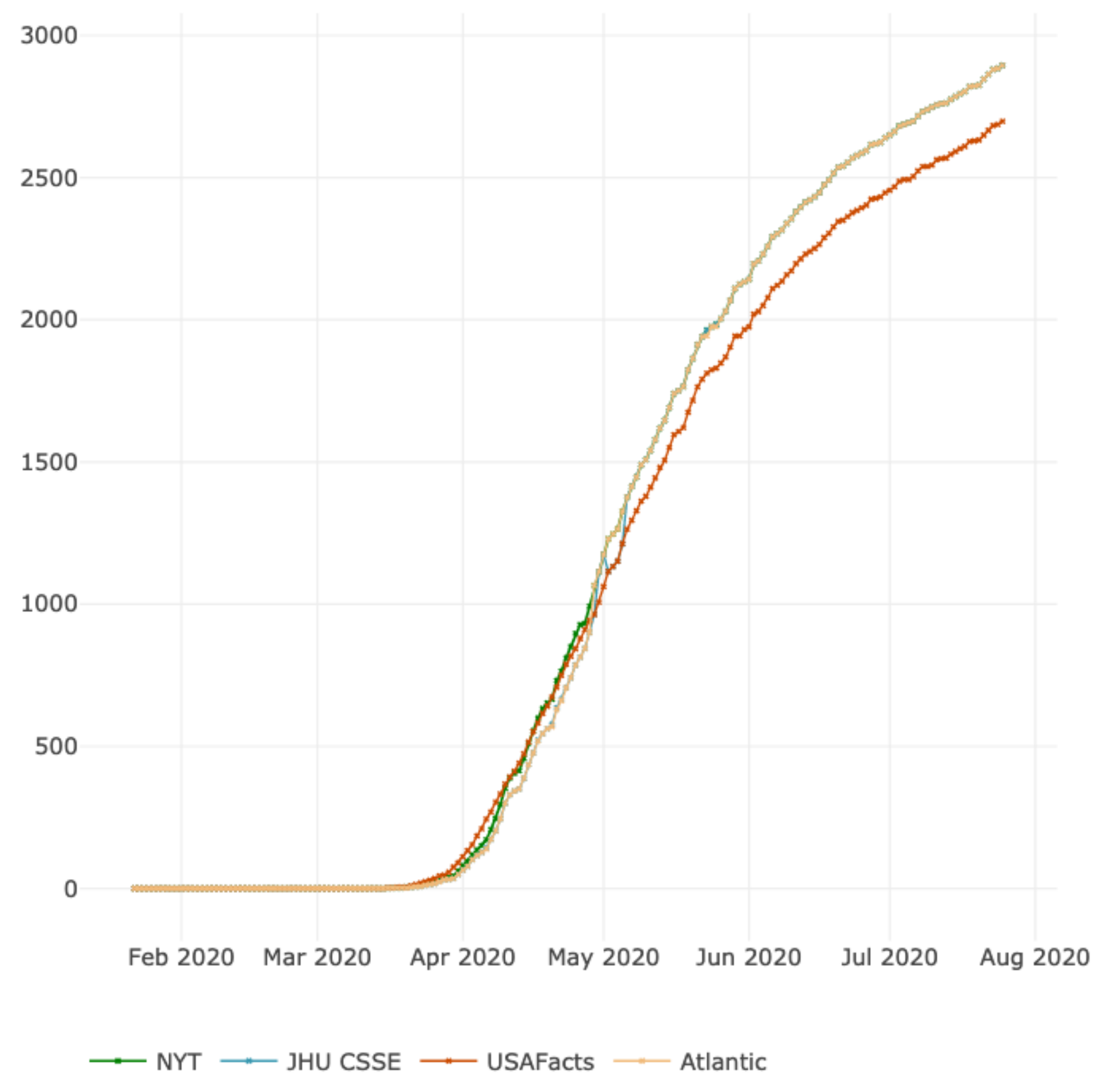}\\
		(c) Indiana infection & (d) Indiana death\\
		\includegraphics[width = .38 
		\textwidth]{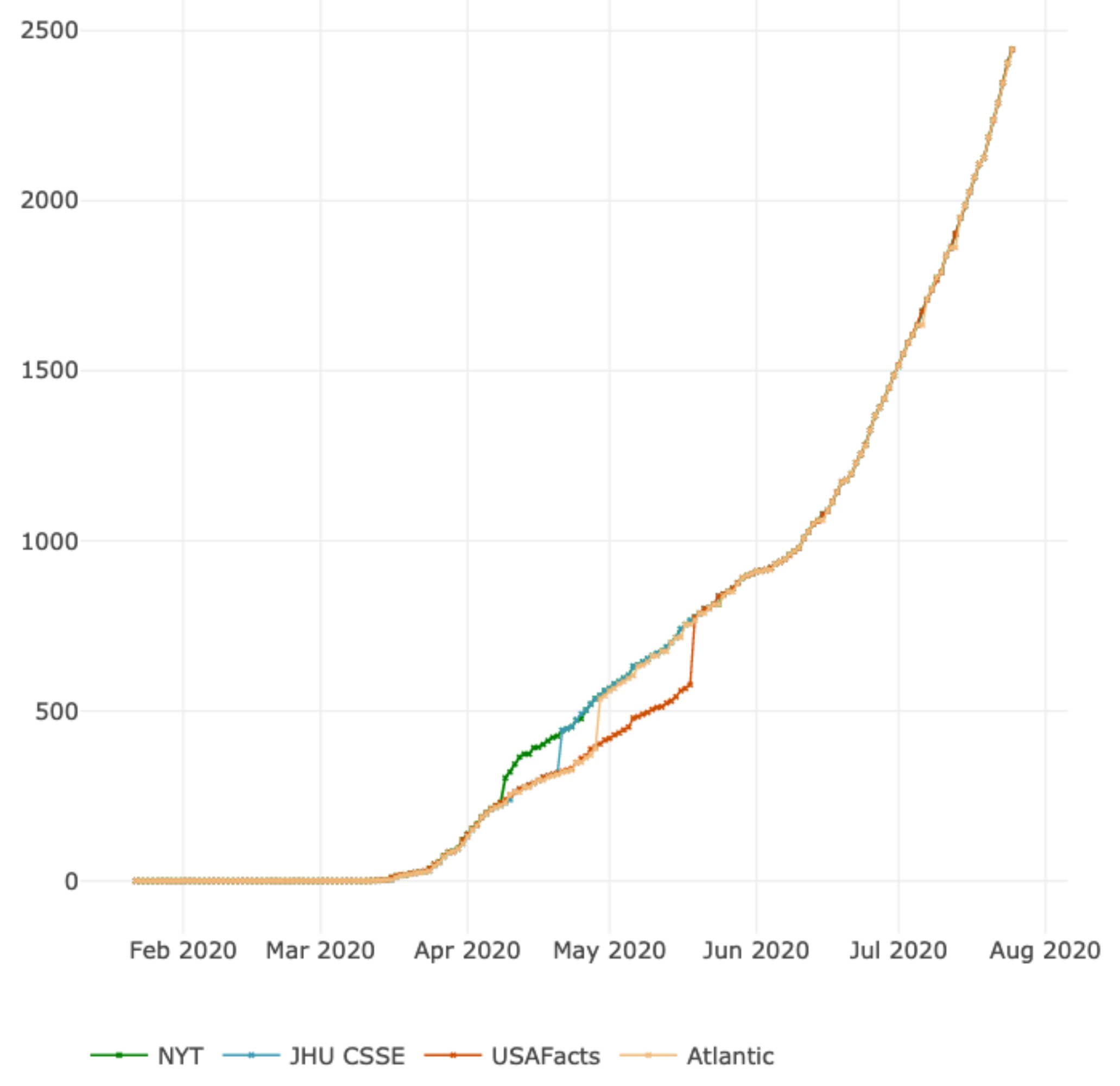} & 
		\includegraphics[width = .38
		\textwidth]{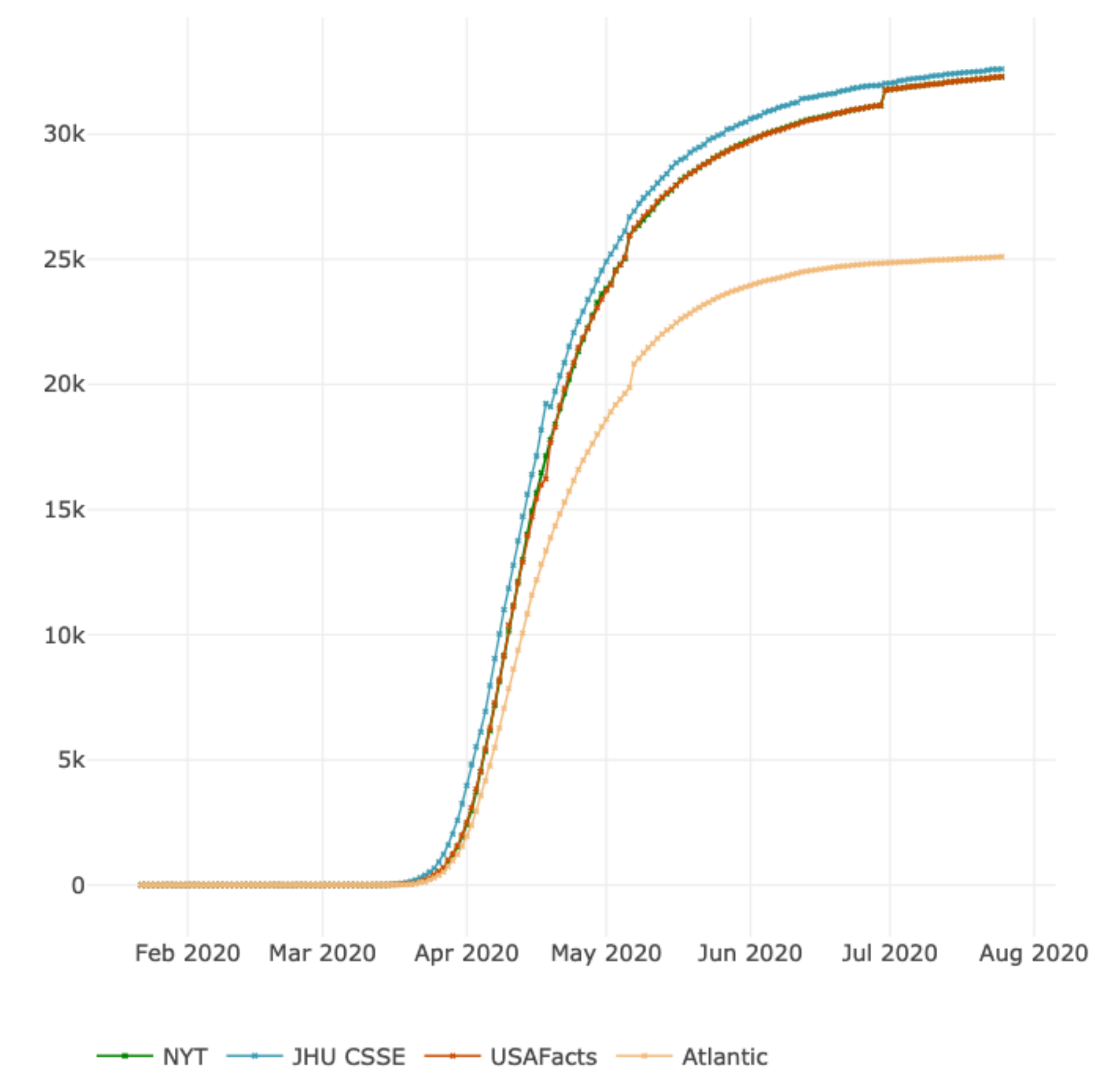} \\
		(e) Wyoming infection & (f) New York death
	\end{tabular}
	\caption{Examples of infection and death time series up until July 25, 2020.}
	\label{fig:ts_4source} 
\end{figure}

Next, we look into the state-level comparison. According to our measure, using data up until July 25, 2020, states that show dissimilar infection data are illustrated in Figure \ref{fig:state.tstat.compare}. Here we list out a few examples about how the dissimilarities arise. On the one hand, different responses to the change of probable cases reporting mechanisms in infections and/or deaths lead to discrepancies in the reported cases between the four sources. For instance, Wyoming started to include probable cases in their infected cases reporting during the week of April 6th. Each of the four sources responded to the change at different dates, as indicated by the jumps in data shown in Figure 3, with NYT being the first source to respond to the change; see Figure \ref{fig:ts_4source} (e). Similarly, Michigan started reporting probable cases and deaths after April 5. This resulted in higher infection counts in Atlantic for the following three months, probably due to a correction for probable cases; see Figure \ref{fig:ts_4source} (a) and (b). As demonstrated in the time series plots of Indiana in Figure \ref{fig:ts_4source} (c) and (d), inclusion or exclusion of probable cases also caused differences in reported cases of both infections and deaths among the four sources. Another difference among sources is caused by whether cases are reported according to the residence or the place of infection/death. For example, when reporting the death cases of New York, the Atlantic uses New York State reported deaths, while the other three sources use New York City reported deaths, which also reports deaths of residents that occur outside New York City. Starting August 6, NYT switched to reporting deaths by residence to make New York State death data consistent with the other states, which led to discontinuity on August 6. To summarize, the state-level dissimilarities are mainly caused by different report mechanisms to probable cases and varied choices of the geographical assignment.

Based on these examples, it is safe to conclude that the differences in reported cases do not indicate the inferiority or superiority of the source per se. No matter which source we use, we need to be clear and careful about the processing behind it. Generally speaking, despite the geographical rules, USAFacts tends to be more conservative because it reports confirmed counts instead of the sum of confirmed and probable counts in several states, and NYT tends to report higher county-level counts by including non-resident data and prison inmates data.

\setcounter{chapter}{4} \renewcommand{\thesection}{\arabic{section}} %
\renewcommand{\thetable}{{\arabic{table}}}
\renewcommand{\thefigure}{\arabic{figure}}
\renewcommand{\thesubsection}{4.\arabic{subsection}} \setcounter{section}{3}

\section{COVID-19 Time Series: Features and Anomaly Detection}
\label{SEC:features}

\subsection{A Seven-day Cycle in Infection and Death Cases} 

We observe a seven-day cycle in (i) reported COVID-19 new cases, and in (ii) reported COVID-19 new deaths at the national and state level. To rigorously test the seven-day cycle, we conduct the hypothesis test using the \texttt{R} package \texttt{seastests} (function \texttt{isSeasonal}) \cite{seastests}. By default, it implements the ``WO-test'', an overall seasonality test scheme proposed in \cite{Webel:Ollech:18}. Given a set of various seasonality tests, the WO-test first conducts a recursive feature elimination algorithm in conditional random forests to identify the most informative candidate tests. Then the $p$-values from the selected tests are used again as predictors to grow a single conditional inference tree. The candidate tests may include a variety of seasonality tests tailored to particular manifestations, such as the modified QS test, the Friedman test, the Kruskal-Wallis test, the periodogram test, and the Welch test.

We first conduct seasonality tests on the time series of national confirmed cases and deaths. Both time series show a seven-day seasonal behavior with $p$-values less than $10^{-7}$ from a variety of tests, including the QS test, the Friedman test, and the Kruskal-Wallis test. Then, we compare the different tests on state-level infected cases and deaths. The results are summarized in Table \ref{TAB:season_pvalue}, where we use a checkmark to signify a significant test result at significance level of $0.05$. For the infected cases, all the tests suggest a significant seven-day cyclical pattern in 22 states. Meanwhile, we observe $27$ states for which all the tests suggest a significant seven-day cyclical pattern on the death time series.

\begin{table}[]
	\caption{The detection of the seven-day cycle on the state-level times series using different tests, including the modified QS test (QS), the Friedman test (Fried), the Kruskal-Wallis test (KW), the Welch test (Welch), and the WO-test (WO) (``$\surd$'' indicates that the $p$-value is less than 0.05).}
	\label{TAB:season_pvalue}
	\centering
	\scalebox{0.9}{
		\begin{tabular}{l|ccccc|ccccc}
			\hline
			\multirow{2}{*}{State} & \multicolumn{5}{c}{Infection} &  \multicolumn{5}{c}{Death}  \\
			\cline{2-6} \cline{7-11}
			& QS & Fried & KW & Welch & WO & QS & Fried & KW & Welch & WO \\ 
			\hline
			Alabama & $\surd$ &   &   &   & $\surd$ & $\surd$ & $\surd$ & $\surd$ & $\surd$ & $\surd$ \\ 
			Arizona & $\surd$ & $\surd$ & $\surd$ & $\surd$ & $\surd$ & $\surd$ & $\surd$ & $\surd$ & $\surd$ & $\surd$ \\ 
			Arkansas & $\surd$ &   & $\surd$ & $\surd$ &   & $\surd$ &   & $\surd$ & $\surd$ & $\surd$ \\ 
			California & $\surd$ & $\surd$ & $\surd$ & $\surd$ & $\surd$ & $\surd$ & $\surd$ & $\surd$ & $\surd$ & $\surd$ \\ 
			Colorado &   & $\surd$ & $\surd$ & $\surd$ & $\surd$ & $\surd$ & $\surd$ & $\surd$ & $\surd$ & $\surd$ \\ 
			Connecticut & $\surd$ &   &   &   & $\surd$ & $\surd$ & $\surd$ & $\surd$ &   & $\surd$ \\ 
			Delaware &   &   &   &   &   &   & $\surd$ &   &   &   \\ 
			Florida & $\surd$ & $\surd$ & $\surd$ & $\surd$ & $\surd$ & $\surd$ & $\surd$ & $\surd$ & $\surd$ & $\surd$ \\ 
			Georgia & $\surd$ & $\surd$ & $\surd$ & $\surd$ & $\surd$ & $\surd$ & $\surd$ & $\surd$ & $\surd$ & $\surd$ \\ 
			Idaho & $\surd$ & $\surd$ & $\surd$ & $\surd$ & $\surd$ & $\surd$ &   & $\surd$ & $\surd$ & $\surd$ \\ 
			Illinois &   & $\surd$ & $\surd$ & $\surd$ & $\surd$ & $\surd$ & $\surd$ & $\surd$ & $\surd$ & $\surd$ \\ 
			Indiana & $\surd$ & $\surd$ & $\surd$ & $\surd$ & $\surd$ & $\surd$ & $\surd$ & $\surd$ & $\surd$ & $\surd$ \\ 
			Iowa & $\surd$ & $\surd$ & $\surd$ & $\surd$ & $\surd$ & $\surd$ & $\surd$ & $\surd$ & $\surd$ & $\surd$ \\ 
			Kansas & $\surd$ & $\surd$ & $\surd$ & $\surd$ & $\surd$ & $\surd$ &   &   &   &   \\ 
			Kentucky &   & $\surd$ & $\surd$ &   & $\surd$ & $\surd$ & $\surd$ & $\surd$ & $\surd$ & $\surd$ \\ 
			Louisiana &   & $\surd$ & $\surd$ &   &   & $\surd$ & $\surd$ & $\surd$ & $\surd$ & $\surd$ \\ 
			Maine & $\surd$ & $\surd$ & $\surd$ & $\surd$ & $\surd$ &   & $\surd$ & $\surd$ & $\surd$ &   \\ 
			Maryland & $\surd$ & $\surd$ & $\surd$ & $\surd$ & $\surd$ & $\surd$ & $\surd$ & $\surd$ & $\surd$ & $\surd$ \\ 
			Massachusetts & $\surd$ & $\surd$ & $\surd$ & $\surd$ & $\surd$ & $\surd$ & $\surd$ & $\surd$ & $\surd$ & $\surd$ \\ 
			Michigan &   & $\surd$ & $\surd$ & $\surd$ & $\surd$ & $\surd$ & $\surd$ & $\surd$ & $\surd$ & $\surd$ \\ 
			Minnesota & $\surd$ & $\surd$ & $\surd$ & $\surd$ & $\surd$ & $\surd$ & $\surd$ & $\surd$ & $\surd$ & $\surd$ \\ 
			Mississippi &   & $\surd$ & $\surd$ & $\surd$ & $\surd$ & $\surd$ & $\surd$ & $\surd$ & $\surd$ & $\surd$ \\ 
			Missouri & $\surd$ &   & $\surd$ & $\surd$ & $\surd$ & $\surd$ & $\surd$ & $\surd$ & $\surd$ & $\surd$ \\ 
			Montana & $\surd$ &   &   &   & $\surd$ &   & $\surd$ &   &   &   \\ 
			Nebraska & $\surd$ & $\surd$ & $\surd$ & $\surd$ & $\surd$ & $\surd$ & $\surd$ & $\surd$ & $\surd$ & $\surd$ \\ 
			Nevada &   & $\surd$ & $\surd$ & $\surd$ &   & $\surd$ & $\surd$ & $\surd$ & $\surd$ & $\surd$ \\ 
			New Hampshire &   & $\surd$ & $\surd$ & $\surd$ & $\surd$ & $\surd$ & $\surd$ & $\surd$ & $\surd$ & $\surd$ \\ 
			New Jersey &   &   &   &   &   &   & $\surd$ & $\surd$ & $\surd$ & $\surd$ \\ 
			New Mexico &   & $\surd$ & $\surd$ & $\surd$ & $\surd$ &   & $\surd$ & $\surd$ & $\surd$ & $\surd$ \\ 
			New York & $\surd$ & $\surd$ & $\surd$ & $\surd$ & $\surd$ &   &   &   &   &   \\ 
			North Carolina & $\surd$ & $\surd$ & $\surd$ & $\surd$ & $\surd$ & $\surd$ & $\surd$ & $\surd$ & $\surd$ & $\surd$ \\ 
			North Dakota &   & $\surd$ & $\surd$ & $\surd$ & $\surd$ & $\surd$ & $\surd$ & $\surd$ &   &   \\ 
			Ohio & $\surd$ & $\surd$ & $\surd$ & $\surd$ & $\surd$ & $\surd$ & $\surd$ & $\surd$ & $\surd$ & $\surd$ \\ 
			Oklahoma & $\surd$ & $\surd$ & $\surd$ & $\surd$ & $\surd$ & $\surd$ & $\surd$ & $\surd$ & $\surd$ & $\surd$ \\ 
			Oregon &   & $\surd$ &   &   &   & $\surd$ & $\surd$ & $\surd$ & $\surd$ & $\surd$ \\ 
			Pennsylvania & $\surd$ & $\surd$ & $\surd$ & $\surd$ & $\surd$ & $\surd$ & $\surd$ & $\surd$ & $\surd$ & $\surd$ \\ 
			Rhode Island & $\surd$ &   &   & $\surd$ & $\surd$ & $\surd$ &   &   &   & $\surd$ \\ 
			South Carolina &   & $\surd$ & $\surd$ & $\surd$ & $\surd$ &   & $\surd$ & $\surd$ & $\surd$ & $\surd$ \\ 
			South Dakota & $\surd$ & $\surd$ & $\surd$ & $\surd$ & $\surd$ &   & $\surd$ & $\surd$ & $\surd$ & $\surd$ \\ 
			Tennessee & $\surd$ &   &   & $\surd$ & $\surd$ & $\surd$ & $\surd$ & $\surd$ & $\surd$ & $\surd$ \\ 
			Texas & $\surd$ & $\surd$ & $\surd$ & $\surd$ & $\surd$ & $\surd$ & $\surd$ & $\surd$ & $\surd$ & $\surd$ \\ 
			Utah & $\surd$ & $\surd$ & $\surd$ & $\surd$ & $\surd$ & $\surd$ &   & $\surd$ & $\surd$ & $\surd$ \\ 
			Vermont &   &   &   &   &   &   & $\surd$ &   &   &   \\ 
			Virginia & $\surd$ & $\surd$ & $\surd$ &   &   & $\surd$ & $\surd$ & $\surd$ & $\surd$ & $\surd$ \\ 
			Washington & $\surd$ & $\surd$ & $\surd$ & $\surd$ & $\surd$ &   & $\surd$ & $\surd$ & $\surd$ & $\surd$ \\ 
			West Virginia & $\surd$ & $\surd$ & $\surd$ &   & $\surd$ &   &   &   &   &   \\ 
			Wisconsin & $\surd$ & $\surd$ & $\surd$ & $\surd$ & $\surd$ & $\surd$ & $\surd$ & $\surd$ & $\surd$ & $\surd$ \\ 
			Wyoming &   &   & $\surd$ &   &   &   &   &   &   &   \\ 
			District of Columbia & $\surd$ &   &   &   &   &   &   & $\surd$ &   &   \\ 
			\hline
	\end{tabular}}
\end{table}

The cyclical pattern is less evident at the county level. After applying the Friedman test to more than 3,000 counties, $701$ counties show the seven-day cyclical pattern in infected cases with $p$-value smaller than $0.05$. For the deaths, only $200$ counties exhibit a cyclical behavior with $p$-value smaller than $0.05$. 

\begin{figure}[!ht]
	\begin{center}
		\includegraphics[width = .95\textwidth, height  = 2.8in]{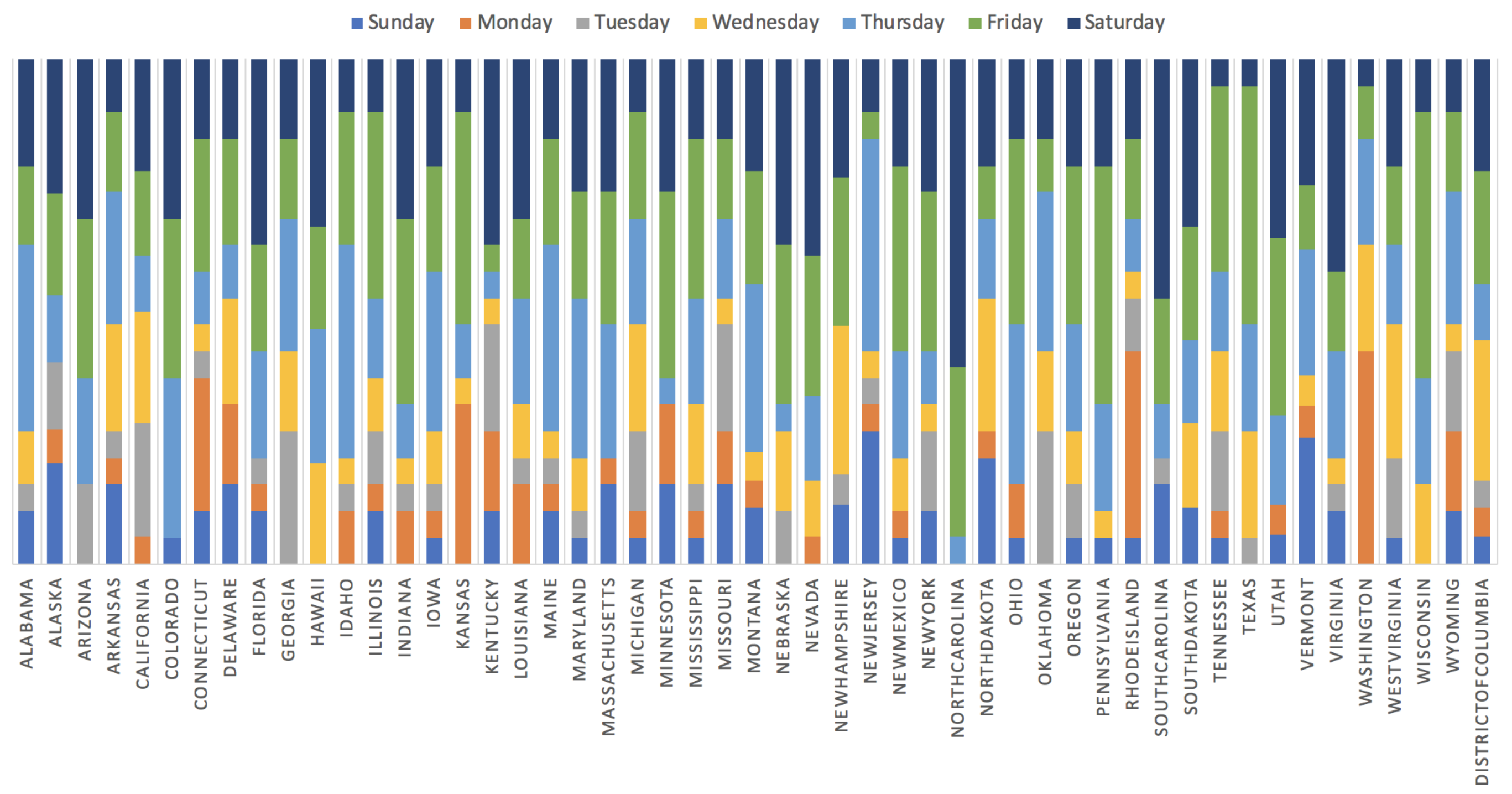} \\
		(a) Infection\\
		\includegraphics[width = .95\textwidth, height  = 2.8in]{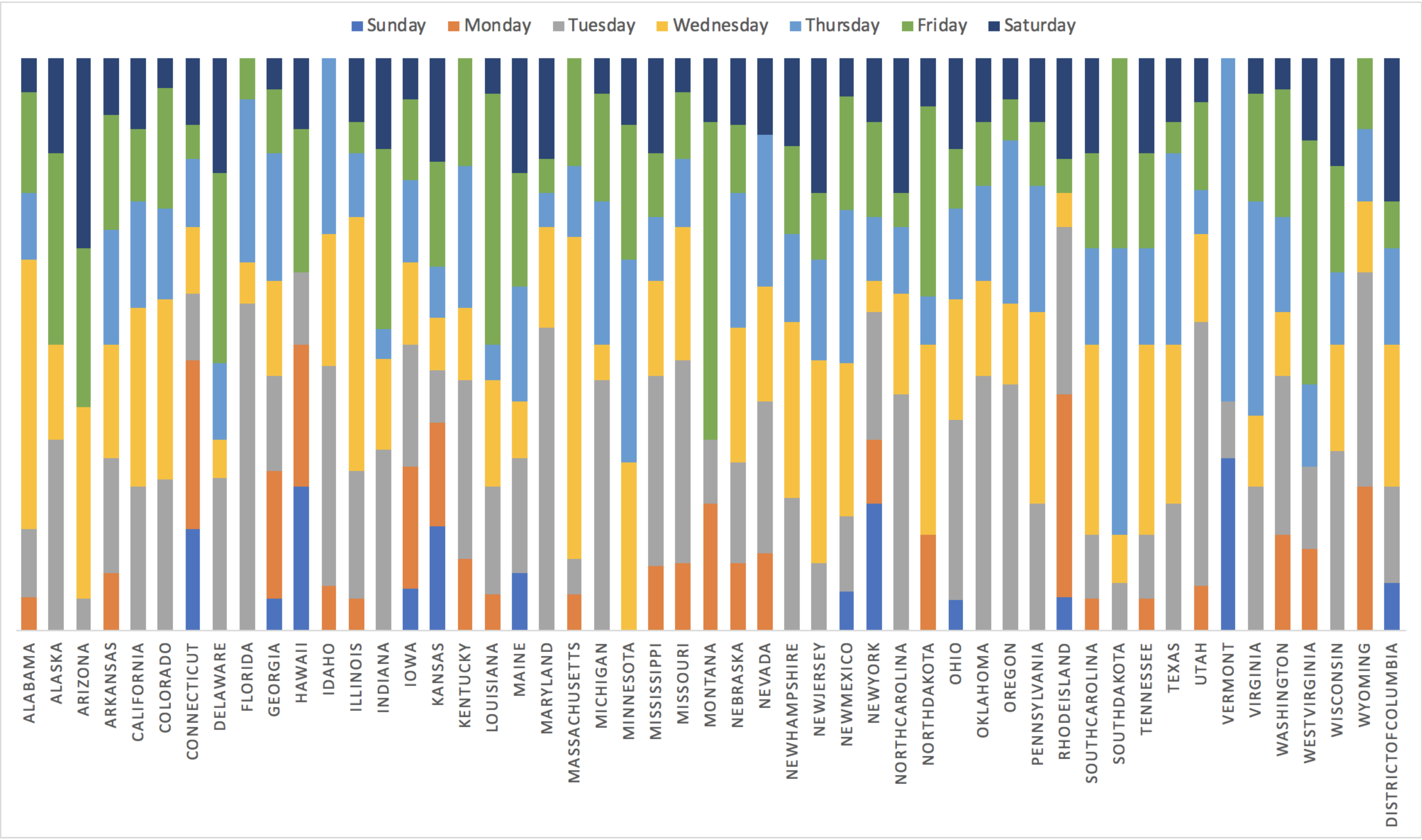} \\
		(b) Death
	\end{center}
	\caption{The 100\% stacked column bar plot of the number of weeks that reaches the weekly maximum of the infection or death counts across days of the week in different states.}
	\label{FIG:case_max} 
\end{figure}

The stacked column bar plots in Figure \ref{FIG:case_max} show the day on which the most infection/death cases are reported for each week in each state. From Figure \ref{FIG:case_max} (a), we can observe that in the states, such as North Carolina, South Carolina and Virginia, Saturdays usually report the most infections. States such as Pennsylvania, Tennessee, and Wisconsin, report more infection cases on Fridays. For death cases, many states reach their peak on Tuesdays; see Figure \ref{FIG:case_max} (b) for more details.

\begin{figure}[!ht]
	\begin{center}
		\includegraphics[width = .95\textwidth, height  = 2.8in]{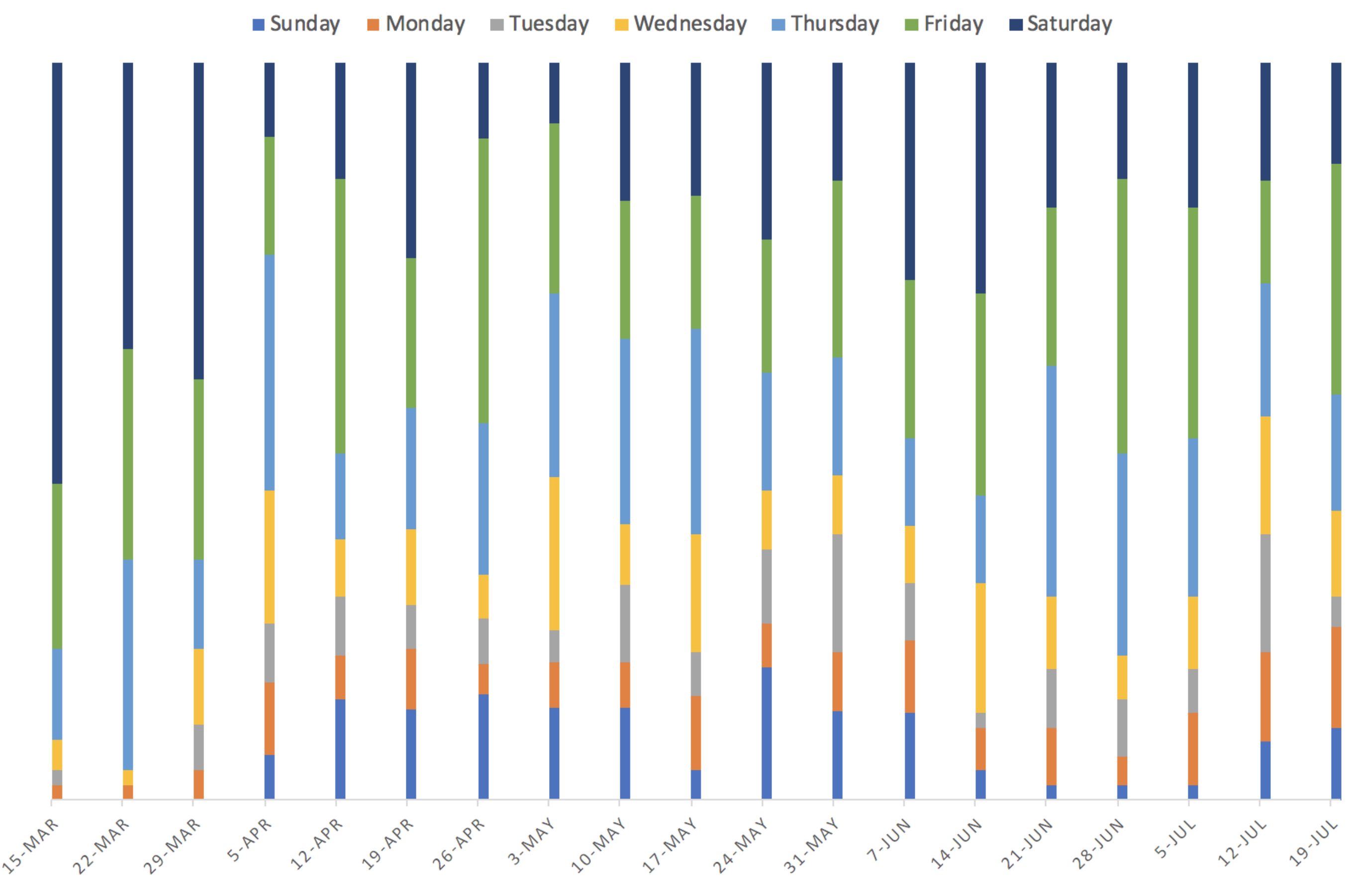} \\
		(a) Infection\\
		\includegraphics[width = .95\textwidth, height  = 2.8in]{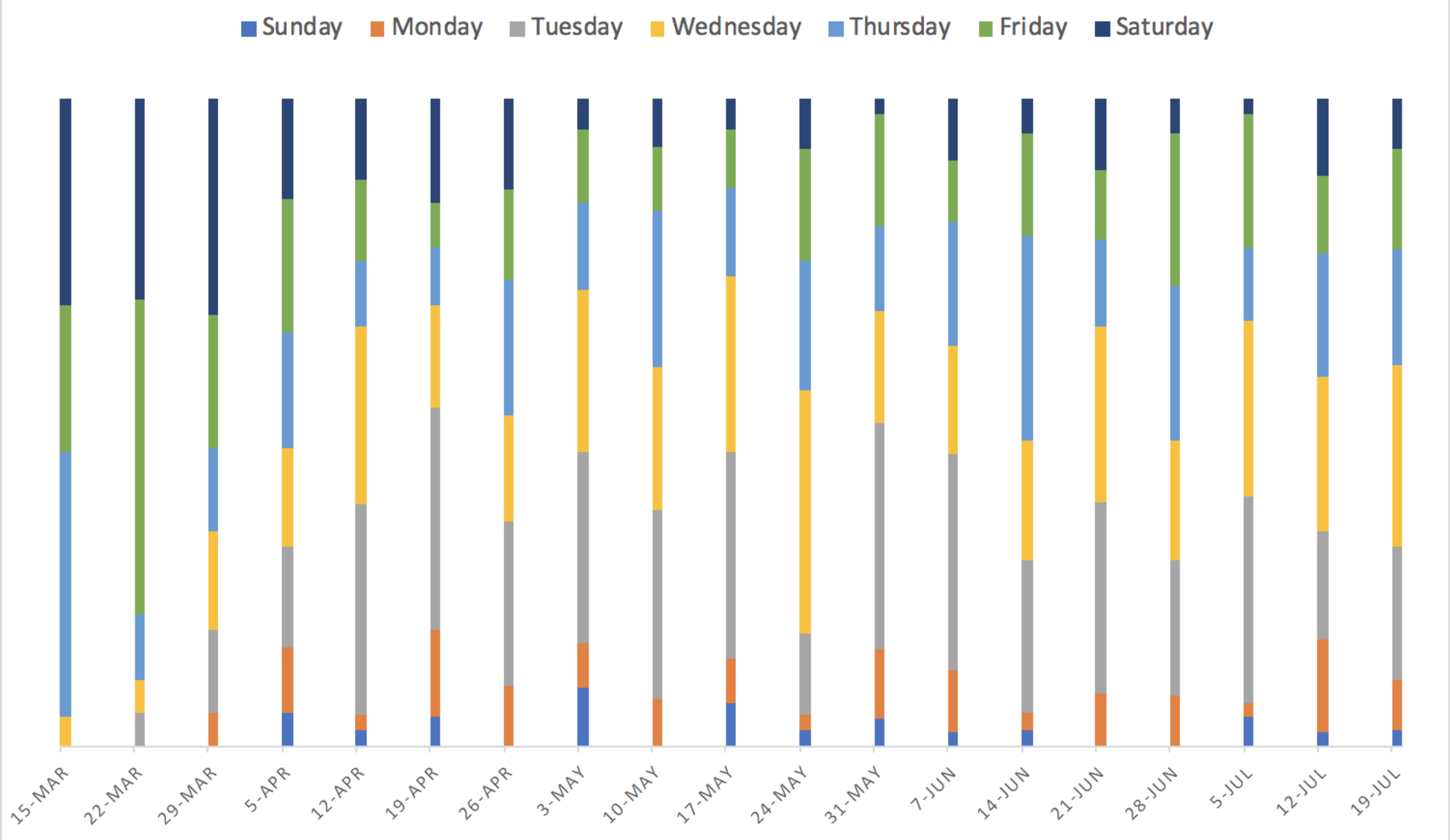} \\
		(b) Death
	\end{center}
	\caption{The 100\% stacked column bar plot of the number of states that reaches the weekly maximum of the infection or death counts across days of the week in different weeks.}
	\label{FIG:case_max_state} 
\end{figure}

The stacked column bar plots in Figure \ref{FIG:case_max_state} illustrate the day on which the highest infection/death cases are reported for each week from March 15 to July 25. From Figure \ref{FIG:case_max_state} (a), one can see that in the early stage of the pandemic (before April), Saturdays usually had the highest number of infections. Later on, Fridays often had the most infections. As seen in Figure \ref{FIG:case_max_state} (b), Sundays and Mondays typically reported a smaller number of death counts than the other days. Meanwhile, the peaks often occured on either Tuesdays or Wednesdays. In time series analysis, for a short-term forecast, a few approaches can be considered to remove the weekly cycle: simple differencing ($Y_t-Y_{t-7}$); seven-day moving average; using dummy variables  control for day-to-day variation; harmonic function (series of sine/cosine functions).

\subsection{Anomaly Detection} \label{ssec:detection}

\begin{figure}[!ht]
	\begin{center}
		\begin{tabular}{cc}
			\includegraphics[width = .48\textwidth]{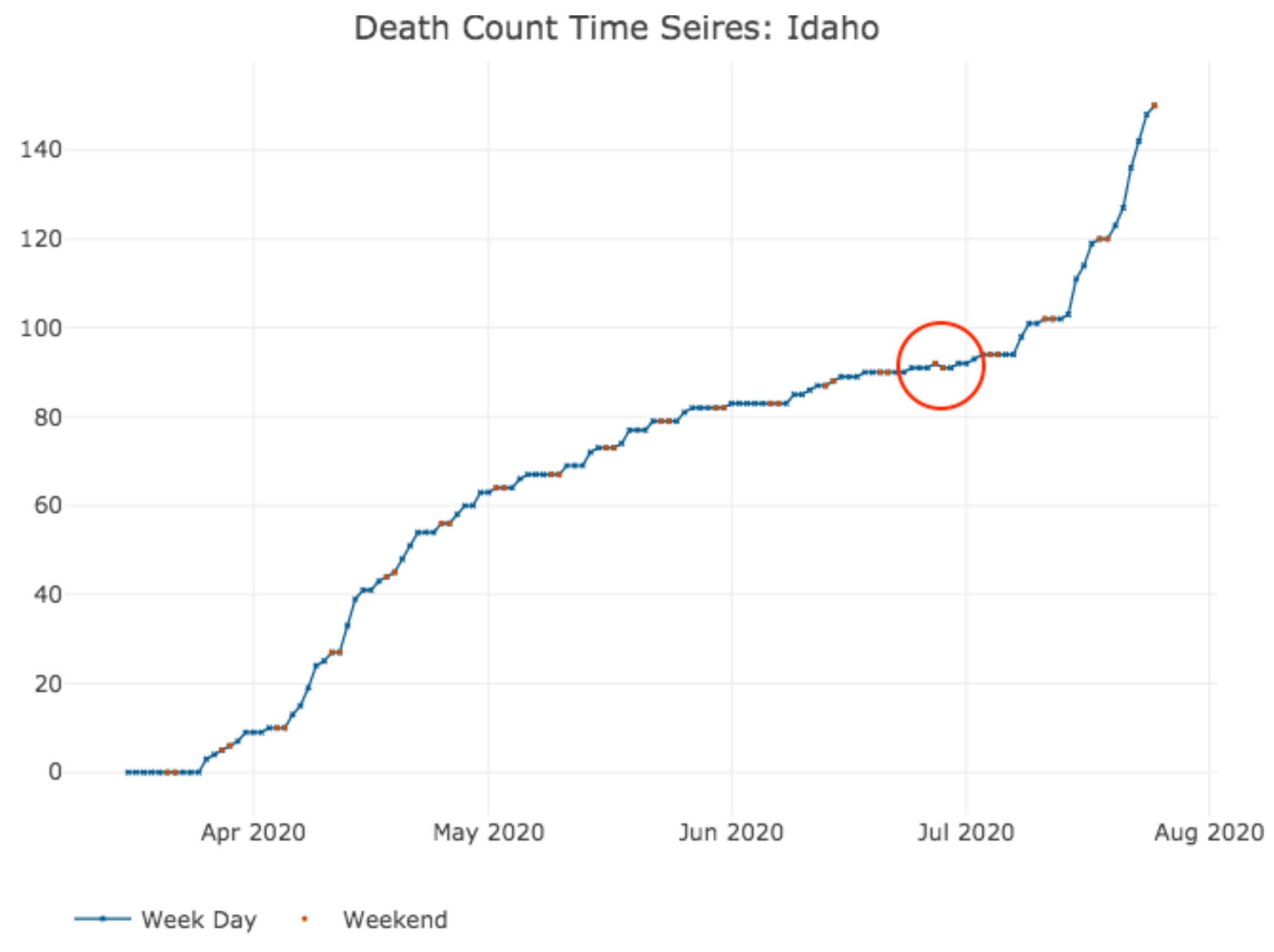} &\includegraphics[width = .48\textwidth]{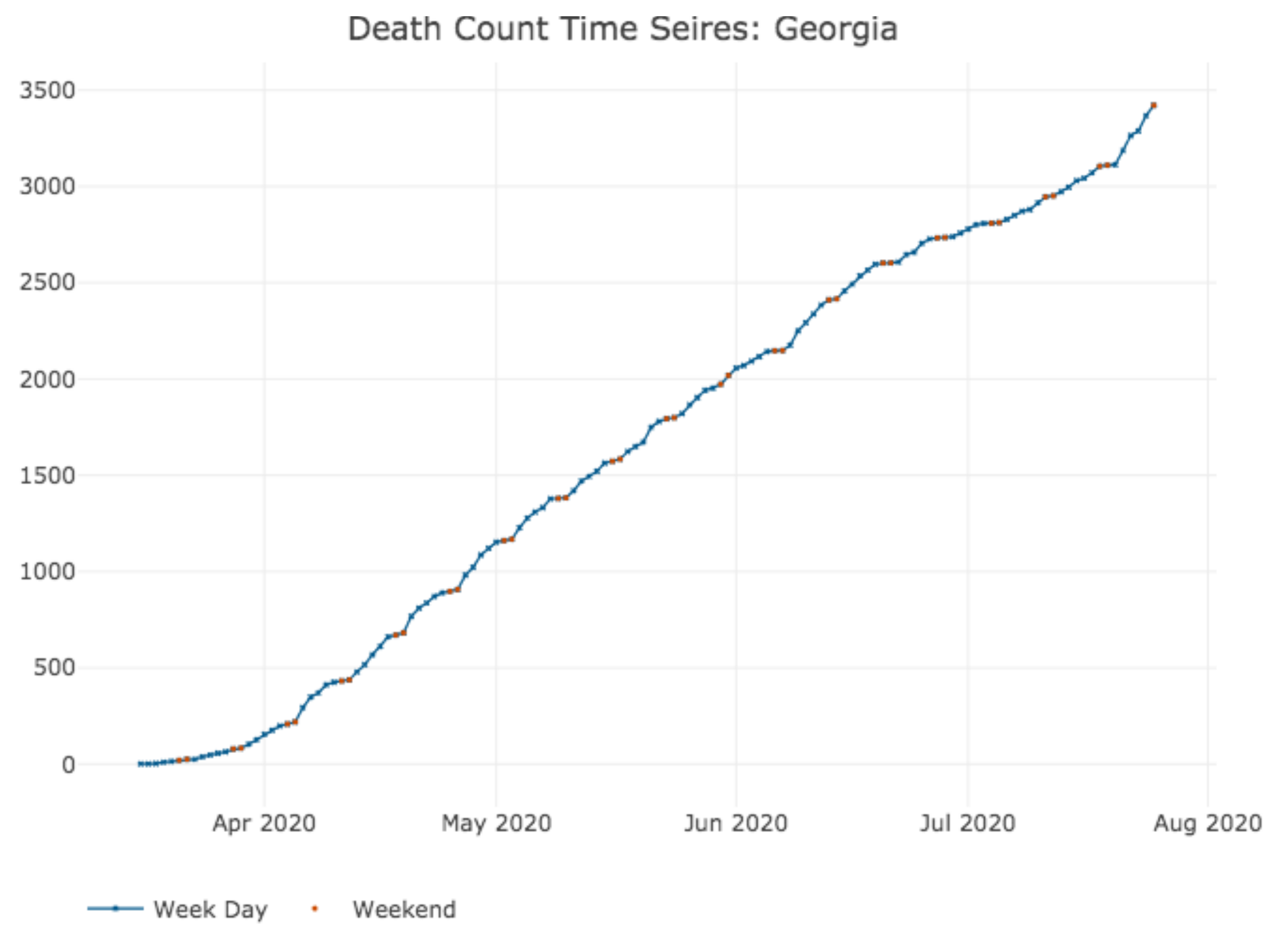}\\
			(a) & (b)\\[6pt]
			\includegraphics[width = .48\textwidth]{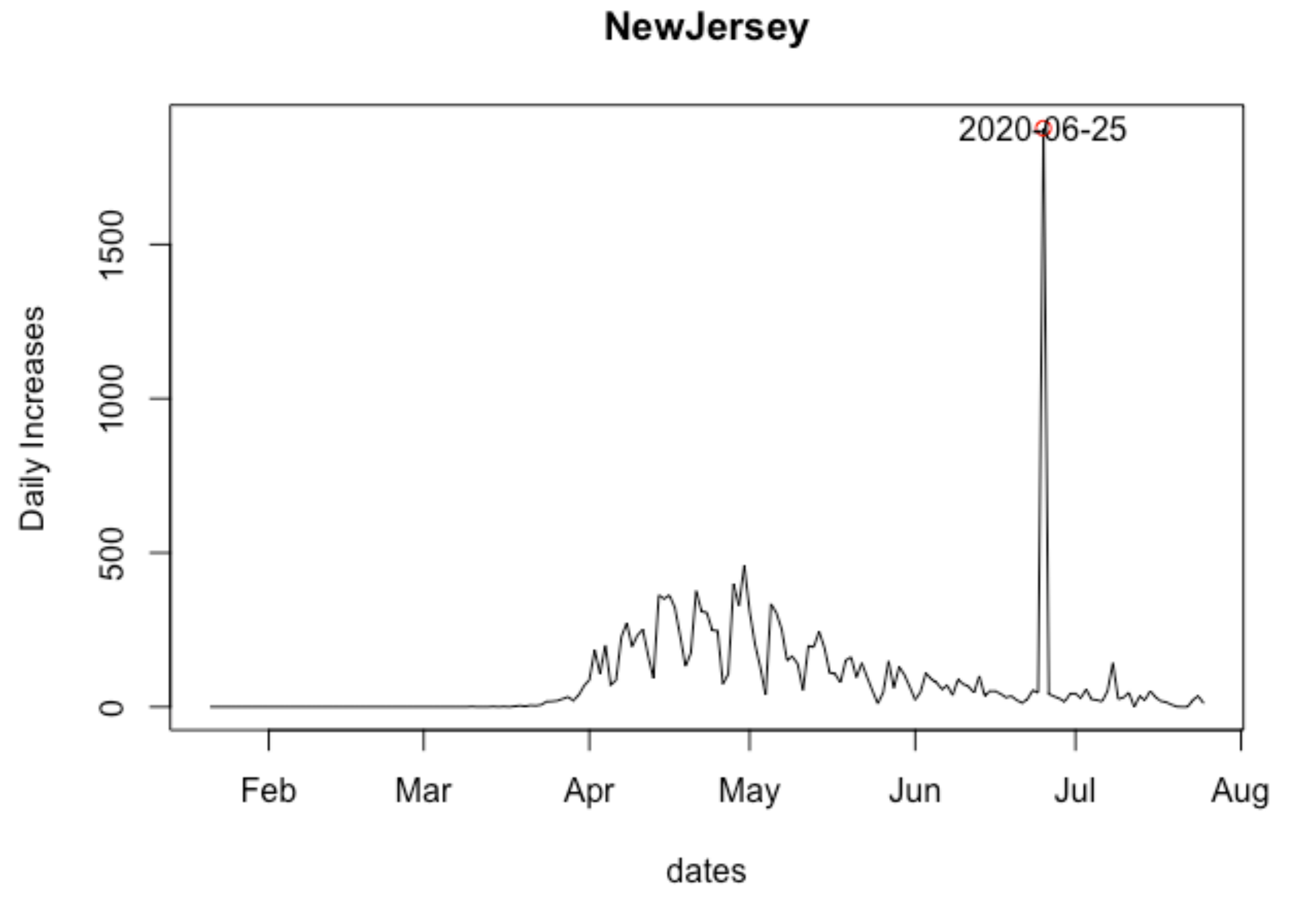} &\includegraphics[width = .48\textwidth]{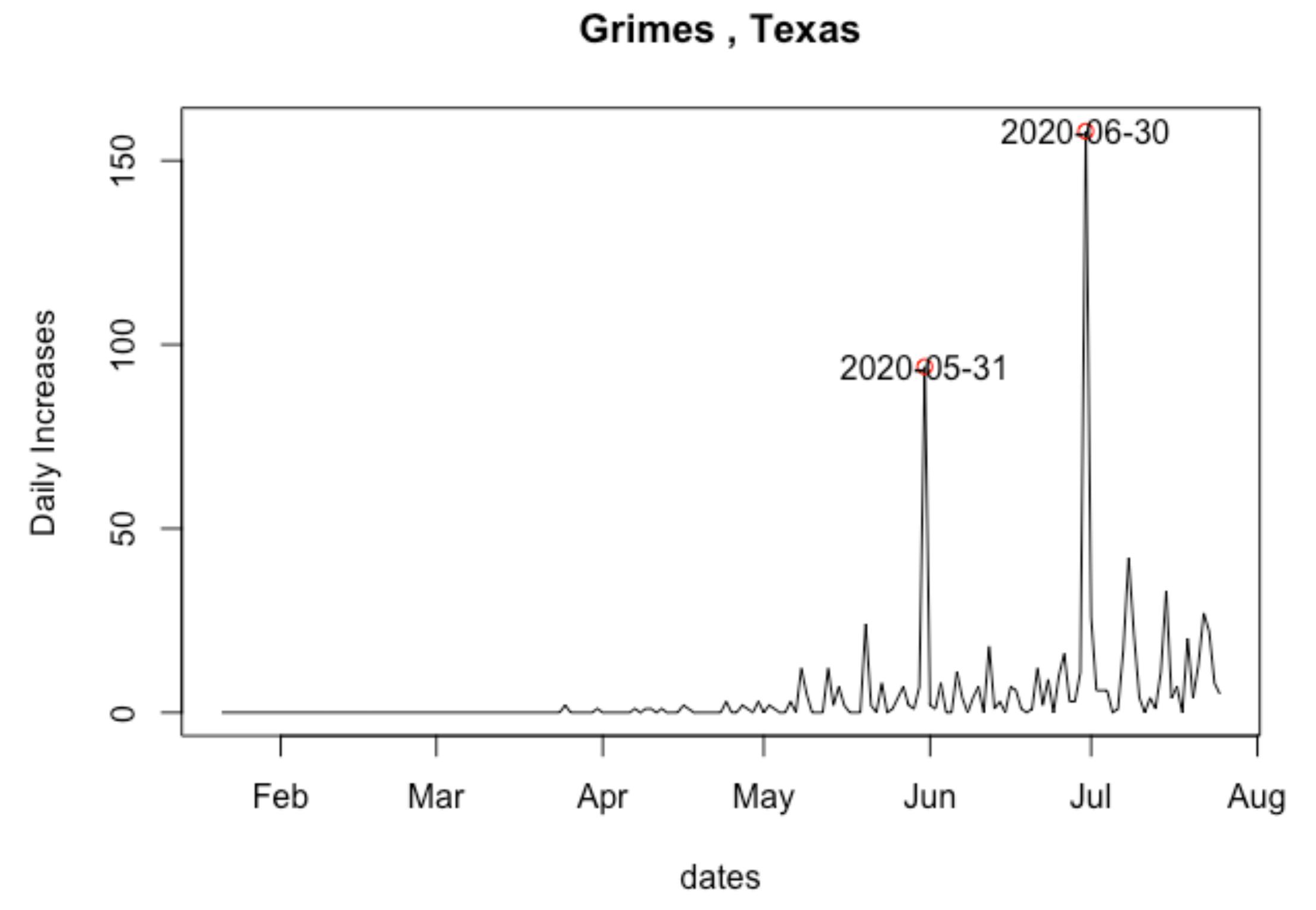} \\
			(c) & (d)\\[6pt]
		\end{tabular}
	\end{center}
	\caption{An illustration of different types of anomalies. (a) order dependency; (b) weekend/holiday delay-reported issue; (c) single point anomaly; (d) two point anomalies.}
	\label{FIG:anomaly_eg} 
\end{figure}

In addition to the exciting findings aforementioned in the raw data collection, we observe two major types of anomalies in the data: (I) order dependencies violation, and (II) point anomalies. Examples of these two issues are illustrated in Figure \ref{FIG:anomaly_eg}. Before conducting any analysis of the epidemic data, one might need to account for these issues. In this section, we use the epidemic data from NYT as an illustration, but all four data sources exhibit similar issues.

\subsubsection{4.2.1 Order Dependencies Violation} \label{sssec:orderdep}

Order dependency (OD) is widely used in the relational database. In this project, we incorporate this concept into the anomaly detection and data repairing process of cumulative time series. To be more specific, the OD for the cumulative time series can be defined as follows: for any two time points, $t_1$ and $t_2$, if $t_1 < t_2$, then $Y_{t_1}\leq Y_{t_2}$, where $Y_t$ represents the cumulative infection/death count on day $t$. Obviously, the time series in Figure \ref{FIG:anomaly_eg} (a) violates the OD.

\subsubsection{4.2.2 Point Anomalies} \label{sssec:pointanomal}

A point anomaly refers to the situation where there is one day of an abrupt increase in the cumulative or daily new time series. This can also be considered as a violation of speed constraints. The anomaly can be caused by a number of factors, including (1) the result of a large batch of tests was released and (2) the change of reporting standard, such as some states starting to report probable cases from a specific date. For example, the anomaly in Figure \ref{FIG:anomaly_eg} (c) is due to New Jersey reporting 1,854 probable deaths that may date back to earlier in the outbreak. The anomaly in Figure \ref{FIG:anomaly_eg} (d) might be related to the reported cases in the Texas Department of Criminal Justice (TDCJ). On May 31, at least 82 of the active cases in the county were TDCJ-related. To detect this type of anomaly, we exam the increasing speed of the time series. Specifically, if the time series $Y_{t_i}$ satisfies some speed constraint (SC), then,
\begin{equation*}
(Y_{t_2} - Y_{t_1})/(t_2 - t_1) < SC_1~\mathrm{and}~\frac{\left(Y_{t_1 + 1} - Y_{t_1}\right)}{\left(Y_{t_2} - Y_{t_1}\right)/(t_2 - t_1)} < SC_2,
\end{equation*}
where $SC_1$ and $SC_2$ are two predetermined thresholds.

\subsection{Change Points in Time Series}

Sometimes, we may experience a pattern change in the time series, which can be referred to as the period when the increasing speed is significantly different from the previous period. We apply the function \texttt{segmented} in \texttt{R} package \texttt{segmented} \cite{segmented} to detect the change points. This function implements the segmented models in which the relationship between response and covariate(s) is modeled as piecewise linear segments connected at some joint points (or change points). Once a change point is detected, we will provide a warning message to let the user decide whether any repair is necessary. 

\begin{figure}[!ht]
	\begin{center}
		\begin{tabular}{ccc}
			\includegraphics[width=.35\textwidth]{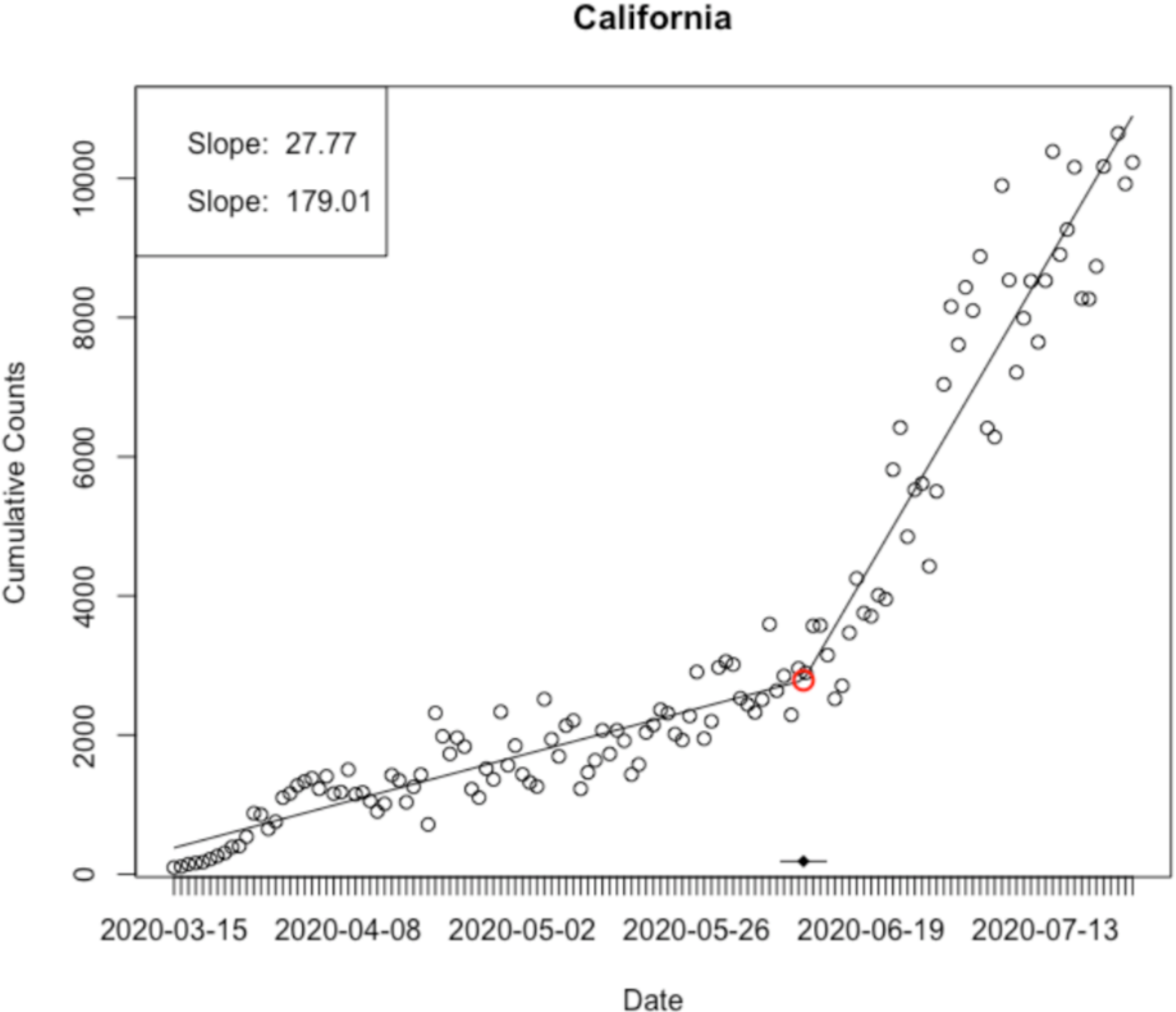} &
			\includegraphics[width=.35\textwidth]{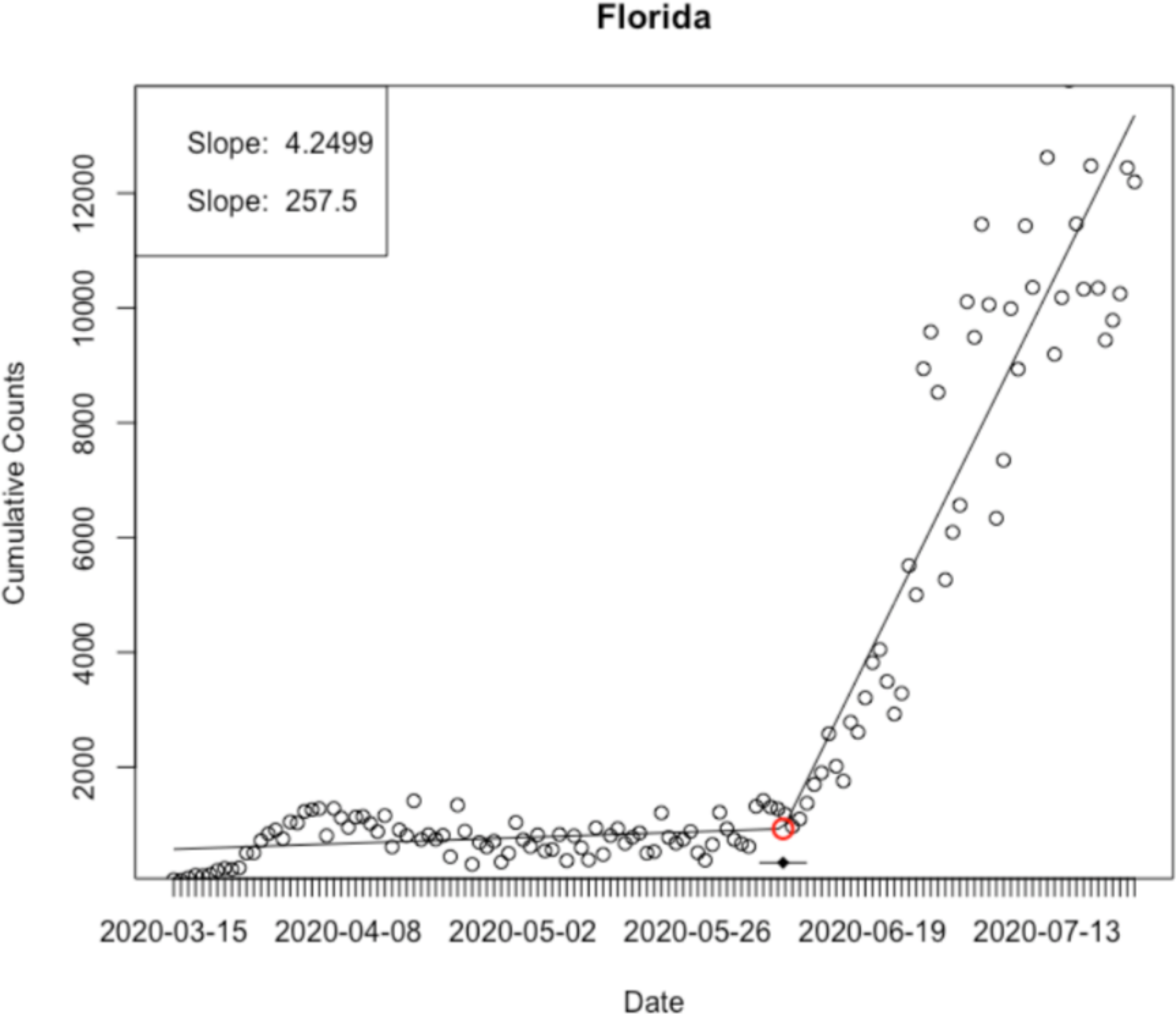}\\
			(a) & (b) \\
			\includegraphics[width=.35\textwidth]{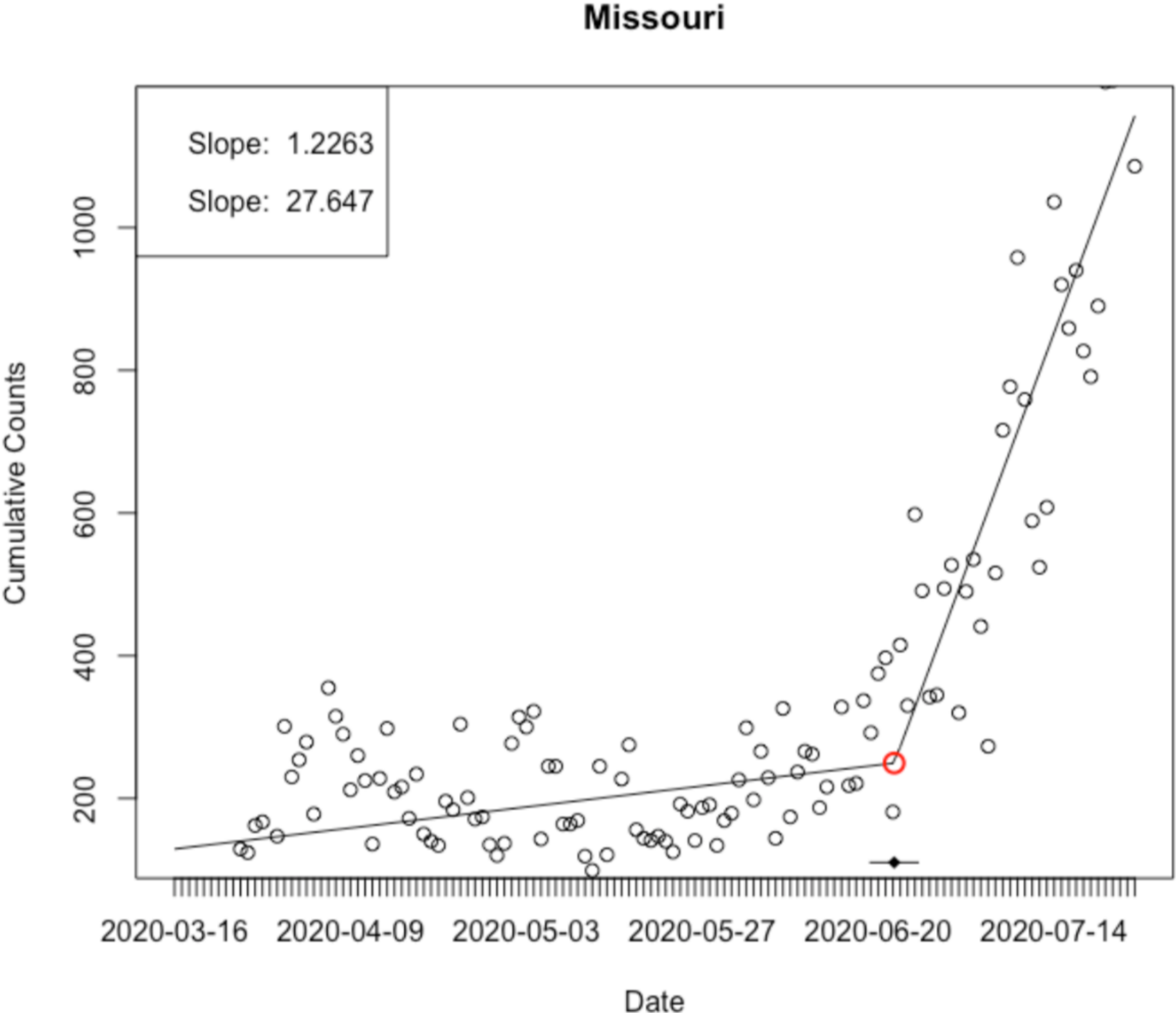}
			&\includegraphics[width=.35\textwidth]{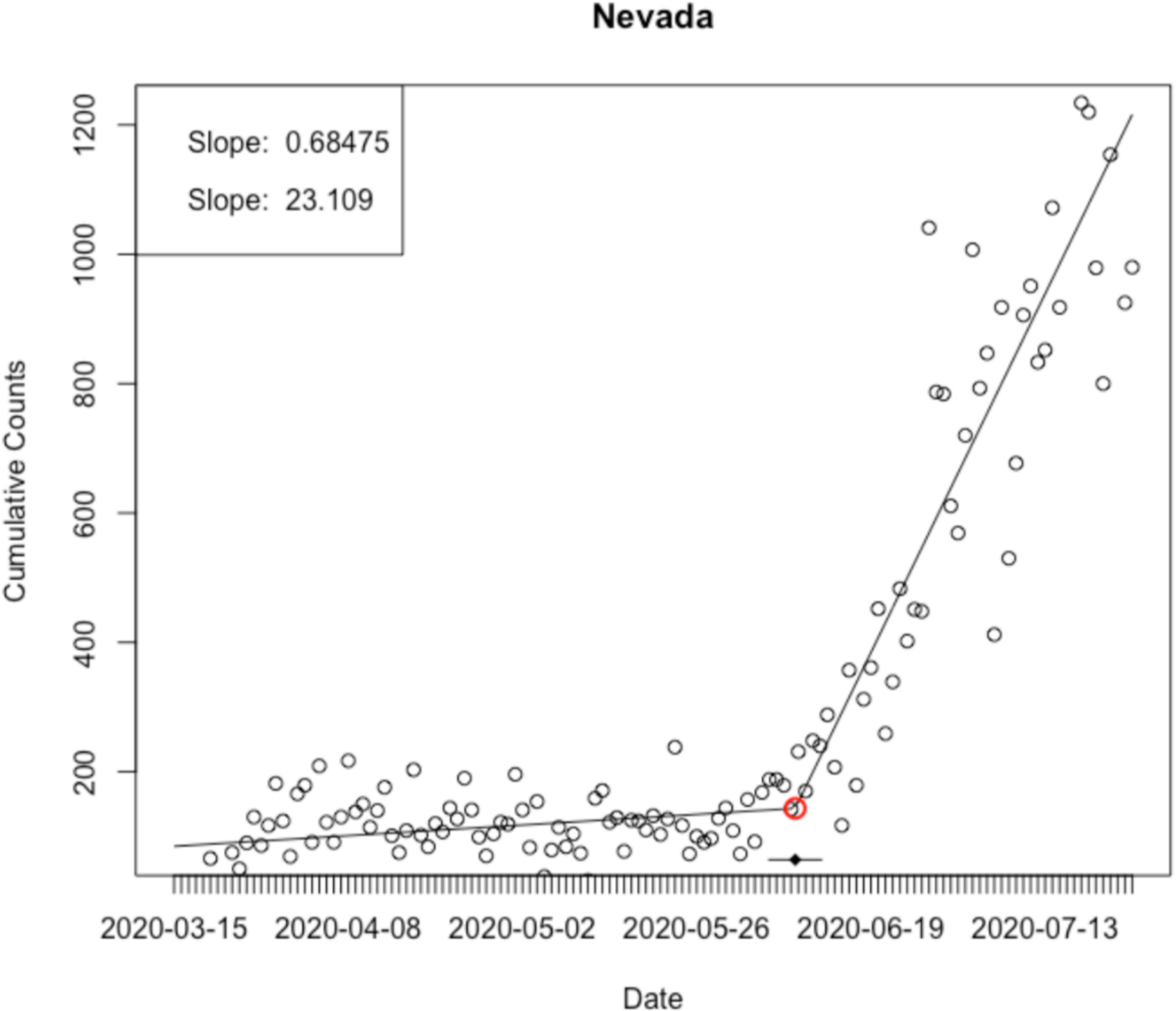} \\
			(c) & (d) \\
			\includegraphics[width=.35\textwidth]{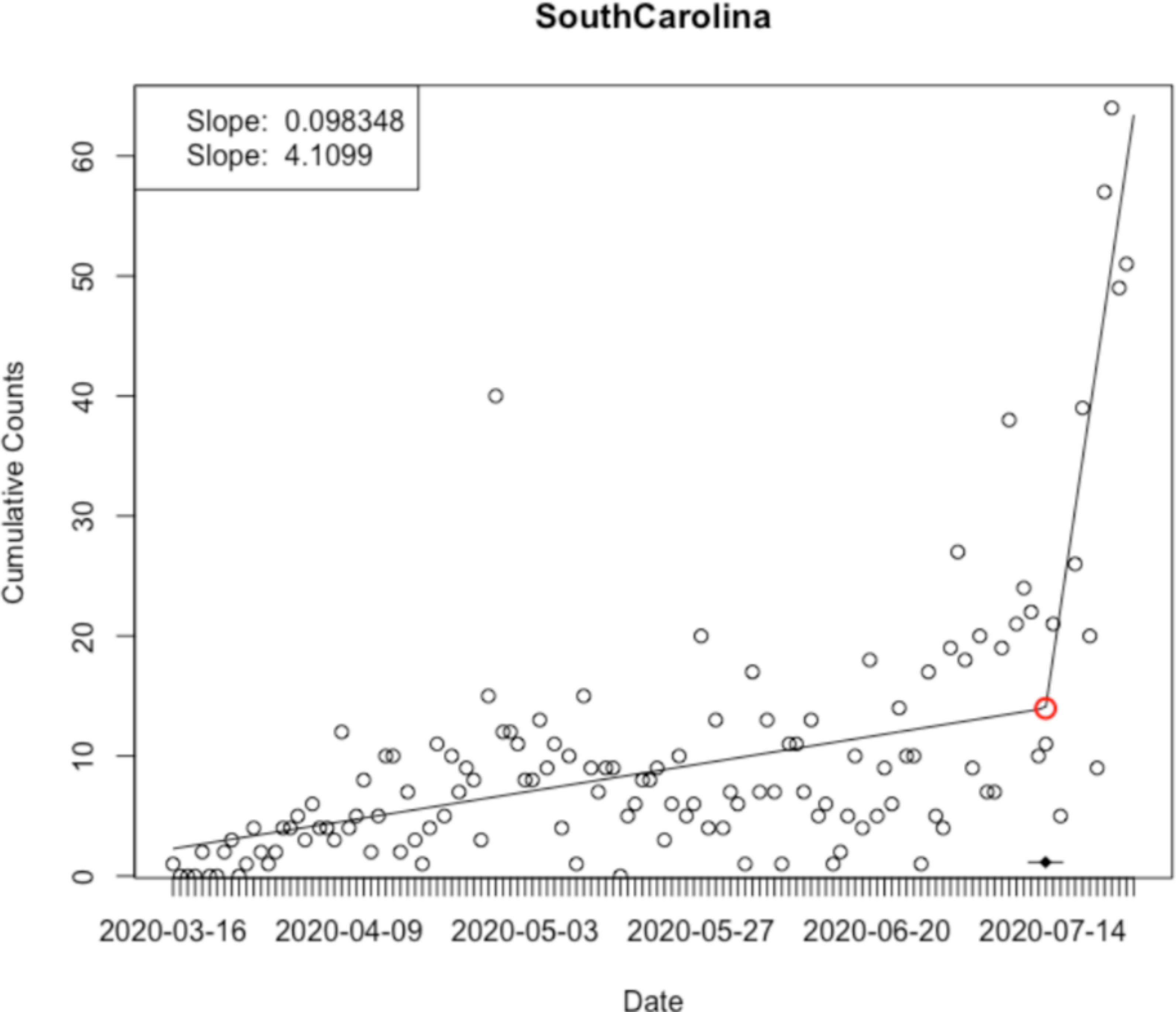}
			&\includegraphics[width=.35\textwidth]{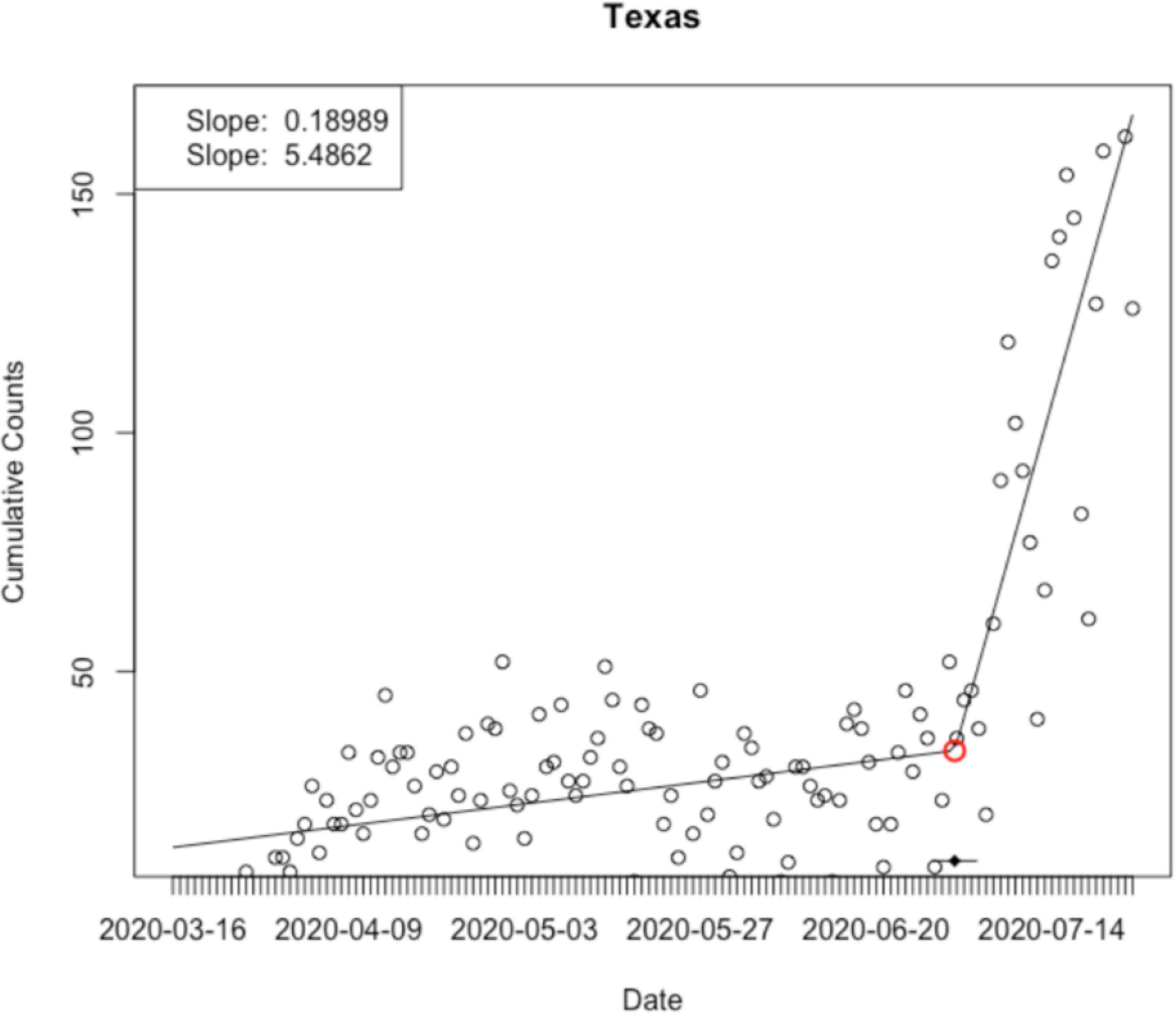}\\
			(e) & (f)
		\end{tabular}
	\end{center}
	\vspace{-0.15in}\caption{Time series plots of the states with change points identified, where the black circle represents the daily observed value, and the red circle indicates the change point detected, and the segment. shows the linear regression line before and after the change point. (a) new infection cases in California; (a) new infection cases in California; (b) new infection cases in Florida; (c) new infection cases in Missouri; (d) new infection cases in Nevada; (e) new deaths in South Carolina; (f) new deaths in Texas.} 
	\label{FIG:ChangePoint} 
\end{figure}

For simplicity, we consider a segmented relationship between the response and the time. Let $\mu_i=E(Y_{t_i})$, and the variable (time) $t_i$ is modeled by $g(\mu_i) =\beta_1 t_i+\beta_2(t_i-\phi)_{+}$, where $g(\cdot)$ is the link function, $(t_i-\phi)_{+} = (t_i-\phi)I(t_i>\phi)$, and $I(\cdot)$ is the indicator function. Here, $\beta_1$ illustrates the slope of left line segment and $\beta_2$ represents the difference-in-slopes. The main idea lies in testing whether $|\beta_2|>0$. If a break-point does not exist, the difference-in-slopes parameter has to be zero. Table \ref{TAB:ChangePoint} presents the states in which change points are detected in daily new infections and deaths, and the dates of the change points are identified. Figure \ref{FIG:ChangePoint} visualizes the identified change points together with the time series of daily new infections and deaths. Based on the change point analysis, most of the changes occurred in June and July with sudden increases in incident cases and death.

\begin{table}
	\caption{A list states with change-point identified in the daily new infected cases and deaths.}
	\label{TAB:ChangePoint}
	\centering
	\begin{tabular}{lcclc}
		\hline
		\multicolumn{2}{c}{Infection} & & \multicolumn{2}{c}{Death}\\ \cline{1-2}\cline{4-5}
		State  & Change Point & & State  & Change Point\\
		\hline
		California & 2020-06-10 & & South Carolina & 2020-07-13\\
		Florida & 2020-06-07 & & Texas & 2020-07-01\\
		Missouri & 2020-06-23 \\
		Nevada & 2020-06-09\\\hline
	\end{tabular}
\end{table}

\setcounter{chapter}{5} \renewcommand{\thesection}{\arabic{section}} %
\renewcommand{\thetable}{{\arabic{table}}}
\renewcommand{\thefigure}{\arabic{figure}}
\renewcommand{\thesubsection}{5.\arabic{subsection}} \setcounter{section}{4}

\section{Data Repairing}
\label{SEC:repair}

Once raw data is collected, we start with the OD violation detection and repairing. Next, we check for the point anomalies, and let the user decide whether to repair it. Last, we investigate the weekly cycles and pattern changes in the time series. In this section, we propose several data repairing methods to handle the issues mentioned in Section \ref{SEC:features}. To resolve these issues, we focus on the daily new infected/death cases instead of the cumulative infected/death cases. In the following, we let $Z_t = Y_t-Y_{t-1}$ be the increase at time point $t$. 

\subsection{Anomaly Repairing}
\label{SUBSEC:ODrepair}

First of all, the daily reported infected/death count could be considered as a count time series by nature. Therefore, when repairing a count time series, we need to take into account that the observations are nonnegative integers, and we should utilize the dependence structure among observations. Furthermore, in the study of the infectious disease, the population is usually assigned to compartments such as Susceptible ($S$), Infectious ($I$), or Recovered ($R$), and people may progress between compartments. Therefore, different compartments are usually considered as an entire system and are studied together; see for example, the SIR models \cite{Brauer:08, Pfeiffer:08, Lawson:16}. Third, the spread of the disease also has a spatial pattern. In general, once a point anomaly is detected, we let $\mathcal{A}=\{t\in \mathcal{T}: \hbox{$Z_t$ is identified as a point anomaly}\}$. For $t\in \mathcal{A}$, the user can decide whether a correction is necessary. If so, the user can choose from the following methods to obtain a more reliable value, $\widehat{Z}_t$, to replace the point anomaly $Z_t$.

\subsubsection{5.1.1 Time Series Model for Count Data}
\label{sssec:timeseriesmodel}

One of the conventional methods to deal with these challenges is the generalized linear model (GLM), which models the observations conditionally on past information. In this project, we consider both Poisson and Negative Binomial as the conditional distribution. The second important class for analyzing count time series is the integer autoregressive moving average models, and a comprehensive review is given by \cite{Wei:08}. The state-space is another type of count time series models. Comparing with the GLM, it allows a more flexible data generating process. However, it requires a more complicated model specification. Due to the explicit formulation, the GLM-based models yield a more convenient way to make predictions. Thus, in this project, we focus on the GLM-based method.

To repair the dataset, we model the conditional mean $\mu_t=\mathrm{E}(Z_t|Z_{t-1}, \mu_{t-1})$ in the following form
\begin{equation*}
\nu_t = \beta_0 + \sum_{k=1}^p \beta_k Z_{t-k} + \sum_{l=1}^q \alpha_l \nu_{t-l},
\end{equation*}
where $\nu_t=\log{(\mu_t)}$. 

For this type of data repairing, we use the \texttt{R} package \texttt{tscount} \cite{Liboschik:etal:17}, which conducts a model estimation by the quasi-conditional maximum likelihood method (function \texttt{tsglm}).

\subsubsection{5.1.2 Combined Linear and Exponential Predictors}
\label{sssec:ComblinExp}

The second method we consider is the combined linear and exponential predictors models proposed in \cite{altieri:20}, which assembles the following three different models.

\begin{enumerate}
	\item An individual county-/state-level exponential predictor: model (\ref{EQN:M1}) uses a series of separate predictors for each county to capture the reported exponential growth of COVID-19 infected and death counts, and we assume
	\begin{equation}
	\log{\{\mathrm{E}(Z_t | t)\}} = \beta_0 + \beta_1 t, 
	\label{EQN:M1}
	\end{equation}
	where the parameters $\beta_0$ and $\beta_1$ are the coefficients in the generalized linear model (GLM) using \texttt{glm} function in \texttt{R} with a log link function.
	
	\item An individual county-/state-level linear predictor: model (\ref{EQN:M5}) fits a linear version of the separate county predictors; specifically, we assume that:
	\begin{equation}
	\mathrm{E}(Z_t | t) = \beta_0 + \beta_1 t.
	\label{EQN:M5}
	\end{equation}
	
	\item An individual county-/state-level exponential epidemic predictor: model (\ref{EQN:M2}) uses a series of disease related factors to capture the reported exponential growth of COVID-19 infectious and death counts. We assume that
	\begin{equation}
	\log{\{\mathrm{E}(Z_t | Z_{t-1})\}} = \beta_0 + \beta_1 \log{(Z_{t-1} +1)}.
	\label{EQN:M2}
	\end{equation}
\end{enumerate}

\subsubsection{5.1.3 Spatio-temporal Epidemic Model}
\label{sssec:stem}

Based on the idea of the SIR models, this paper \citep{Wang:etal:arxiv:20} proposes the discrete-time spatial epidemic model, which combines the susceptible state, infectious state, and removed state together. In the following, we denote $I_{it}$, $D_{it}$, and $R_{it}$ the cumulative number of cases in infected, death and recovered states, respectively, in county $i$ observed on day $t$. Let $\mu_{i,t}^I$, $\mu_{i,t}^D$, $\mu_{i,t}^R$ be the conditional mean value of daily new positive cases, deaths and recovery cases, respectively, which can be modeled via a link function $g$ as follows: 
\begin{eqnarray*}
	g(\mu_{i,t}^I)&=&\beta_{0t}^I (\mathrm{lon}_i, \mathrm{lat}_i) + \beta_{1t}^I (\mathrm{lon}_i, \mathrm{lat}_i) \log{(I_{i,t-1})},\\
	g(\mu_{i,t}^D)&=&\beta_{0t}^D (\mathrm{lon}_i, \mathrm{lat}_i) + \beta_{1t}^D (\mathrm{lon}_i, \mathrm{lat}_i) \log{(I_{i,t-1})},\\
	\mu_{i,t}^R&=&\beta_{0t}^R+\beta_{1t}^R I_{i,t-1}.
\end{eqnarray*}

In practice, one can use the bivariate spline over triangulation to approximate the spatially varying coefficient functions, $\beta_{0t}(\mathrm{lon}_i, \mathrm{lat}_i)$ and $\beta_{1t}(\mathrm{lon}_i, \mathrm{lat}_i)$. The triangulation can be obtained through various software packages; see for example, the \texttt{Matlab} code \texttt{DistMesh}, and the \texttt{R} package \texttt{Triangulation} \cite{Triangulation}. Based on a triangulation, the bivariate spline basis can be generated via the \texttt{R} package \texttt{BPST} \cite{BPST}. The entire estimation procedure is completed using a quasi-likelihood approach via the penalized spline approximation and an iteratively reweighted least-squares technique; see details in \cite{Wang:etal:arxiv:20}. 

\subsection{Outlier Correction}

The consequences of outliers may result in reduced forecast accuracy due to (1) bias in the estimates of model parameters and (2) a carry-over effect of the outlier on the prediction. Consequently, the reported data (i.e., infection, death, and recovery) should undergo a preprocessing step to lessen the impact of inaccurate data or anomalies. Therefore, outlier detection and correction are vital for time series analysis based on the reported COVID-19 data. 

\begin{figure}[!ht]
	\begin{center}
		\begin{tabular}{ccc}
			\includegraphics[width=.4\textwidth]{Death_ts_NewJersey.pdf} &
			\includegraphics[width=.4\textwidth]{Infec_ts_Grimes_Texas.pdf}\\
			(a) & (b) \\
			\includegraphics[width=.4\textwidth]{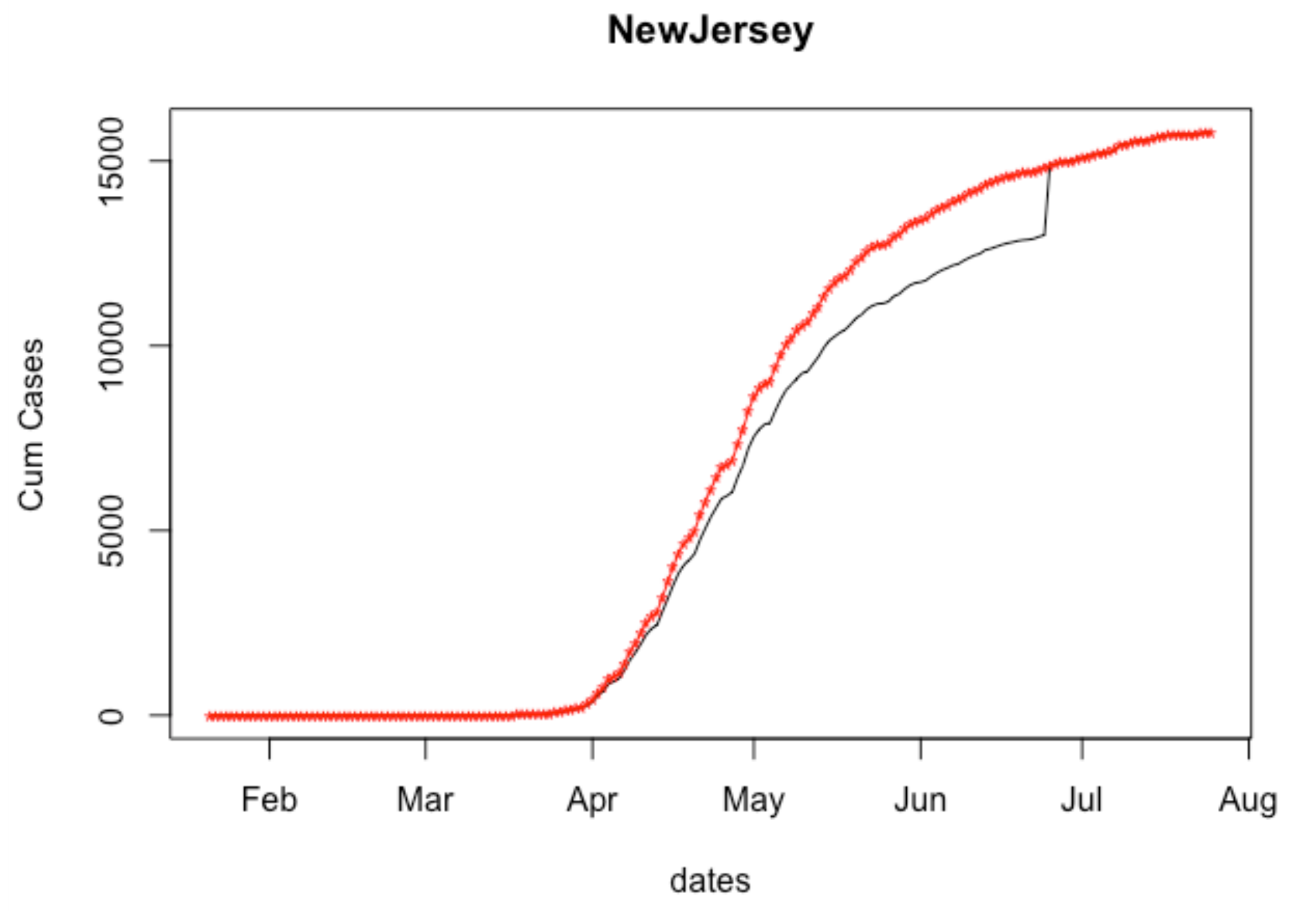}
			&\includegraphics[width=.4\textwidth]{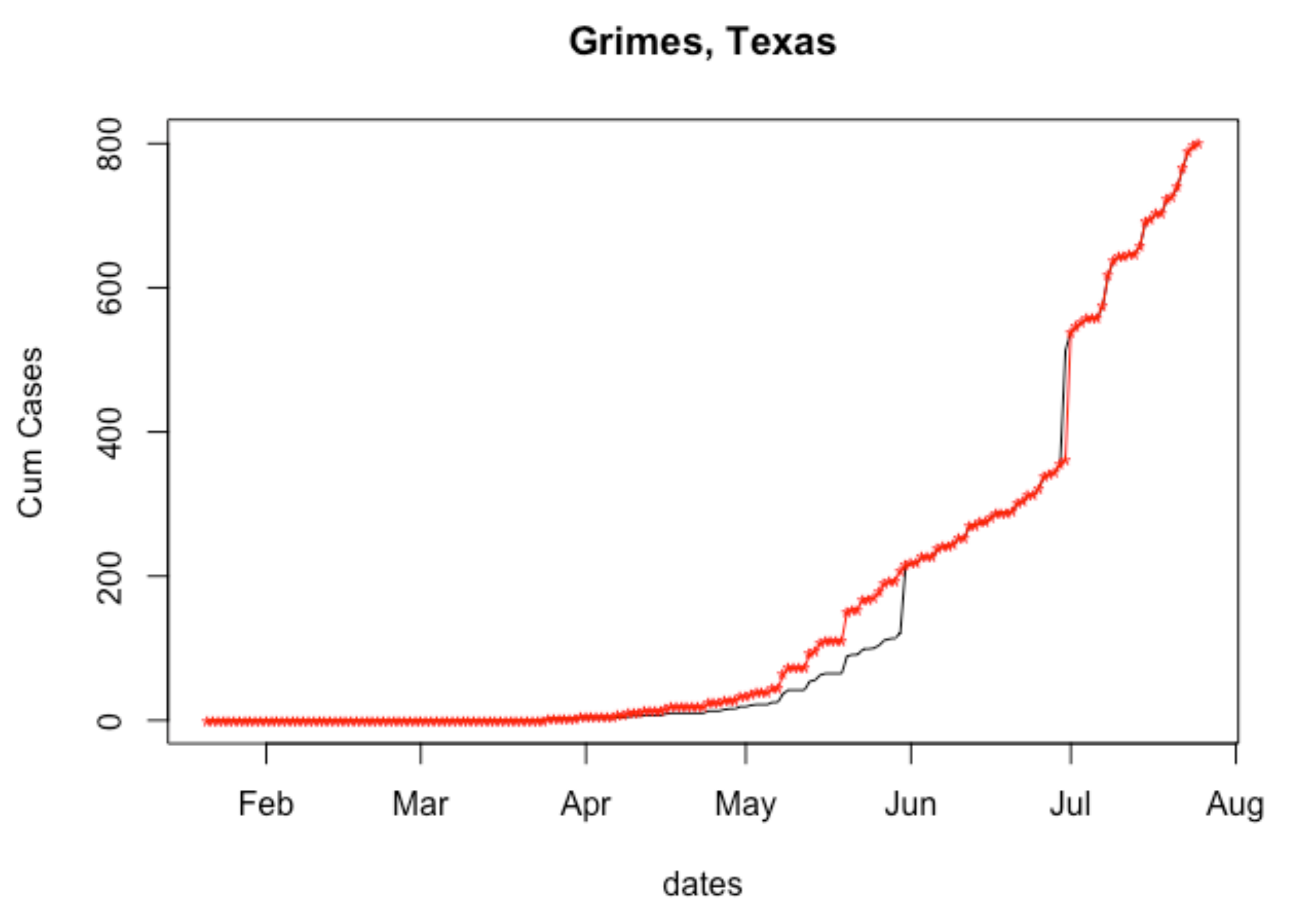} \\
			(c) & (d) \\
			&\includegraphics[width=.4\textwidth]{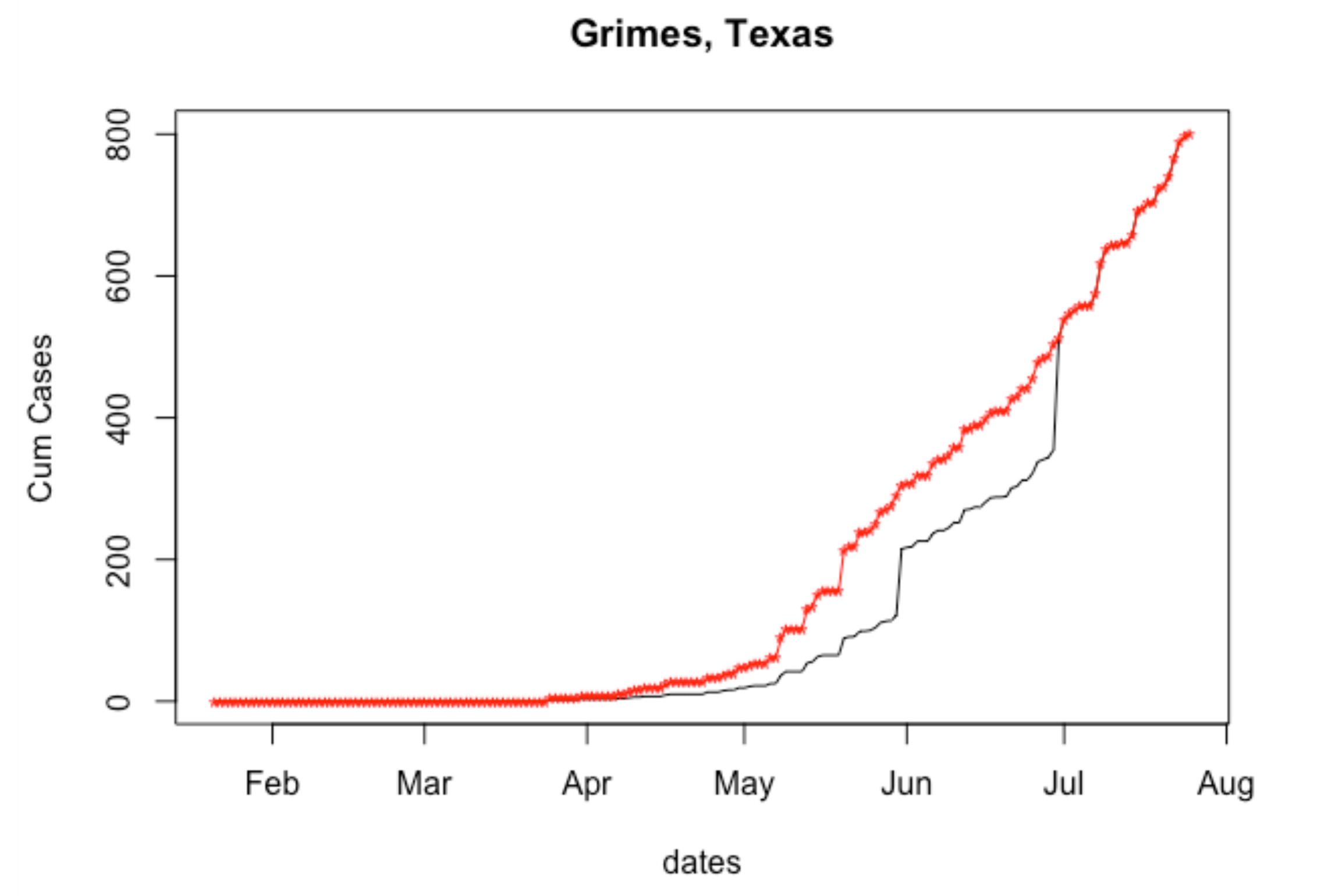}\\
			& (e)
		\end{tabular}
	\end{center}
	\caption{Point anomaly repairing. (a) new death count with one point anomaly; (b) new infected count with two individual point anomalies; (c) point anomalies repairing on the cumulative death count time series; (d) the first point anomaly repairing on the cumulative infected count time series; (e) the second point anomaly repairing on the cumulative infected count time series.} 
	\label{FIG:anomaly_repair} 
\end{figure}

A simple solution to lessen the impact of an outlier is to replace the outlier with a more typical value before generating the forecasts. This process is often referred to as ``Outlier Correction''. Traditionally, an outlier usually occurs due to measurement variability, data entry, or experimental error. Once an outlier is detected, especially for the ones resulted from the experimental error, we sometimes exclude them from the dataset.  However, the outliers in COVID-19 epidemic data typically occur for specific reasons, such as the release of a large batch of tests and the change of reporting standards. Simply excluding extreme values solely due to their extremeness can distort the data analysis. Therefore, the outlier correction procedure should be different from the traditional ones. In this subsection, we describe an automated procedure for ``correcting'' the history before forecasting. For a count time series $\{Z_t\}_{t=1}^T$, we assume that $M$ outliers, $\{Z_{t_m}\}_{m=1}^M$, have been detected, then we implement Algorithm \ref{ALGO:PIRLS} for the repairing procedure. Figure \ref{FIG:anomaly_repair} illustrates some examples of the outlier correction. 

\normalem{
	\begin{algorithm}
		\footnotesize
		\KwData{Count time series with outliers.}
		\KwResult{Count time series with outliers repaired.}
		\SetKwBlock{Begin}{}{}
		\caption{The outlier repairing algorithm.}
		\BlankLine
		\textbf{Step 0.} Sort the $M$ outliers based on the time, i.e. $t_1 < t_2 < \cdots < t_M$.
		
		\For {$m \gets 1$ to $M$}
		{\textbf{Step 1.} Implement the data repairing methods discussed in Section \ref{SUBSEC:ODrepair} to obtain a reasonable estimate $\widehat{Z}_{t_m}$ of the observed data point $Z_{t_m}$.
			
			\textbf{Step 2.} \If {$|Z_{t_m} - \widehat{Z}_{t_m}| > \delta$}
			{(i) Manually investigate the causes and the problematic period caused by the point anomaly $Z_{t_m}$, denoted as $\mathcal{A}_m$.\\
				(ii) Distribute the residual, $Z_{t_m} - \widehat{Z}_{t_m}$, proportional to the value of the time series within the problematic period
				\[
				Z_{t}^{\ast}=Z_{t}+(Z_{t_m} - \widehat{Z}_{t_m}) Z_{t}/\sum_{t\in \mathcal{A}_m} Z_{t},~ t\in \mathcal{A}_m
				\]}}
		\label{ALGO:PIRLS}
	\end{algorithm}
	\ULforem}

\setcounter{chapter}{6} \renewcommand{\thesection}{\arabic{section}} %
\renewcommand{\thetable}{{\arabic{table}}} 
\renewcommand{\thefigure}{\arabic{figure}}
\renewcommand{\thesubsection}{6.\arabic{subsection}} \setcounter{section}{5}

\section{Technical Validation and Usage Notes} \label{SEC:tech}

\begin{figure}
	\centering
	\includegraphics[width = 1\textwidth]{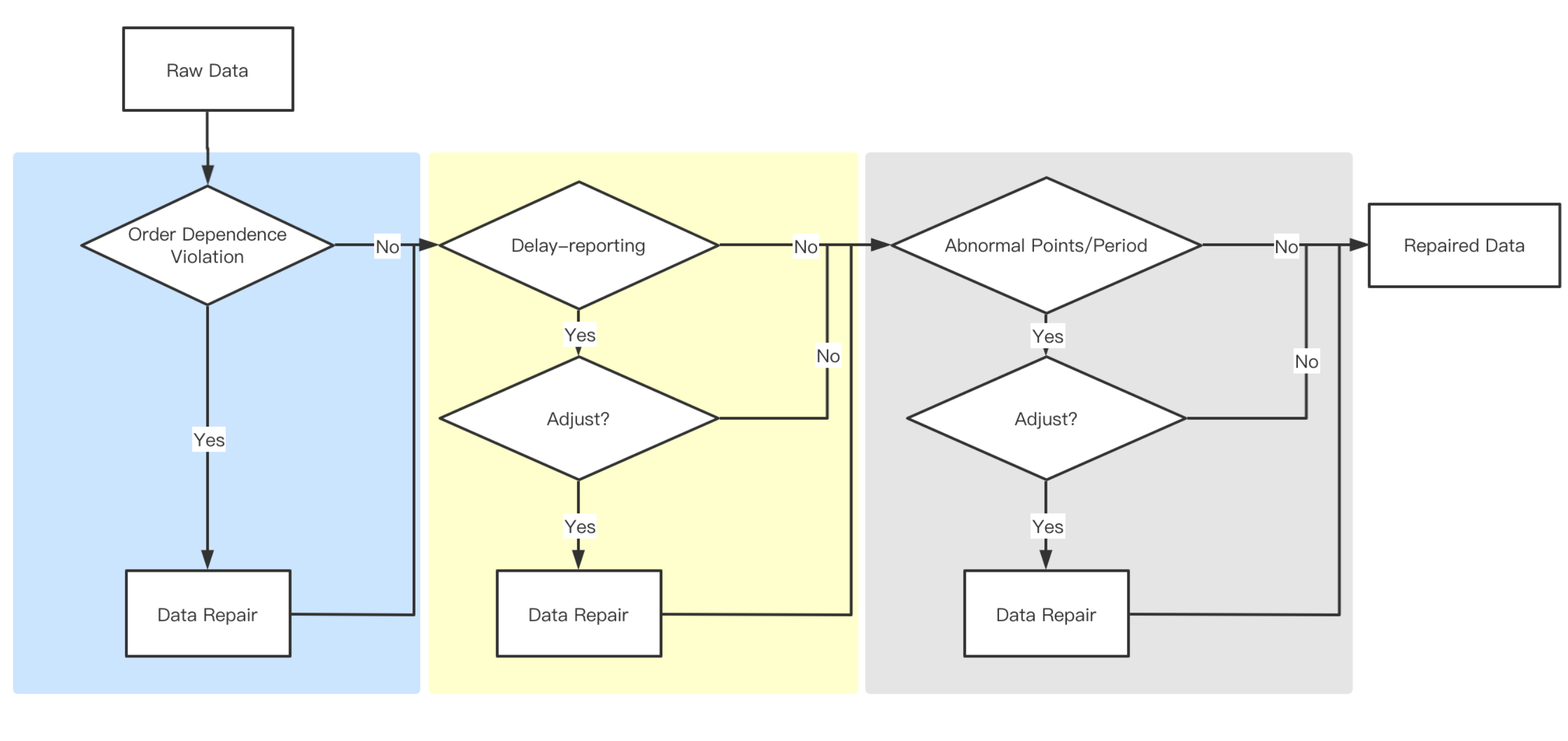}
	\caption{Data curation flowchart.}
	\label{fig:flow} 
\end{figure}

The entire detection and repairing procedure is illustrated in Figure \ref{fig:flow}. First of all, we obtained the data from all of the four data sources, and use the dissimilarity measure proposed in the above to compare them. We visualize and check the difference at the state level among different data sources based on the comparison results. For the county-level data, we calculate the measure and report the top 10 counties, which are the most different pairwisely. Then, all the data are processed with all types of anomaly detection discussed in Section \ref{ssec:detection}. Once an anomaly has been detected, a warning will be given automatically by \texttt{R} package \texttt{cdcar}. We handle different types of anomalies depending on the circumstances. For example, if an order dependency violation is detected, we will repair that point using our data repairing algorithms proposed in Section \ref{SUBSEC:ODrepair}. If a point anomaly is detected, we first manually check possible legitimate reasons based on news and social media. If correction is necessary, we will repair the point anomalies using the proposed algorithm, see Algorithm \ref{ALGO:PIRLS}.

The integrated data are openly available to assist researchers to investigate the spread of COVID-19 in the US. We will continue to provide the cleaned data as the pandemic progresses. Despite the fact we try our best in the data curation, there are two issues without a perfect solution now, but requires attention from the users when they try to draw conclusive statements using the data.
\begin{enumerate}
	\item \textbf{Probable versus Confirmed Cases.} Excluding the population with symptoms but not confirmed by tests leads to the under-reported issue of infectious counts. On April 5, 2020, the Council of State and Territorial Epidemiologists released an interim \cite{CDC:20} related to the COVID-19 reporting. It requires the local or state public health authority to submit a report of a condition to the Centers for Disease Control and Prevention (CDC) within 24 hours, and CDC should publish data for both ``Confirmed'' and ``Probable'' cases in the CDC Print Criteria. Before the interim was released, most of the states primarily reported confirmed cases. For the states and counties which started to report probable cases, thereafter, the count of the cases would incur an unavoidable jump after including the probable cases.
	
	\item \textbf{Antibody Test versus Virus Test.} In general, there are two types of tests on infection\red{;} one is an antibody test, and the other is a virus test (also referred to as the PCR test). Unfortunately, in many datasets, the type of reported tests is not specified. Those tested positive for the virus are infected at the moment and suggested to be quarantined to avoid infecting others. Meanwhile, those tested positive on an antibody test must have been exposed to the virus, but there is no indication of whether they are still infectious or recovered \cite{CDC:20:b}.  In addition, the antibody tests are known to be much less accurate. Mixing these two tests makes positive cases uninterpretable. Some states and counties have started to separate antibody tests from virus tests \cite{Atlantic:20}, while states such as Pennsylvania, Texas, Georgia, and Vermont did not specify the type of tests. 
\end{enumerate}

We will continue to keep close track of the data sources we depend on, and update our datasets regularly. We strongly encourage users of our datasets to contact us if there is any anomaly or error. You can reach us either by submitting a request on the Github repository (\url{https: //github.com/covid19-dashboard-us/cdcar}) or emailing the corresponding author.

\setcounter{chapter}{7} \renewcommand{\thesection}{\arabic{section}} %
\renewcommand{\thetable}{{\arabic{table}}} \setcounter{table}{0} %
\renewcommand{\thefigure}{\arabic{figure}} \setcounter{figure}{0} %
\renewcommand{\thesubsection}{7.\arabic{subsection}} \setcounter{section}{6}

\section{Conclusion and Discussion} \label{SEC:discussion}

The COVID-19 pandemic is generating enormous amounts of data. Open-access data with high quality are critical for the COVID-19 scientific research and response efforts. This paper compares and integrates different data sources and provides a semiotic-based framework for understanding the reported cases' data quality and the techniques for anomaly detection and anomaly repairing.

Correcting the history for severe outliers or anomalies will often improve the forecast; however, if the outlier is not genuinely severe, corrections might make the history smoother than it actually was, which will change the forecasts and narrow the confidence intervals. If the correction was not necessary, it might lead to poor forecasts and unrealistic confidence intervals. We suggest using a high threshold for anomaly detection. In addition, the detected outliers should ideally be individually reviewed by the forecaster, and the reasons for the outliers should be investigated to determine whether a correction is appropriate.

For public usage, all code regarding the proposed anomaly detection and repairing algorithms is built in \texttt{R} package \texttt{cdcar}. The package and the cleaned data that are regularly updated can be found in the Github repository (\url{https://github.com/covid19-dashboard-us/cdcar}). 

Some aspects of our data comparison and curation methods are constrained by the official information released, and we will continuously investigate. First of all, some data are not retrievable at the county level, such as the recovery data and the face mask data. Second, the data reporting protocols are not consistent, especially for the recoveries. Third, although we discover the cyclical pattern in the epidemic data, the related reason still needs to be examined. Fourth, the unassigned issue and under-reported issue might be some other important problems under the pandemic without detailed discussion. In the future, we plan to overcome the data sharing barrier and extend our US COVID-19 database to a worldwide database.

\section*{Acknowledgements}

The authors are grateful to the referee, Associate Editor and Editor for careful reading of the paper and valuable suggestions and comments. Zhiling Gu and Li Wang's research was partially supported by National Science Foundation award DMS-1916204. Shan Yu's research was partially supported by the Iowa State University Plant Sciences Institute Scholars Program. Myungjin Kim's research was partially supported by National Science Foundation award DMS-1934884 and Laurence H. Baker Center for Bioinformatics \& Biological Statistics.

\section*{Disclosure statement}

No potential conflict of interest was reported by the authors.

\newpage
\renewcommand{\thetable}{{\arabic{table}}} \setcounter{table}{0} %
\renewcommand{\thefigure}{\arabic{figure}} \setcounter{figure}{0} %
\renewcommand{\thesubsection}{A.\arabic{subsection}} 
\section*{Appendix.  Data Records}
\renewcommand{\thesection}{A} 

\subsection{Epidemic Data}
Using the algorithm discussed in Section \ref{SEC:repair}, we aggregate the reported COVID-19 infected, death, and recovered cases from January 22, 2020 from (1) the NYT \cite{NYT:20}, (2) the Atlantic \cite{COVIDTrack}, (3) the COVID-19 Data Repository from the JHU \cite{JHUCSSE}, and (4) the USAFacts \cite{USAFact:20}. These daily updated epidemic datasets are available on Github repository \url{https://github.com/covid19-dashboard-us/cdcar}. 

\noindent In the state level epidemic data, we include the following variables. Among those variables, the variable \textbf{State} can be used as the key for data merge.
\begin{enumerate}
	\item \textbf{State} -- Name of state. There are 48 mainland US states and the District of Columbia.
	
	\item \textbf{XYYYY.MM.DD} -- Cumulative infection or death cases related to the date of \textbf{YYYY.MM.DD}. \textbf{YYYY}, \textbf{MM}, and \textbf{DD} represent year, month and day, respectively. It starts from \textbf{X2020.01.22}. For example, the variable \textbf{X2020.01.22} is either infection or death cases in a certain state (\textbf{State}) on 01/22/2020.
\end{enumerate}

\noindent  For county-level data, two more county-specific variables are included. As the key of this table, variable \textbf{ID} can be used for future data merge.
\begin{enumerate}
	\item \textbf{ID} -- County-level Federal Information Processing System (FIPS) code, which uniquely identifies the geographic area. The number has five digits, of which the first two are the FIPS code of the state to which the county belongs.
	
	\item \textbf{County} -- Name of county matched with \textbf{ID}. There are about 3,200 counties and county-equivalents (e.g. independent cities, parishes, boroughs) in the US.
	
	\item \textbf{State} -- Name of state matched with \textbf{ID}. There are 50 states and the District of Columbia in the US.
	
	\item \textbf{XYYYY.MM.DD} -- Cumulative infection or death cases related to the date of \textbf{YYYY.MM.DD}. \textbf{YYYY}, \textbf{MM}, and \textbf{DD} represent year, month and day, respectively. It starts from \textbf{X2020.01.22}. For example, the variable \textbf{X2020.01.22} is either infection or death cases in a certain (\textbf{County}) on 01/22/2020.
\end{enumerate}

\subsection{Other Factors}

\subsubsection{A.2.1 Policy Data}
We release two datasets for the ``stay-at-home/shelter-in-place'' order and the declaration of ``state of emergency'' from (1) Business Insider \cite{BusiInsi:20}, (2) New York Times \cite{NYTSAH2020}, and additional local news. 

\begin{enumerate}
	\item \textbf{ID} -- County-level Federal Information Processing System (FIPS) code, which uniquely identifies the geographic area. The number has five digits, of which the first two are the FIPS code of the state to which the county belongs.
	
	\item \textbf{County} -- Name of county matched with \textbf{ID}. There are about 3,200 counties and county-equivalents (e.g. independent cities, parishes, boroughs) in the US.
	
	\item \textbf{State} -- Name of state matched with \textbf{ID}. There are 50 states and the District of Columbia in the US.
	
	\item \textbf{XYYYY.MM.DD} -- Indicators for whether the policy is in effect on the date of \textbf{YYYY.MM.DD}, $1$ indicates the policy is in effect, $0$ otherwise. \textbf{YYYY}, \textbf{MM}, and \textbf{DD} represent year, month and day, respectively. It starts from \textbf{X2020.01.22}. For example, the variable \textbf{X2020.01.22} represents whether the policy is in effect in a certain (\textbf{County}) on 01/22/2020.
\end{enumerate}

\subsubsection{A.2.2 Demographic Characteristics}
In the demographic characteristics category, we consider the factors describing racial, ethnic, sexual, and age structures. Specifically, we include the following six variables. Among these six variables, \textbf{AA\_PCT} and \textbf{HL\_PCT} are obtained from the 2010 Census \cite{DC10}. The other four variables are extracted from the 2010--2018 American Community Survey (ACS) Demographic and Housing Estimates \cite{ACS1018}.
\begin{enumerate} 
	\item \textbf{AA\_PCT} -- The percent of the population who identify as African American;
	
	\item \textbf{HL\_PCT} -- The percent of the population who identify as Hispanic or Latino; 
	
	\item \textbf{Old\_PCT} -- The percent of aged people (age $\geq 65$ years); 
	
	\item \textbf{Sex\_ratio} -- The ratio of male over female; 
	
	\item \textbf{PD\_log} -- The logarithm of the population density per square mile of land area;
	
	\textbf{Pop\_log} -- The logarithm of local population;
	
	\item \textbf{Mortality} -- The five-year (1998-2002) average mortality rate, measured by the total counts of deaths per $100,000$ population in a county.
\end{enumerate}

\subsubsection{A.2.3 Healthcare Infrastructure}
We incorporated three features related to the healthcare infrastructure at the county level in the datasets. Among these variables, \textbf{NHIC\_PCT} is available in the USA Counties Database \cite{uscounties}, \textbf{EHPC} is obtained from Economic Census 2012 \cite{EC12}, and \textbf{TBed} is compiled from Homeland Infrastructure Foundation-level Data \cite{HIFLD}.
\begin{enumerate}
	\item \textbf{NHIC\_PCT} -- the percent of persons under $65$ years without health insurance;
	
	\item \textbf{EHPC} -- the local government expenditures for health per capita;
	
	\item \textbf{TBed} -- total bed counts per $1,000$ population.
\end{enumerate}

\subsubsection{A.2.4 Socioeconomic Status}
We consider diverse socioeconomic factors in the county level datasets. All of these factors collected from 2005--2009 ACS 5-year estimates \cite{ACS0509}. We also calculate the Gini coefficient based on the household income data from the 2005--2009 ACS \cite{ACS0509} to measure the income inequality.
\begin{enumerate}
	\item \textbf{Affluence} -- Social affluence generated by factor analysis from \textbf{HighIncome}, \textbf{HighEducation}, \textbf{WCEmployment} and \textbf{MedHU}; 
	
	\item \textbf{HIncome\_PCT} -- The percent of families with annual incomes higher than \$75,000;
	
	\item \textbf{HEducation\_PCT} -- The percent of the population aged 25 years or older with a bachelor's degree or higher;
	
	\item \textbf{MedHU} -- The median value of owner-occupied housing units;
	
	\item \textbf{Disadvantage} -- Concentrated disadvantage obtained by factor analysis from \textbf{HHD\_PAI\_PCT}, \textbf{HHD\_F\_PCT} and \textbf{Unemployment\_PCT};
	
	\item \textbf{HHD\_PAI\_PCT} -- The percent of the households with public assistance income;
	
	\item \textbf{HHD\_F\_PCT} -- The percent of households with female householders and no husband present;
	
	\item \textbf{Unemployment\_PCT} -- Civilian labor force unemployment rate;
	
	\item \textbf{Gini} -- The Gini coefficient, a measure for income inequality and wealth distribution in economics.
\end{enumerate}

\subsubsection{A.2.5 Environmental Factor}
We also collect some environmental factors that might affect the spread of epidemics significantly, such as the urban rate and crime rate.
\begin{enumerate}
	\item \textbf{UrbanRate} -- Urban rate \cite{DC10};
	
	\item \textbf{ViolentCrime} -- The total number of violent crimes per $1,000$ population \cite{uscounties};
	
	\item \textbf{PropertyCrime} -- The total number of property crimes per $1,000$ population \cite{uscounties};
	
	\item \textbf{ResidStability} -- The percent of the population residence in the same house for one year and over \cite{DC10}.
\end{enumerate}

\subsubsection{A.2.6 Mobility} 
The mobility data are collected from the US Department of Transportation, Bureau of Transportation Statistics. It describes the daily number of trips within each county, which are produced from an anonymized national panel of mobile device data from multiple sources. Trips are defined as movements that include a stay of longer than 10 minutes at an anonymized location away from home.

\begin{enumerate}
	\item  \textbf{Number of trips X -- XX} -- Number of trips by residents greater than X miles and shorter than XX miles. There are 10 different trip ranges: `` $\leq$ 1''	``1 -- 3'', ``3 -- 5'', ``5--10'', ``10 -- 25'', ``25 -- 50''	``50 -- 100'', ``100 -- 250'', ``250 -- 500'', and `` $\geq$ 500''.
	\item  \textbf{Population Stay at Home} -- Number of residents staying at home, that is, persons who make no trips with a trip end more than one mile away from home. 
\end{enumerate}

\subsection{Geographic Information}
The \textbf{longitude} and \textbf{latitude} of the geographic center for each county in the US are available in Gazetteer Files \cite{gazetteer}.

\bibliographystyle{asa}
\bibliography{references}


\end{document}